\documentclass[12pt]{article}
\pdfoutput=1

\usepackage{putex}
\usepackage{amsmath,amssymb,amsfonts}
\usepackage{psfrag}
\usepackage{footmisc}
\usepackage{url}
\usepackage{mathtools}
\usepackage{color}

\usepackage{tikz}
    \usepackage{amssymb,amsfonts,amsmath}
    \usepackage{tkz-euclide}
        \usetikzlibrary{arrows,calc,patterns}
\usepackage{pgfplots}

\definecolor{darkblue}{rgb}{0.1,0.1,.7}
\definecolor{purple}{rgb}{0.6,0,0.6}
\definecolor{orange}{rgb}{0.9,0.6,0}
\usepackage[colorlinks, linkcolor=darkblue, citecolor=darkblue, urlcolor=darkblue, linktocpage]{hyperref} 
\usepackage[square, comma, compress,numbers]{natbib}
\usepackage[]{amsmath}
\usepackage[]{graphicx}
\usepackage[]{latexsym}
\usepackage[utf8]{inputenc}
\usepackage{geometry}
\usepackage{amscd}
\usepackage[all,cmtip]{xy}
\usepackage{mathrsfs}

\usepackage[margin=10pt,font=small,labelfont=bf]{caption}
\usepackage{changepage}
\usepackage{setspace}
\usepackage{tikz}
\usepackage{subcaption}
\usepackage{comment}
\usepackage[titles]{tocloft}
\usetikzlibrary{arrows,positioning}

\def\SL2{\widetilde{SL}(2,\mathbb R)}

\numberwithin{equation}{section}

\newcommand{\grp}[1]{\mathrm{#1}}

\newcommand {\bes} {\begin {equation*}}
\newcommand {\ees} {\end {equation*}}
\newcommand {\beq} {\begin {equation}}
\newcommand {\eeq} {\end {equation}}
\newcommand {\bea} {\begin {eqnarray}}
\newcommand {\ea} {\end {eqnarray}}
\newcommand {\eea} {\end {eqnarray}}

\usepackage{capt-of}

\numberwithin{equation}{section}

\def\<{\langle}
\def\>{\rangle}

\tikzset{
    >=stealth',
    punkt/.style={
           rectangle,
           rounded corners,
           draw=black, very thick,
           text width=15em,
           minimum height=2em,
           text centered},
    pil/.style={
           ->,
           thick,
           shorten <=2pt,
           shorten >=2pt,}
}

 \def\ie{\begin{equation}\begin{aligned}}
\def\fe{\end{aligned}\end{equation}}

\begin{document}

\institution{IAS}{${}^1$  Institute for Advanced Study, Princeton, NJ 08540, USA}
\institution{PU}{${}^2$ Joseph Henry Laboratories, Princeton University, Princeton, NJ 08544, USA \\
}

\title{
Phases of $\mathcal{N}=2$ Sachdev-Ye-Kitaev models
}

\authors{M.  Heydeman,${}^{1,2}$~~ G. J. Turiaci,${}^1$~~ W. Zhao${}^2$}
\vspace{1cm}

\abstract{
We study $\mathcal{N}=2$ supersymmetric Sachdev-Ye-Kitaev (SYK) models with complex fermions at non-zero background charge. Motivated by multi-charge supersymmetric black holes, we propose a new $\mathcal{N}=2$ SYK model with multiple $U(1)$ symmetries, integer charges, and a non-vanishing supersymmetric index, realizing features not present in known SYK models. In both models, a conformal solution with a super-Schwarzian mode emerges at low temperatures, signalling the appearance of nearly AdS$_2$/BPS physics. However, in contrast to complex SYK, the fermion scaling dimension depends on the background charge in the conformal limit. For a critical charge, we find a high to low entropy phase transition in which the conformal solution ceases to be valid. This transition has a simple interpretation-- the fermion scaling dimension violates the unitarity bound.  We offer some comments on a holographic interpretation for supersymmetric black holes.
}
\vspace{2cm}
\date{{\small Email:~\texttt{mheydeman@ias.edu, turiaci@ias.edu, wz10@princeton.edu}}}

\maketitle
\tableofcontents


\section{Introduction}
The Sachdev-Ye-Kitaev (SYK) models \cite{Sachdev:1992fk, kitaevTalks} of randomly interacting fermions are the simplest known quantum mechanical systems without quasi-particles that display maximal chaos. In a seemingly unrelated arena of gravitational physics, a charged black hole near the extremal ``mass = charge'' limit develops an approximate two-dimensional anti-de Sitter (AdS$_2$) region. The physics of AdS$_2$ displays chaos, scrambling, and an approximate conformal symmetry. In light of the AdS/CFT correspondence, these features (among others) make SYK models candidate toy descriptions of quantum black holes, following some earlier insight given by \cite{Sachdev:2010um, Sachdev:2015efa}. For applications to charged black holes, one may consider of SYK models with $N$ complex fermions and at least one conserved $U(1)$ symmetry. A model of this kind is the complex SYK model (cSYK), where a set of four or more fermions are coupled through a random interaction drawn from a fixed Gaussian ensemble. See \cite{Gu:2019jub} for a detailed exposition of this model.

In this paper we will focus on a simple (but ultimately drastic) modification of the complex SYK model, where the Gaussian random interaction is roughly replaced by its covariance matrix. A precise version of this modification, which we will review below, was introduced by Fu, Gaiotto, Maldacena, and Sachdev in \cite{Fu:2016vas}. The seemingly small modification of the random Hamiltonian allows one to write it as the square of a complex (Gaussian random) operator-- the combination of the conserved supercharges $\mathcal{Q}+\bar{\mathcal{Q}}$, which leads to very different physics. In particular, the model in fact possesses $\mathcal{N}=2$ supersymmetry; the $U(1)$ phase rotation of the fermions does not commute with supersymmetry and thus becomes an R-symmetry. Supersymmetry typically relates bosons and fermions via the action of $\mathcal{Q}$, however, as noted in \cite{Fu:2016vas}, a supersymmetric model consisting only of fermions may be achieved as long as supersymmetry is non-linearly realized. 

While the proposal in \cite{Fu:2016vas} seems to be the simplest generalization of SYK to incorporate $\mathcal{N}=2$ supersymmetry, for reasons that will be explained shortly, we will sometimes refer to this model as $\mathcal{N}=2$ SYK with fractional charges. Unlike cSYK and even $\mathcal{N}=1$ versions, the model with extended supersymmetry possesses states annihilated by one complex supercharge; these BPS states have a large exact degeneracy which survives at strong coupling (this feature of the strong coupling partition function was observed in \cite{Stanford:2017thb, Mertens:2017mtv}). It is natural to conjecture there is some relationship between a large BPS degeneracy in supersymmetric SYK and the degeneracy of supersymmetric AdS$_2$ black holes \cite{Heydeman:2020hhw, Boruch:2022tno}. However, the $\mathcal{N}=2$ SYK model with fractional charges does not have all of the expected features of these near-BPS black holes-- a point we will come back to later.

The first goal of this paper will be to perform a more exhaustive study of this $\mathcal{N}=2$ SYK model, in particular at non-zero R-charge, or equivalently a non-zero chemical potential. For example, in \cite{Fu:2016vas} and subsequent work, the focus was placed on analyzing the model in an ensemble of zero $U(1)$ R-charge \cite{Stanford:2017thb, Mertens:2017mtv, Peng:2017spg, Kanazawa:2017dpd, Marcus:2018tsr, Bulycheva:2018qcp, Peng:2020euz, Berkooz:2020xne}. Because the background R-charge does not commute with supersymmetry, this choice of zero charge ensemble preserves supersymmetry and dramatically simplifies the solution of the model. We generalize this analysis and will uncover an interesting phase structure as a function of the charge when supersymmetry is broken.

We will contrast the results obtained in a non-zero R-charge sector of $\mathcal{N}=2$ SYK with the solution of complex SYK. In particular we are interested in the strong coupling limit, or equivalently temperatures smaller than the scale set by the variance of the couplings. In this limit we expect an emergent conformal symmetry at the level of correlation functions between the fermions; this is the infrared reparametrization symmetry. It is spontaneously broken to $SL(2,\mathbb{R})$ by the associated ``conformal solution" of the Schwinger-Dyson equations which determines the two-point functions. In complex SYK the scaling dimension of the fermions is simply obtained by dimensional analysis arguments and has a universal value 
independent of the charge density. 

In $\mathcal{N}=2$ SYK, the model possesses not just time translation and $U(1)_R$ symmetries, but also a complex supercharge which is charged under the $U(1)_R$. In the infrared, we will show that there is an emergent super-reparametrization symmetry, that is spontaneously broken to $SU(1,1|1)$ superconformal symmetry by the conformal solution. The superconformal group $SU(1,1|1)$ extends the $SL(2,\mathbb{R})$ with bosonic subgroups $SL(2,\mathbb{R}) \times U(1)_R \subset SU(1,1|1)$. This superconformal symmetry is further broken spontaneously by the background R-charge, and because of this, the scaling dimension of the fermion is no longer fixed by dimensional analysis. Instead, one needs to look more carefully at the solution of the model to derive a constraint that uniquely determines the scaling dimension $\Delta$. The result gives a non-trivial function of the charge which can be found in equation \eqref{eq:FuDeltaEqn} (this expression was independently derived by S. Sachdev \cite{SSN2}). Another feature is that the coefficient of the fermion two-point function in the conformal limit is undetermined from the IR at non-zero charge (when the charge is zero it is fully determined by supersymmetry). We will verify numerically that this parameter of the solution is fixed once the UV behavior of the model is incorporated, consistent with the idea that there is a unique two-point function at strong coupling. We also derive a Luttinger-Ward relation given in equation \eqref{eq:LWFuetalmt} and the grand potential in \eqref{eq:fullGE} (by proposing a ``spooky fermion" picture for $\mathcal{N}=2$ SYK similar to \cite{Gu:2019jub}), and verify numerically its validity. Finally, we analyze the spectrum of bilinear operators within the conformal solution at any charge.

This takes us to the second departure between complex and $\mathcal{N}=2$ SYK. While the conformal solution of complex SYK is well-behaved for any charge, we find a critical charge in $\mathcal{N}=2$ SYK for which the scaling dimension of the fermion and the overall coefficient of the two-point function vanish. At higher values of the charge, the scaling dimension becomes either negative or complex, signaling the fact that this solution is no longer physical (see for example figure \ref{fig:DeltaFuEtAl}). Instead, for these charges the model develops a gap and behaves as a set of massive complex free fermions. We note that the phenomenon is distinct from the complex modes already observed in the literature such as in \cite{Klebanov:2018fzb,Kim:2019upg,Klebanov:2020kck}, where some bilinear operator becomes complex. In our case, the fundamental fermion directly gets a dimension that violates unitarity. 

This phase transition we found in $\mathcal{N}=2$ SYK is analogous to a similar phenomena observed in \cite{Azeyanagi:2017drg,Ferrari:2019ogc,Louw:2022njq} for complex SYK. Those references pointed out that even in complex SYK there is a phase transition at non-zero charge where the conformal solution stops dominating and a gap develops. This leads to a new phase referred to as a ``low entropy phase". This is in contrast to the conformal phase which is ``high entropy", in the sense that there is a large order $N$ zero temperature entropy (at least when the temperature is taken to zero after taking the large $N$ limit, in this order). Even though there is a phase transition, it is an open problem to understand analytically the origin of this phase transition since nothing seems to go wrong with the conformal solution. Some observations on this direction were made in \cite{Gu:2019jub} which show some unusual features in the four-point function kernel when the charge is too large, but the analysis is not conclusive and does not determine the precise value of the critical charge. In contrast we can understand this transition much better in $\mathcal{N}=2$ SYK. Now there is a sharp issue with the conformal solution at finite charge since the fermion itself develops an instability. We give some numerical evidence that the charge at which the fermion becomes unstable corresponds to the same charge at which there is a high- to low-entropy phase transition.

\begin{figure}[h!]
    \centering
    \includegraphics[scale=0.5]{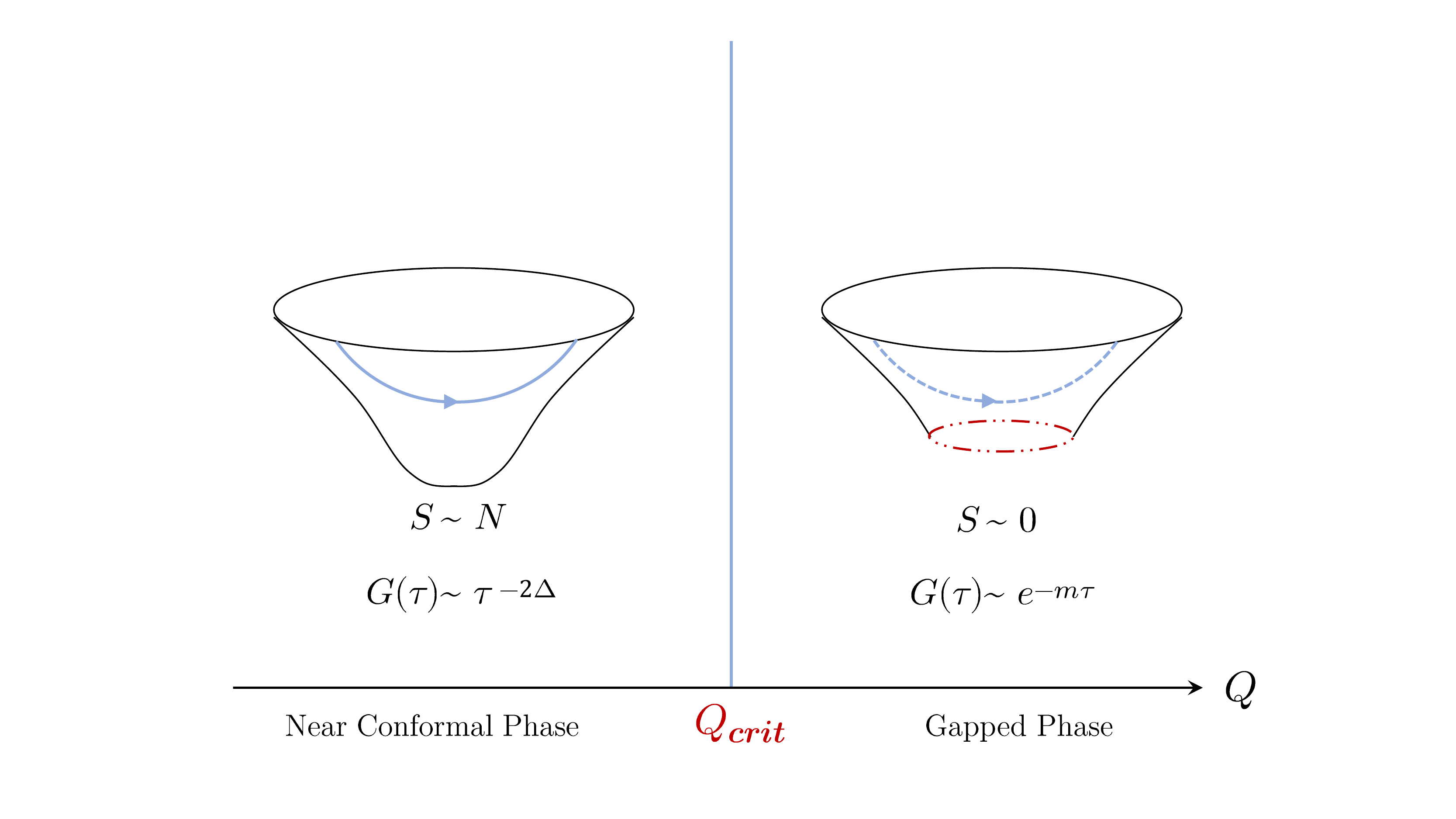}
    \caption{A depiction of the transition of $\mathcal{N}=2$ SYK, along with a schematic dual bulk geometries in each phase. At background charge $Q=0$, it was shown in \cite{Fu:2016vas} that the model possesses a superconformal solution. We show that for $Q<Q_{crit}$ supersymmetry may be lost (depending on which model is under consideration), but a conformal solution exists. This system is dual to a (possibly supersymmetric) AdS$_2$ geometry with a Euclidean horizon and a large extremal entropy. Schematically, a single boundary fermion two point function is computed by a charged fermion in AdS in which the semi-classical scaling dimension depends on the background charge. However, we find above a critical background charge, the dimension of the fermion violates the unitarity bound and the conformal solution ceases to be valid. In SYK, we find a new massive phase with vanishing large $N$ entropy. We conjecture the bulk dual is a horizon-less geometry. }
    \label{fig:PhaseDiag}
\end{figure}

There is a nice analogy between this phase transition and black hole physics. The phase diagram described above can be compared with the one for a near extremal charged black hole, which develops an AdS$_2$ throat close to its horizon. This black hole is unstable towards discharging when a charged fermion is present with a mass (in AdS$_2$ units) smaller than the electric field, even though the black hole is perfectly stable in the vacuum. This happens through Schwinger pair production, see for example \cite{Pioline:2005pf}. A similar phenomenon can also occur with scalar fields when their effective mass is below their Breitenlohner-Freedman (BF) bound close to the horizon \cite{Gubser:2008px}. The supersymmetric model seems to have more in common with the candidate bulk description than complex SYK, since we find an electric field dependent scaling dimension and an instability. Thus, we can interpret the appearance of a new boundary phase at a critical charge as the endpoint of the black hole stability. Since conformal symmetry is lost, the bulk spacetime will be drastically changed and we can no longer use a dual description as a disc of Euclidean AdS$_2$ with $SL(2,\mathbb{R})$ isometry. The fact that the entropy is low in the new phase suggests that the horizon completely disappears. A schematic picture of the proposal for the bulk is in figure \ref{fig:PhaseDiag}.

 We can now move to the second part of the paper. So far, we have discussed the conformal solution (and when it becomes unstable). However, even though the solution of the SYK model develops a conformal symmetry in the IR, this symmetry is both spontaneously and explicitly broken. The breaking of the conformal symmetry dominates the dynamics of the model at these scales, explaining the maximal chaos for example \cite{kitaevTalks}. Similarly, in $\mathcal{N}=2$ SYK there is a supersymmetric Schwarzian mode controlling the breaking of superconformal symmetry. We verify a similar statement may be made at finite charge before the phase transition.
 
However, there is an essential restriction on the $\mathcal{N}=2$ SYK model thusfar considered in the literature, and this restriction continues to the $\mathcal{N}=2$ Schwarzian theory one derives from this model. As we mentioned previously, the model of \cite{Fu:2016vas} has the particular feature that the fermionic charge is fractional compared to the R-charge of the supercharge\footnote{In $\mathcal{N}=2$ SYK, $q$ labels the number of fermions appearing in the supercharge, and not the number of fermions appearing in the Hamiltonian. If we normalize the supercharge to have unit R-charge, the fundamental fermion has charge $1/q$. This is consistent with charge quantization for abelian groups.}. This leads to a Schwarzian theory with fractional charges. An interesting feature of these models of fractional charge fermions is that the BPS states come with a range of charges. This may be seen directly in the grand partition function $Z(\beta, \mu) = \textrm{Tr}[e^{-\beta H + \beta \mu Q}]$, and in particular can lead to cancellations in the supersymmetric index $\mathcal{I}=\textrm{Tr}[(-1)^F]$ which counts BPS states weighted by their fermion number. For example, in the large $N$ limit one obtains the super-Schwarzian theory which may be solved exactly-- \cite{Stanford:2017thb, Mertens:2017mtv} gives a density of states of the BPS sector for even $N$ given by
\beq\label{eq:introdosbps}
Z(\beta,\mu) = \sum_{Q \in \mathbb{Z},~|Q|<\frac{q}{2}} e^{\beta \mu Q}~ \frac{2e^{N s_0}\cos( \frac{\pi Q}{q} )}{q} + \ldots 
\eeq
where $s_0 = \log ( 2 \cos \frac{\pi}{2q})$. (For odd $N$ the same expression is valid but now the charge ranges over half-integers instead of integers.) The dots denote the non-BPS contributions to the spectrum. For even $N$ they are separated on average by a gap given by $E_{\rm gap}= 0.412 J/N$ (obtained for the $q=3$ model using the Schwarzian coupling computed in section \ref{ssec:fNumericalSD}). For odd $N$ the gap is exponentially small in $N$ instead. This is written with respect to the charge $Q$ that assigns unit charge to the fermions. The R-charge $Q_R$ assigning unit charge to the supercharge is related to it by $Q_R = Q/q$. The formula \eqref{eq:introdosbps} happens to be exact for $q=3$ and $N$ even or $N=3~{\rm mod}~4$, but for $q>3$ is only approximate in the large $N$ limit. We would like to emphasize that even though we are describing the density of BPS states, their R-charge distribution \eqref{eq:introdosbps} is not protected by supersymmetry. For example, because of the spread in charges the Witten index of this theory vanishes, something that can be reproduced by neglecting the interaction between the fermions (the index being independent of the temperature and the couplings).

 As has been touched upon already, the Schwarzian theory also appears in the context of near extremal black holes in higher dimensions, describing both the classical dynamics of excitations above extremality \cite{Sachdev:2015efa, Almheiri:2016fws, Anninos:2017cnw, Turiaci:2017zwd, Nayak:2018qej, Moitra:2018jqs, Hadar:2018izi, Castro:2018ffi, Larsen:2018cts,  Moitra:2019bub, Sachdev:2019bjn, Hong:2019tsx, Castro:2019crn, Charles:2019tiu,Larsen:2020lhg} and quantum effects that become large as we take the extremal limit \cite{Ghosh:2019rcj, Iliesiu:2020qvm, Maxfield:2020ale, Heydeman:2020hhw, Boruch:2022tno}. In particular, it was recently shown that a black hole in AdS$_5 \times S^5$ develops an emergent $SU(1,1|1)$ symmetry in the BPS limit in \cite{Boruch:2022tno}. However, the Schwarzian theory describing its near-BPS dynamics is not the same as the ones constructed in \cite{Fu:2016vas} (Some interesting questions are being raised regarding how to understand the BPS black hole microstates in $\mathcal{N}=2$ JT gravity in \cite{Lin:2022zxd, Lin:2022rzw}). Instead, the charge of the fundamental fermion is equal to the charge of the supercharge. One can then use the same formula for the large $N$ spectrum \eqref{eq:introdosbps} with $q=1$, so that the BPS states have only zero R-charge and the index and degeneracy match. 
 
Naively setting $q=1$ does not make sense from the SYK perspective-- with only fundamental fermions and a supercharge linear in them, the model would become trivial and not display any emergent conformal symmetry. To circumvent this we will instead construct models with multiple fermions\footnote{We thank E. Witten for this suggestion.} and show one can then define interacting theories described at low energies by unit fundamental R-charge\footnote{In three spacetime dimensions, related models with and without supersymmetry and multiple fields were constructed in \cite{Chang:2021wbx, Chang:2021fmd}. The presence of dynamical bosons in higher dimensions can lead to different conclusions about the IR effective solution compared to what is discussed in this work.}. We study in detail a prototypical model with these features, but more general multi-fermion theories are possible. One interesting feature of $\mathcal{N}=2$ SYK models with multiple fermions is the presence of flavor symmetries that commute with the supercharge. In this case we obtain the Schwarzian with integer R-charge and a non-zero index, in a sector of fixed flavor charge (in the simplest realization we can even gauge the flavor charge). The partition function is $Z(\beta,\mu) = e^{N s_0} +Z_{\rm non-BPS}(\beta,\mu)$ for some order one number $s_0$, in the Schwarzian approximation. Our new $\mathcal{N}=2$ SYK model is thus the first of its kind to realize the $q=1$ Schwarzian, and passing to a sector of fixed flavor charge is somewhat analogous to passing to a particular charge sector of the AdS$_5$ black holes in \cite{Boruch:2022tno}.

For completeness we study these models with multiple fermions at non-zero flavor and R-charges. We find a similar phase structure as in the models of \cite{Fu:2016vas} with the fundamental fermion becoming unstable at large charges. When charges are too large the conformal solution breaks down and the system is driven to a low-entropy phase, although we leave a more exhaustive analysis for future work of the phase structure of these models. We also find a potential instability as we raise the flavor charge, without breaking supersymmetry (see figure \ref{fig:SusyMulFer}). Using the Luttinger-Ward relation we show that the conformal ansatz breaks down precisely where the charge of one of the fermions becomes maximal saturation value of $\pm N/2$. 

\vspace{0.5cm}

The rest of the paper is organized as follows. In section \ref{sec:N2SYKFu} we analyze the model of \cite{Fu:2016vas} at non-zero charge. After reviewing the definition of the model and the derivation of the mean field action we solve the IR Schwinger-Dyson equations in section \ref{sec:FuConfSol}. We discuss the breakdown of the conformal ansatz in section \ref{sec:FuBreakConf}. We also solve the full Schwinger-Dyson equations numerically to verify our claims, derive a Luttinger-Ward relation for $\mathcal{N}=2$ SYK relating the charge to the spectral asymmetry present in the IR limit of the two-point functions, find the grand potential, and compute the Schwarzian coupling numerically. In section \ref{sec:FuOpSpe} we analyze the operator spectrum by looking at the four-point function. In section \ref{Sec:FuHoloIn} we elaborate on the holographic interpretation. In section \ref{sec:N2SYKNew} we carry out most of the same analysis for a model with multiple fermions. We analyze the behavior of the Witten index, solve the IR Schwinger-Dyson equations, study the phase structure at non-zero charge, and analyze the operator spectrum. We conclude in section \ref{sec:Conclusions} pointing out some future directions.

\section{$\mathcal{N}=2$ SYK with fractional charge}\label{sec:N2SYKFu}
We begin with the $\mathcal{N}=2$ supersymmetric SYK model of \cite{Fu:2016vas}.  We will study this model at non-zero charge (some of the results were independently derived in \cite{SSN2}). This model consists of $N$ complex fermions which obey the standard Dirac algebra:
\begin{equation}
    \{\psi^i, \bar{\psi}_j \} = \delta^i_j \, , \qquad \{\psi^i, \psi^j \} = 0 \, , \qquad \{\bar{\psi}_i, \bar{\psi}_j \} = 0 \, .
\end{equation}
In terms of these fermions, the model is defined by a complex supercharge $\mathcal{Q}$ with the property that it involves $q$-fermion interactions\footnote{In contrast to \cite{Fu:2016vas}, we employed Einstein summation rather than an explicit ordered sum $    \mathcal{Q}= i^{\frac{q-1}{2}} \!  \sum_{1\leq i_1<\ldots<i_q\leq N}C_{i_1 i_2 \ldots i_q}~\psi^{i_1}\psi^{i_2}\ldots \psi^{i_q}$. This leads to some different numerical factors.}:
\begin{equation}\label{superchargeQ}
    \mathcal{Q}= i^{\frac{q-1}{2}} C_{i_1 i_2 \dots i_q}\psi^{i_1}\psi^{i_2}\ldots \psi^{i_q},
\end{equation}
where we take the couplings $C_{i_1 i_2 \ldots i_q}$ to be totally antisymmetric. We introduce disorder by integrating out these couplings with Gaussian statistics and we will fix the normalization of the random variables $\langle C_{i_1 i_2 \ldots i_q} \bar{C}^{i_1 i_2 \ldots i_q}\rangle \sim J/N^{q-1}$, with a $q$-dependent constant we will fix later to match the mean field action and equations of motion \eqref{eq:FuSD1}. Additionally, we take $q$ to be odd so $\mathcal{Q}$ has Fermi statistics. We see that \eqref{superchargeQ} has $\mathcal{Q}^2 = \bar{\mathcal{Q}}^2=0$, and the Hamiltonian is 
 \begin{align}
 H=\{ \mathcal{Q},\bar{\mathcal{Q}} \} \, ,
 \label{eq:fHamiltonian}
 \end{align}
as dictated by the supersymmetry algebra. When the Hamiltonian is expanded as a sum of interactions between $2q-2$ fermions, it has the same form as complex SYK \cite{Sachdev_1993, Georges_2001, Sachdev:2010um,  Sachdev:2015efa, Davison:2016ngz,Gu:2019jub, Chowdhury:2021qpy, Li:2017hdt}, but the complex SYK couplings are not Gaussian independent variables, given instead in terms of quadratic combinations of $C_{i_1\ldots i_q}$.

Supersymmetry relates bosons and fermions, but with supersymmetric SYK models we realize the supersymmetry nonlinearly. For example take 
\begin{align}
    \{ \mathcal{Q}, \psi^i \} = 0 \, , \quad \{ \mathcal{Q}, \bar{\psi}_i \} =  i^{\frac{q-1}{2}} q \, C_{i,i_1,\ldots,i_{q-1}} \psi^{i_1}\ldots \psi^{i_{q-1}} \sim  \bar{b}_i \, ,
    \label{eq:susymults}
\end{align}
where the second equation defines the bosonic composite complex field $b^i$, up to a choice of normalization. This means we do not need to include fundamental bosons in the theory in order for it to be supersymmetric. It is convenient nevertheless to do a Hubbard-Stratonovich transformation integrating-in a fundamental boson $b^i$ such that on-shell its given by the expression above. This has two advantages. The first is that now supersymmetry transformations \eqref{eq:susymults} are linearly realized. The second is that the theory with only fermions is not melonic and is not described by a mean field action, but the equivalent theory with the boson integrated-in is melonic and solvable at low temperatures \cite{Fu:2016vas}.

The Hamiltonian constructed above has an additional symmetry that corresponds to a $U(1)_R$ phase rotation of the fundamental fermions. The complex fermions transform in conjugate representations, and therefore the supercharge is purely (anti)holomorphic with respect to the rotation, which guarantees for example that $\mathcal{Q}^2=0$. This implies that the $U(1)_R$ symmetry is the $R$-symmetry of the one-dimensional $\mathcal{N}=2$ Poincare algebra. The explicit generator is: 
\beq
Q=\sum_j \bar{\psi}_j \psi^j - \frac{N}{2},~~~~Q_R =\frac{1}{q}~ Q.
\eeq
The first expression defines the fermion charge $Q$, with a shift of $N/2$ to enforce charge conjugation symmetry. This is the standard definition with a fermion having unit charge. On the right we give the natural definition of the $R$-charge $Q_R$, such that the R-charge of the supercharge is one. In these units the fundamental fermion has a fractional R-charge $1/q$. For this reason we will refer to these modes as fractional charge $\mathcal{N}=2$ SYK, as opposed to the models we will study in the next section.

\subsubsection*{Refined Witten Index}

The $\mathcal{N}=2$ SYK model is a particular instance of supersymmetric quantum mechanics, and therefore one may compute the Witten index \cite{Witten:1982im} which counts the number of bosonic and fermionic ground states with a $(-1)^F= e^{-i \pi Q_R}$, with $F$ the fermion number. This is a protected quantity which may be computed in the free field limit with all $C_{i_1 \dots i_q}$ set to zero. However, due to the presence of fractionally charged ground states with opposite bose-fermi statistics, the Witten index of this $\mathcal{N}=2$ SYK model vanishes. As explained in \cite{Fu:2016vas}, this model possesses a $\mathbb{Z}_q$ global symmetry under which the fermions transform with a $q$-th root of unity, and in particular commutes with the supercharge. Turning on a (discrete) chemical potential for this symmetry allows one to define a non-vanishing refined Witten index. This refined Witten index $\mathcal{I}(r) \equiv {\rm Tr}\left[ (-1)^F e^{2 \pi i r Q_R}\right]$ for this theory is given by
\beq\label{eq:wittenindex}
\mathcal{I}(r) =  e^{\pi i N (\frac{1}{2}-\frac{r}{q})}(1-e^{\frac{2\pi i r}{q}})^N = \left(2 \sin \frac{\pi r}{q}\right)^N,
\eeq
which vanishes for $r=0$. In general, the index and the number of ground states do not agree because there may be cancellation in the index. A bound on the number of ground states by computing the maximum absolute value of the index. The answer is 
\beq\label{EqnIndexS0}
{\rm max}_r \log |\mathcal{I}(r)| = N \log \left( 2 \cos \frac{\pi}{2q}\right)
\eeq
The maximum value is given for $r_{\rm max}=(q\pm1)/2$. We will match this later at large $N$ to the zero temperature entropy using the low temperature solution of the model.

\subsection{Mean field and the conformal solution}\label{sec:FuConfSol}

We now review the mean field formulation of this theory. The supersymmetry can be made manifest by introducing superspace coordinates $Z\equiv (\tau,\theta,\bar{\theta})$. In terms of these coordinates, the infinitesimal supersymmetry and R-symmetry transformations are realized respectively as:
\begin{align}
    \tau \rightarrow \tau + \theta \bar{\eta} + \bar{\theta}\eta \, , \qquad
    \theta \rightarrow \theta + \eta \, , \qquad
    \bar{\theta} \rightarrow \bar{\theta} + \bar{\eta} \, 
\end{align}
for an infinitesimal complex Grassmann parameter $\eta$ and,
\begin{align}
    \theta \rightarrow e^{i a} \theta \, , \qquad
    \bar{\theta} \rightarrow e^{-i a} \bar{\theta} \, ,
    \label{eq:fThetaCharge}
\end{align}
for a phase $a$. The fermion and boson \eqref{eq:susymults} are encoded in a chiral fermionic superfield $\Psi^i(\tau, \theta, \bar{\theta})$, where the chirality is defined by a suitable supercovariant derivative,
\begin{align}
\label{eq:chiralsuperfield}
    D_{\bar{\theta}} \equiv(\partial_{\bar{\theta}}+\theta \partial_\tau) \, , \quad D_{\bar{\theta}} \Psi^i(\tau, \theta, \bar{\theta}) = 0 \, \, \Longrightarrow \, \, \Psi^i(\tau, \theta, \bar{\theta}) = \psi^i(\tau+\theta\bar{\theta})+\sqrt{2}~\theta ~b^i(\tau) \, .
\end{align}
A similar set of expressions hold for the anti-chiral superfield $\bar{\Psi}_i(\tau, \theta, \bar{\theta})$ annihilated by $D_\theta$ defined by conjugation\footnote{In our conventions, conjugation always reverses the order of the fermions.}.

The supersymmetric Lagrangian density corresponding to the Hamiltonian \eqref{eq:fHamiltonian} may be written in terms of the chiral superfields as 
\begin{equation}
    \mathcal{L}=\frac{1}{2}\!\int \! d^2\theta \, \bar{\Psi}_i  \Psi^i+i^{\frac{q-1}{2}}\int \! d\theta \, C_{i_1 i_2 \ldots i_q} \Psi^{i_1} \Psi^{i_2} \ldots \Psi^{i_q}+i^{\frac{q-1}{2}}\int\!  d\bar{\theta} \, \bar{C}^{i_1 i_2 \ldots i_q} \bar{\Psi}_{i_1} \bar{\Psi}_{i_2} \ldots \bar{\Psi}_{i_q}.
    \label{eq:fLagrangian}
\end{equation}
In component form, the Lagrangian becomes after integrating by parts
\begin{equation}
    \mathcal{L} = \bar{\psi}_i \partial_\tau \psi^i -  \bar{b}_i b^i + i^{\frac{q-1}{2}}  \sqrt{2}q C_{i_1 i_2 \ldots i_q} b^{i_1} \psi^{i_2} \dots \psi^{i_{q}} + i^{\frac{q-1}{2}} \sqrt{2} q \bar{C}^{i_1 i_2 \ldots i_q} \bar{b}_{i_1} \bar{\psi}_{i_2} \dots \bar{\psi}_{i_{q}}
\end{equation}

Following \cite{Fu:2016vas} we introduce the superspace two-point function $\mathcal{G}(Z_1,Z_2) = \frac{1}{N} \langle \bar{\Psi}_i(Z_1) \Psi^i (Z_2) \rangle $. The components of this anti-chiral-chiral superfield contains the fermion and boson two point functions (noting the sum over $i$ is implicit):
\beq
G_{\psi\psi}(\tau_1,\tau_2) \equiv  \frac{1}{N}  \langle \bar{\psi}_i (\tau_1) \psi^i (\tau_2) \rangle,~~~~G_{bb}(\tau_1,\tau_2) \equiv  \frac{1}{N}  \langle \bar{b}_i(\tau_1) b^i (\tau_2)\rangle.
\eeq
The full two-point function in superspace also includes fermionic correlators mixing fermions and bosons. The complete expansion is
\begin{eqnarray}\label{eqn:defsuperG}
    \mathcal{G}(Z_1,Z_2)&=&G_{\psi\psi}(\tau_1-\theta_1\bar{\theta}_1,\tau_2+\theta_2\bar{\theta}_2)+2 \bar{\theta}_1\theta_2 G_{bb}(\tau_1,\tau_2) \nonumber \\
    &&+\sqrt{2}\bar{\theta}_1G_{b\psi}(\tau_1,\tau_2+\theta_2\bar{\theta}_2)-\sqrt{2}\theta_2 G_{\psi b}(\tau_1-\theta_1\bar{\theta}_1,\tau_2).
\end{eqnarray}
In the large $N$ limit we will see that the classical solution of the mean field equations we derive satisfies $G_{\psi b}=0$ and we can consistently set the mixed correlators to zero for now. It will be necessary to include them later when considering the kernel giving quadratic fluctuations around the mean field classical solution. 

Starting with the action as in \eqref{eq:fLagrangian}, our next goal is to produce the aforementioned mean-field equations of motion which are valid at large $N$. We will follow a series of now standard steps for SYK models, working largely in superspace:
\begin{itemize}
    \item Integrate out the random couplings to generate a bi-local Lagrangian:  
\begin{equation}
    \mathcal{L}= \frac{1}{2}\!\int \! d^2\theta \, \bar{\Psi}_i  \Psi^i+\int \! d\bar{\theta}_1 d\theta_2 \, \bar{\Psi}_{i_1} \bar{\Psi}_{i_2} \ldots \bar{\Psi}_{i_q} \langle \bar{C}^{i_1 i_2 \ldots i_q} C_{j_1 j_2 \ldots j_q} \rangle \Psi^{j_1} \Psi^{j_2} \ldots \Psi^{j_q} .
    \label{eq:fLagrangian_bilocal}
\end{equation}
    \item Integrate in the field $\mathcal{G}(Z_1,Z_2)$ which is then fixed to be the superspace two-point function by further integrating in an anti-chiral-chiral bilocal field $\Sigma(Z_1,Z_2)$. The components of $\Sigma(Z_1,Z_2)$ contain the fermionic and bosonic self energies, $\Sigma_{\psi\psi}(\tau_1,\tau_2)$  and $\Sigma_{bb}(\tau_1,\tau_2)$, respectively. This amounts to inserting a superspace identity
    \begin{equation}
    1 = \int D \mathcal{G} D\Sigma\exp{\left (-N \int d \bar{Z}_1 d Z_2 \, \Sigma(Z_1, Z_2)\left(\mathcal{G}(Z_1, Z_2) - \frac{1}{N}\bar{\Psi}_i(Z_1)\Psi^i(Z_2)\right) \right )} \, ,
    \end{equation}
    with the order of indices chosen to match the fact that $\mathcal{G}$ and $\Sigma$ have the same chirality.
    \item Integrate out the fundamental fermions to obtain an action in terms of the collective variables only. We will not write the action explicitly, but instead focus on the equations of motion.
    \item Vary the action with respect to the bilocal fields. To write the equations of motion in superspace, we will introduce chiral and anti-chiral integrations
\begin{align}
    \int dZ \equiv\int d\tau d\theta \, , \qquad \int d\bar{Z} \equiv \int d\tau d\bar{\theta} \, ,
\end{align}
which manifest supersymmetry when acting on superfields of the appropriate chirality. Because the mean field action is bi-local, we will sometimes encounter several such integrations over distinct superspace coordinates.
\end{itemize}

The equations of motion resulting from this procedure can be written as
\bea
\Sigma(Z_2, Z_3) &=&\frac{1}{2} J~\mathcal{G}(Z_2, Z_3)^{q-1},\nonumber\\
\frac{1}{2}D_{\theta_3} \mathcal{G}(Z_1, Z_3) + \int dZ_2 \, \mathcal{G}(Z_1, Z_2) ~\Sigma(Z_3, Z_2) &=& \delta(Z_1-Z_3),\label{eqn:SDinsuperspace}
\ea
where we define a supersymmetric delta function $\delta(Z_1-Z_2)\equiv (\bar{\theta}_1-\bar{\theta}_2)\delta(\tau_1-\theta_1\bar{\theta}_1-\tau_2+\theta_2\bar{\theta}_2)$. This equation defines the normalization we chose for $J$ the coupling constant which leads to particularly simple component equations \eqref{eq:FuSD1}. Note also that the second equation is properly anti-chiral-anti-chiral in $Z_1$ and $Z_3$; in particular the superspace integration over $Z_2$ as well as the ordering of the indices inside $\mathcal{G}$ and $\Sigma$ means we do not need to introduce explicit conjugate bilocals $\bar{\mathcal{G}}$ and $\bar{\Sigma}$.

The superspace presentation of the mean field action for the $\mathcal{N}=2$ SYK model is well known, but as already discussed, we will also introduce and analyze the consequences of a chemical potential ($\mu$) for the $U(1)_R$ symmetry. In our conventions the chemical potential $\mu$ appears as $\mu G_{\psi\psi}$ in the action which means it couples to $q Q_R$. Because the R-symmetry does not commute with supersymmetry, or equivalently because the Grassmann variables $(\theta, \bar{\theta})$ are charged under the R-symmetry \eqref{eq:fThetaCharge}, the inclusion of a chemical potential generically corresponds to turning on background fields which break supersymmetry. Therefore, we will study the mean field equations in component form from now on and study the supersymmetric point as a special case. 

In the fermion-number conserving saddle-point we are interested in, the mixed $G_{b \psi} $ correlators vanish and we can focus on $G_{\psi \psi}$, $G_{bb}$, and their self energies. Moreover we will also assume the solution is time translation invariant. Under these assumptions, the Schwinger-Dyson equations derived from the mean field action with chemical potential $\mu$ are given by
\bea\label{DSeqs}
\Sigma_{\psi \psi}(\tau) &=& J(q-1) G_{\psi \psi}(\tau)^{q-2} G_{bb}(\tau),~~~~~\Sigma_{bb}(\tau) = J G_{\psi \psi}(\tau)^{q-1},\label{eq:FuSD1}\\
G_{\psi \psi}(\omega) &=& \frac{1}{-i \omega + \mu + \Sigma_{\psi \psi}(-\omega)},~~~~~G_{bb}(\omega)=\frac{1}{-1-\Sigma_{bb}(-\omega)}.\label{eq:FuSD2}
\ea
The $\tau = \tau_1-\tau_2$ is the time difference and the second line is written in Fourier space. Also, one may note that while the boson $b$ carries $U(1)_R$ charge, there is no chemical potential present in the $G_{bb}$ equation because this is only an auxiliary field.

The full Schwinger-Dyson equations are complicated to solve and are typically studied numerically, as we do in section \ref{ssec:fNumericalSD}. However, they are tractable in the IR limit, meaning long time separations $|J\tau|\gg1$ compared to the scale set by $J$. In this regime we will assume the theory is approximately conformal, and importantly we will see below that this assumption may be violated depending on the background charge. This conformal ansatz determines the fermion and boson two point functions at zero temperature to be 
\beq
G_{\psi \psi}(\tau)= \frac{g_{\psi\psi}}{|\tau|^{2\Delta}} \left( e^{\pi \mathcal{E}} \Theta(\tau) - e^{-\pi \mathcal{E}} \Theta(-\tau) \right),~~G_{bb}(\tau) = \frac{g_{bb}}{|\tau|^{2\Delta_b}} \left( e^{\pi \mathcal{E}_{b}} \Theta(\tau) + e^{-\pi \mathcal{E}_{b}} \Theta(-\tau) \right),
\eeq
where we allow for the possibility of a spectral asymmetry in order to incorporate states with non-zero charge, parametrized by $\mathcal{E}$ for the fermion and $\mathcal{E}_b$ for the boson. The spectral asymmetry defined in the IR should be thought of as related to the chemical potential or the charge through the UV behavior of the correlators. The prefactors $g_{\psi\psi}$ and $g_{bb}$ are coefficients to be determined by the equations of motion. Finally we also introduce the IR scaling dimensions for fermions $\Delta$ and bosons $\Delta_b$. After solving the zero temperature equations, the ansatz can also be put at finite temperature by a reparametrization
\begin{equation}\label{eq:ansatz_ft}
    G_{\psi \psi}(\tau)=g_{\psi\psi}\left(\frac{\beta}{\pi}\sin{\frac{\pi\tau}{\beta}}\right)^{-2\Delta}e^{2\pi \mathcal{E}\left(\frac{1}{2}-\frac{\tau}{\beta}\right)}\, , \, \, \, \, G_{bb}(\tau)=g_{bb} \left(\frac{\beta}{\pi}\sin{\frac{\pi\tau}{\beta}}\right)^{-2\Delta_b}e^{2\pi \mathcal{E}_b\left(\frac{1}{2}-\frac{\tau}{\beta}\right)},
\end{equation}
valid for $0<\tau<\beta$, together with $J\tau \gg 1$ and $J(\beta-\tau)\gg1$ so that the IR solution we are going to find is accurate.

Using the two-point functions outlined above, the next step is to solve the equations \eqref{eq:FuSD2} in the IR strong coupling limit in which the self-energies dominate over bare propagators. This leads to the simple form $\Sigma_{\psi\psi}(-\omega) G_{\psi\psi}(\omega)=-\Sigma_{bb}(-\omega) G_{bb}(\omega)= 1$ (after a shift of self-energies, see \cite{Gu:2019jub}). The solution is obtained by transforming the two point functions to Fourier space, taking the inverse, and then transforming back to time. This determines the self-energies (which for simplicity we write at zero temperature) to be
\bea
\Sigma_{\psi\psi}(\tau) &=& \frac{1}{g_{\psi\psi}}~ \frac{(1-2\Delta)\sin 2 \pi \Delta}{2\pi(\cosh 2 \pi \mathcal{E}+\cos 2\pi \Delta )}~\frac{1}{|\tau|^{2(1-\Delta)}}\left( -e^{\pi \mathcal{E}} \Theta(-\tau) + e^{-\pi \mathcal{E}} \Theta(\tau) \right),\\
\Sigma_{bb}(\tau) &=&-\frac{1}{g_{bb}}~ \frac{(1-2\Delta_{b})\sin 2 \pi \Delta_{b}}{2\pi (\cosh 2 \pi \mathcal{E}_b - \cos 2\pi \Delta_b)}~\frac{1}{|\tau|^{2(1-\Delta_{b})}} \left( e^{\pi \mathcal{E}_{b}} \Theta(-\tau) + e^{-\pi \mathcal{E}_{b}} \Theta(\tau) \right),~~~\nonumber
\ea
Having both the two point functions and self-energies, we can now plug these back into the first set of Schwinger-Dyson equations \eqref{eq:FuSD1}. This places constraints on the dimensions, prefactors and spectral asymmetries. In contrast to the special case studied in \cite{Fu:2016vas} with no spectral asymmetry, we find a different set of constraints. 

The equations \eqref{eq:FuSD1} can be easily solved in steps. We can separately match the spectral asymmetry, overall scaling dimension, and prefactors in the left and right-hand sides. Matching the spectral asymmetry in both equations gives a single constraint
\beq
\mathcal{E}_b =- (q-1) \mathcal{E}
\eeq
which is consistent with the on-shell expression for the boson in term of fermions $b\sim \bar{\psi}^{q-1}$ \eqref{eq:susymults}. Matching scaling dimensions in both equations gives again a single constraint, 
\beq
(q-1) \Delta + \Delta_b = 1,
\label{eq:fDimConstraint}
\eeq
which can be interpreted as demanding the interaction term $C_{i_1\ldots i_q} b^{i_1} \psi^{i_2}\ldots \psi^{i_{q}}$ in the component action  to be marginal. For consistency of the IR ansatz, namely to make sure we can neglect the $i\omega$ and $+1$ terms in the equations \eqref{eq:FuSD2}, we need $\Delta>0$ and $\Delta_b > 1/2$ which, upon using the constraint on dimensions \eqref{eq:fDimConstraint}, implies $0<\Delta<1/(2(q-1))$. This means that a solution outside of this range should be considered inconsistent in the sense that the IR ansatz is invalid.

The final step to solving the equations is to match the prefactors in the left and right hand side of equations \eqref{eq:FuSD1}. The two equations give the following constraint:
\beq\label{eqfuconstra}
\frac{(1-2\Delta)\sin 2 \pi \Delta}{2\pi(\cosh 2 \pi \mathcal{E}+\cos 2\pi \Delta )} = (q-1)J g_{\psi\psi}^{q-1} g_{bb},~~~  \frac{(1-2\Delta_{b})\sin 2 \pi \Delta_{b}}{2\pi (\cosh 2 \pi \mathcal{E}_b - \cos 2\pi \Delta_b)}=Jg_{\psi\psi}^{q-1} g_{bb}
\eeq
By matching these two expressions for $g_{\psi\psi}^{q-1} g_{bb}$, we can completely determine the scaling dimensions as a function of $q$ and the spectral asymmetry $\mathcal{E}$, after using the expression for $\Delta_b$ and $\mathcal{E}_b$. The solution can only be found implicitly through the equation\footnote{This expression was independently derived in \cite{SSN2}.}
\beq\label{eq:FuDeltaEqn}
\frac{(1-2\Delta)\sin 2 \pi \Delta}{\cosh 2 \pi \mathcal{E}+\cos 2\pi \Delta} = (q-1)\frac{(1-2(q-1)\Delta) \sin 2\pi (q-1)\Delta}{ \cosh 2 \pi (q-1) \mathcal{E} - \cos 2\pi (q-1) \Delta}
\eeq
This determines $\Delta$, $\Delta_b$ and $\mathcal{E}_b$, all as a function of $q$ and $\mathcal{E}$. Finally, since now the two equations \eqref{eqfuconstra} are identical we can only determine the combination $g_{\psi\psi}^{q-1} g_{bb}$, but not each prefactor separately. As explained in \cite{Fu:2016vas} this can be traced back to an emergent symmetry in the IR Schwinger-Dyson equations under $G_{\psi\psi}(\tau)\to \lambda G_{\psi\psi}(\tau)$ and $G_{bb}(\tau)\to \lambda^{1-q} G_{bb}(\tau)$. It is expected that this does not generate a relevant mode in the IR and simply the UV boundary conditions determine a precise value of $g_{\psi\psi}$ and $g_{bb}$. As reviewed for example in \cite{Davison:2016ngz} unitarity requires both coefficients to be positive, and this is consistent with the constraint \eqref{eqfuconstra}. We will verify that the full solution determines a consistent coefficient using a numerical solution of the Schwinger-Dyson equations below in section \ref{ssec:fNumericalSD}.

\subsubsection*{Emergent $SU(1,1|1)$ symmetry}

The equation above that determines $\Delta$ cannot be solved explicitly in general. There are special values of parameters where we expect the solution to preserve supersymmetry. In these cases the equations can be exactly solved and we reproduce the results of \cite{Fu:2016vas} (We will see other examples in the next section where even supersymmetry is not powerful enough to fully fix the IR solution). The first observation we can make is that for the following values of the spectral asymmetry the scaling dimensions have a simple expression 
\beq\label{eqn:FuSusyDiscrete}
\mathcal{E}_{\rm susy} = \frac{i r}{q},~~~r\in \mathbb{Z}_q,~~~~\Rightarrow ~~~\Delta= \frac{1}{2q},~~~\Delta_b = \frac{1}{2} + \frac{1}{2q}.
\eeq
The interpretation for this family is that it corresponds to turning on the discrete chemical potential conjugate to the $\mathbb{Z}_q$ global symmetry that commutes with the supercharge. A simple way of deriving constraints from global supersymmetry is to construct the unique superspace two-point function that is anti-chiral in the first variable, chiral in the second, and manifestly invariant under global super-translation $(\tau,\theta,\bar{\theta}) \to (\tau +\epsilon+ \bar{\theta} \eta + \theta \bar{\eta} , \theta+\eta,\bar{\theta}+\bar{\eta})$,  with parameters $(\epsilon,\eta,\bar{\eta})$. The answer is given by
\beq
\mathcal{G}_{\rm susy}(Z_1,Z_2) = f(\tau_1-\tau_2 - \theta_1 \bar{\theta}_1 -\theta_2\bar{\theta}_2 -2 \bar{\theta}_1\theta_2),
\eeq
where the right hand side involves an arbitrary function $f$. Using \eqref{eqn:defsuperG} we can see that this implies $G_{\psi\psi}(\tau_1,\tau_2) = f(\tau_1-\tau_2)$ and $G_{bb}(\tau_1,\tau_2) = - f'(\tau_1-\tau_2)$. Combining these facts we obtain the constraint
\beq
\label{eq:SusyDyson}
G_{bb}(\tau_1,\tau_2)= - \partial_{\tau_1} G_{\psi\psi}(\tau_1-\tau_2),
\eeq
which together with the conformal ansatz implies $\Delta_b = \Delta+1/2$, and replacing this in \eqref{eq:FuSD1} and \eqref{eq:FuSD2} gives $\Delta=1/2q$, as obtained in the equation above \eqref{eqn:FuSusyDiscrete}. Moreover, supersymmetry also implies a relation between coefficients $g_{bb} = 2 \Delta g_{\psi\psi}$, allowing one to fully find the solution for the prefactors as well: 
\beq\label{eq:fu_etal_sol}
g_{\psi\psi}^{\rm susy} = \left(\frac{1}{2\pi J} \frac{\sin \frac{\pi}{q}}{\cos \frac{\pi}{q} + \cos \frac{2\pi r}{q}} \right)^{1/q},~~~~g_{bb}^{\rm susy} =\frac{1}{q} \left(\frac{1}{2\pi J} \frac{\sin \frac{\pi}{q}}{\cos \frac{\pi}{q} + \cos \frac{2\pi r}{q}} \right)^{1/q} .
\eeq
These solutions correspond to a finite two-point functions except when $r=(q \pm 1)/2$. For those two values the denominators vanish and $g_{\psi\psi}^{\rm susy}$ and $g_{bb}^{\rm susy}$ both diverge. It is interesting to note those are precisely the values at which the refined index has maximal absolute value. 

So far we have imposed only the global $\mathcal{N}=2$ supersymmetric conditions but the model at low temperatures enjoys a bigger group of symmetries. In the IR limit we solved the Schwinger-Dyson equations \eqref{eqn:SDinsuperspace} in a regime where we can neglect the first term and get 
\beq
\label{eq:IRFuSD}
\int dZ_2 ~ \mathcal{G}(Z_1,Z_2)~ \left[\frac{J}{2}\mathcal{G}(Z_3,Z_2)^{q-1}\right] = \delta(\bar{Z}_1-\bar{Z}_3).
\eeq
It was pointed out in \cite{Fu:2016vas} that these equations have a symmetry under $\mathcal{N}=2$ supersymmetric reparametrizations which we denote by ${\rm Diff}(S^{1|2})$. This consists of super-reparametrizations $Z=(\tau,\theta,\bar{\theta}) \to Z'=(\tau',\theta',\bar{\theta}')$ that satisfy the constraints
\bea
&&D_\theta \bar{\theta}'=0~~~~D_\theta \tau' = \bar{\theta}' D_\theta \theta',\\
&&D_{\bar{\theta}} \theta'=0~~~~D_{\bar{\theta}} \tau' = \theta' D_{\bar{\theta}} \bar{\theta}'.
\ea
As shown in \cite{Fu:2016vas} the solutions of these constraints can be parametrized by a bosonic reparametrization mode $f(\tau)$, a local $U(1)$ transformation $e^{i a(\tau)}$ and a complex fermionic mode $\eta(\tau)$ which may be grouped into the bosonic and fermionic reparametrizations as:
\begin{align}
  {\rm Bosonic:}~~~  \tau' &= f(\tau) \, , \quad \theta' = e^{i a(\tau)}\sqrt{\partial_\tau f(\tau)} \theta \, \quad \bar{\theta}' = e^{-i a(\tau)}\sqrt{\partial_\tau f(\tau)} \bar{\theta} \, , \\
 {\rm Fermionic:}~~~   \tau' &= \tau + \theta \bar{\eta}(\tau) + \bar{\theta} \eta(\tau) \, , \quad \theta' = \theta + \eta(\tau + \theta \bar{\theta}) \, \quad \bar{\theta}' = \bar{\theta} + \bar{\eta}(\tau - \theta \bar{\theta}) \label{eqn:fermionreparam}\, .
\end{align}
Then its easy to check that the IR Schwinger-Dyson equation is invariant under
\beq\label{eqn:FususytransfIR}
\mathcal{G}(Z_1,Z_2) \to (D_{\theta_1} \theta'_1)^{\frac{1}{q}}~(D_{\bar{\theta}_2} \bar{\theta}'_2)^{\frac{1}{q}}~\mathcal{G}(Z'_1,Z'_2).
\eeq
Even though this is a symmetry of the equations, this is not a symmetry of the solutions. The correlators found above at the supersymmetric point with $\Delta=1/(2q)$ are only invariant under a subgroup $SU(1,1|1)$ of ${\rm Diff}(S^{1|2})$. Therefore we see that the emergent super-reparametrization symmetry is spontaneously broken to the $\mathcal{N}=2$ superconformal group. Of course both of these symmetries are broken by the UV term, giving rise to the $\mathcal{N}=2$ Schwarzian mode. 

In a state with non-zero charge, we saw that $\Delta$ is not given by the supersymmetric solution. It is interesting to notice that the IR equations preserve supersymmetry even at finite charge, when the UV term is ignored. From this point of view, the breaking of supersymmetry that happens from the inclusion of a chemical potential is spontaneous: the solution breaks the symmetry.

From \eqref{eqn:FususytransfIR} one can deduce that the fermion $\psi$ is a superconformal primary. In particular that transformation rule, applied to the solution with zero charge density only, implies the R-charge $1/q$ is twice the scaling dimension $\Delta = 1/(2q)$. This will be important to keep in mind in the next section. 

It is instructive to see how a super-reparametrization explicitly acts on the two-point function. Its enough to do this at the linearized level. Bosonic reparametrizations and $U(1)_R$ transformations look the same as reparametrizations and local gauge transformations in complex SYK, and we will not repeat it here. The new generators are the infinitesimal fermion super-reparametrization, and under them the two-point function changes as
\begin{eqnarray}
\delta G_{\psi\psi} &=& -G_{b\psi}(\tau_1,\tau_2) \bar{\eta}(\tau_1)\\
\delta G_{b\psi} &=& G_{bb}(\tau_1,\tau_2) \bar{\eta}(\tau_1) - \partial_{\tau_2} G_{\psi\psi}(\tau_1,\tau_2) \bar{\eta}(\tau_2)-\frac{1}{q} G_{\psi\psi}(\tau_1,\tau_2) \bar{\eta}'(\tau_2)\\
\delta G_{bb} &=& \partial_{\tau_2} G_{b\psi}(\tau_1,\tau_2) \bar{\eta}(\tau_2)-\frac{1}{q} G_{b\psi}(\tau_1,\tau_2)\bar{\eta}'(\tau_2)  .
\end{eqnarray}
and a similar transformation for $\eta$. To simplify the expressions we rescale $\eta \to \eta/\sqrt{2}$ compared to \eqref{eqn:fermionreparam}. Its clear the first and third equation vanishes on-shell since the fermionic correlators are zero $G_{b\psi}=0$. The second line is more interesting. On a superconformal solution of the type discussed above, the variation of $G_{b\psi}$ also vanishes whenever $\eta = \eta_0 + \eta_1 \tau$ and $\bar{\eta}=\bar{\eta}_0 + \bar{\eta}_1 \tau$. These four real parameters are the fermionic generators of $SU(1,1|1)$. For $\eta \sim \tau^{n}$ with $n\neq 0,1$ the transformation acts non-trivially on the two-point function, and all these modes have $SL(2,\mathbb{R})$ Casimir given by $h=3/2$ at $\mathcal{E}=\mathcal{E}_{\rm susy}$\footnote{The Casimir operator $C_{1+2}$ acting on $\delta G_{b\psi}(\tau_1,\tau_2)$ is defined in the same way as equations (3.54) of \cite{Maldacena:2016hyu}, with the difference that the scaling dimension appearing in the generators of $SL(2,\mathbb{R})$ are different for the first and second insertion. Writing the eigenvalue of the Casimir as $h(h-1)$ defines the parameter $h$.}. We will see these modes later again when we compute fluctuations of the action around the classical Schwinger-Dyson solution. At the same time we see that for $\mathcal{E}\neq \mathcal{E}_{\rm susy}$ the four fermion zero-modes are broken spontaneously since the on-shell correlators no long satisfy any supersymmetry relation.

\subsubsection*{The Zero Temperature Entropy}

We can compute the zero temperature entropy of the model and compare with the index, using the IR asymptotic of the solution of the mean field action. In order to do this it is convenient to take a derivative with respect to $q$ in the action. This simplifies when evaluating on a solution of the equations of motion and gives
\bea
\partial_q \frac{\log Z}{N} &=& J\beta \int G_{\psi\psi}(\tau)^{q-1} G_{bb}( \tau) \log G_{\psi\psi}(\tau),\\
&=& \# \beta +2\pi^2  \Delta ~J g_{\psi\psi}^{q-1}g_{bb} + \mathcal{O}(\beta^{-1})
\ea
In the second line we inserted the conformal solution, keeping terms insensitive to the UV behavior. If we denote the temperature independent term in the partition function by $\log Z = \beta \# + G + \mathcal{O}(\beta^{-1})$, where we introduce the grand potential $G$, then it is given by \footnote{This expression was independently derived in \cite{SSN2}.}
\beq\label{eqn:dGdqfu}
\frac{dG}{dq}= N\frac{\pi \Delta  (1-2\Delta)  \sin 2 \pi \Delta}{(q-1) (\cos 2 \pi \Delta + \cosh 2 \pi \mathcal{E})}.
\eeq
In \cite{Davison:2016ngz} it is explained how to go from this expression to computing $S_0(Q)$. Unfortunately this cannot be done in this model since it is not clear what boundary conditions to use when integrating over $q$. Instead, for now we will focus on the particle-hole symmetric point with $Q_R=0$. In this case the solution is the supersymmetric one with $\mathcal{E}=0$ which implies $\Delta=1/(2q)$. In this case $G=S_0$ and we obtain $\frac{dS_0}{dq}=N\frac{\pi \tan \frac{\pi}{2q}}{2 q^2}$. This can be easily integrated, using the free fermion limit to fix the integration constant, and gives 
\beq
S_0(Q_R=0) = N \log \big( 2 \cos \frac{\pi}{2q}\big),
\eeq
which precisely matches with the maximization of the index in equation \eqref{EqnIndexS0}. We will see a similar phenomenon in the models we study in the next section. 

We will later propose a closed formula for the zero-temperature entropy and grand potential at non-zero charge based on an extension of the spooky propagator of \cite{Gu:2019jub}, and verify these relations numerically.

\subsection{Breakdown of conformal ansatz}\label{sec:FuBreakConf}
\begin{figure}[h!]
    \centering
    \includegraphics[scale=0.3]{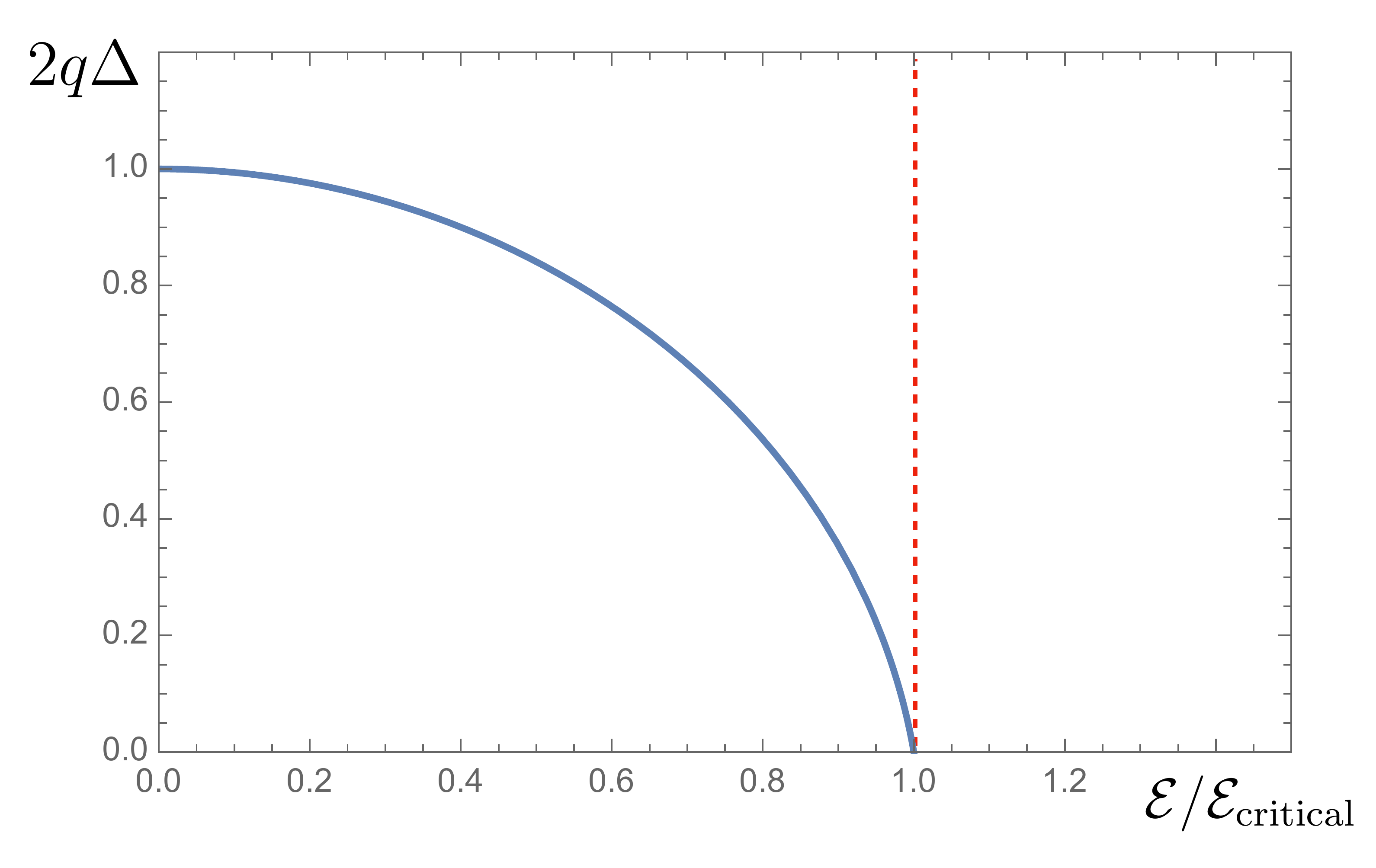}
    \caption{Plot showing the conformal solution for the fermion scaling dimension $\Delta$ as a function of the spectral asymmetry $\mathcal{E}$ for $q=3$. The behavior is similar for other values of $q$. We see the scaling dimension reaches zero at the critical spectral asymmetry, and for larger $\mathcal{E}$, the conformal ansatz is invalid, to the right of the dashed red line.}
    \label{fig:DeltaFuEtAl}
\end{figure}
We will now analyze what happens to the conformal solution of the Schwinger-Dyson equations as we turn on the spectral asymmetry, and therefore the background R-charge. We start with the $\mathcal{E}=0$ case for which we obtain a supersymmetric solution with $\Delta=\frac{1}{2q}$. As we turn on $\mathcal{E}$ we find $\Delta$ is a solution of the transcendental equation given above. The scaling dimension smoothly goes from $\Delta(\mathcal{E}=0)=\frac{1}{2q}$ to $\Delta(|\mathcal{E}|= \mathcal{E}_{\rm critical})=0$. The critical asymmetry $\mathcal{E}_{\rm critical}>0$ is determined by the following equation 
\beq
\frac{\sinh \pi (q-1) \mathcal{E}_{\rm critical} }{ \cosh \pi \mathcal{E}_{\rm critical}} = (q-1)
\eeq
For $|\mathcal{E}|>\mathcal{E}_{\rm critical}$ there are solution for the scaling dimension in the complex $\Delta$-plane indicating the conformal ansatz breaks down\footnote{The point $\Delta=0$ is always a solution but one with $g_{\psi\psi}^{q-1}g_{bb} = 0$. Therefore at least one of the two prefactors has to vanish which is a non-physical solution.}. The value for $q=3$ is given numerically by
\beq
\mathcal{E}_{\rm critical}= \frac{\log (1 + \sqrt{2})}{\pi} = 0.28055..
\eeq
We show the behavior of $\Delta$ as a function of $\mathcal{E}$ in figure \ref{fig:DeltaFuEtAl}. After $\mathcal{E}>\mathcal{E}_{\text{critical}},$ there is no valid conformal solution. In such a region, the Dyson Schwinger equations Eq.(\ref{DSeqs}) can be studied numerically. We will in fact show that the solution exponentially decays once $\mathcal{E}>\mathcal{E}_{\text{critical}}.$ In contrast to known transitions \cite{Azeyanagi:2017drg,Ferrari:2019ogc}, in this case the fundamental fermion $\psi^i$ itself develops an either negative or complex scaling, making the physical interpretation of this transition more transparent.

\subsubsection*{Numerical Solution to Schwinger Dyson equations}
\label{ssec:fNumericalSD}
\begin{figure}[t!]
    \centering
    \includegraphics[width=0.32\textwidth]{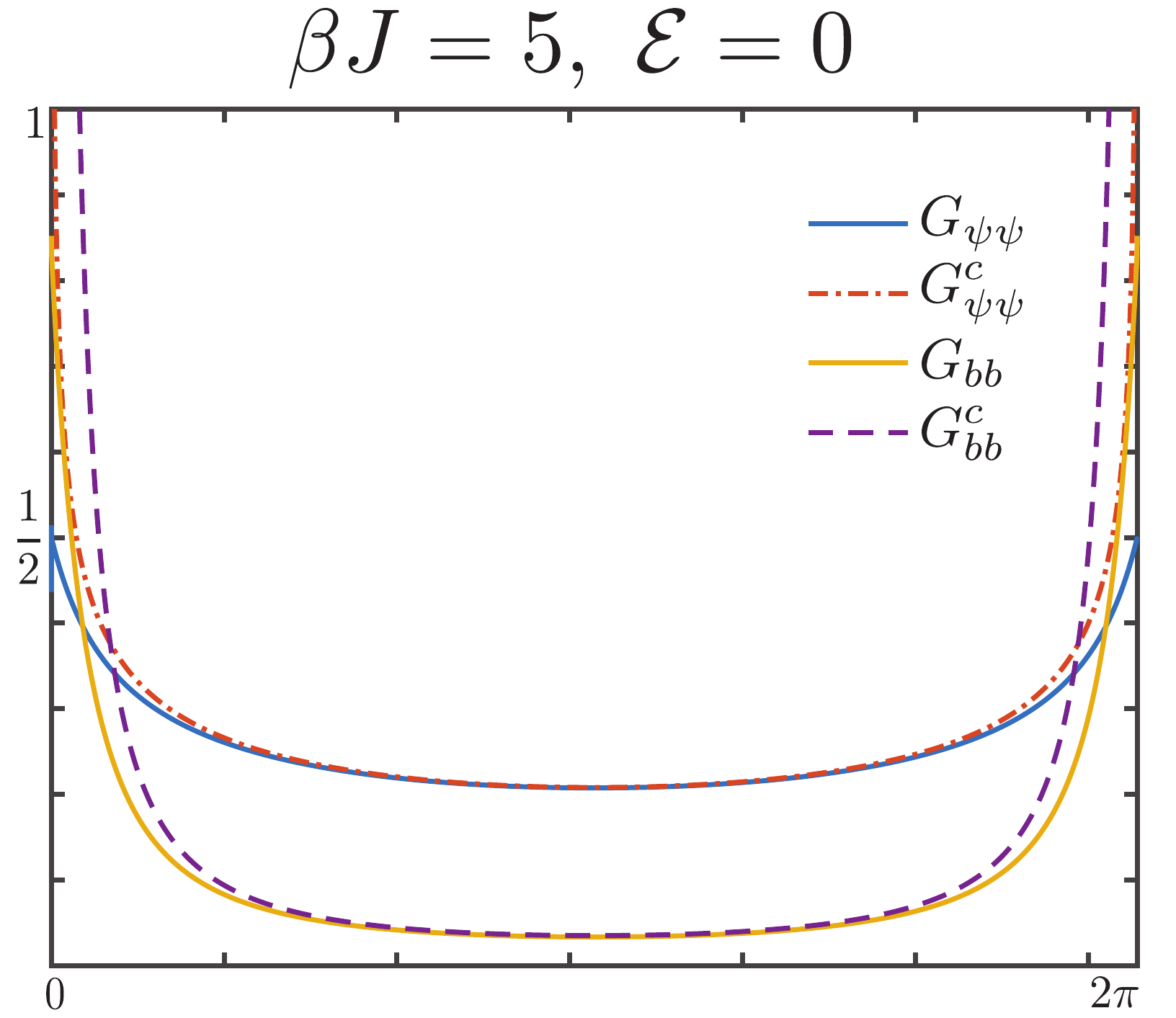}
   \includegraphics[width=0.32\textwidth]{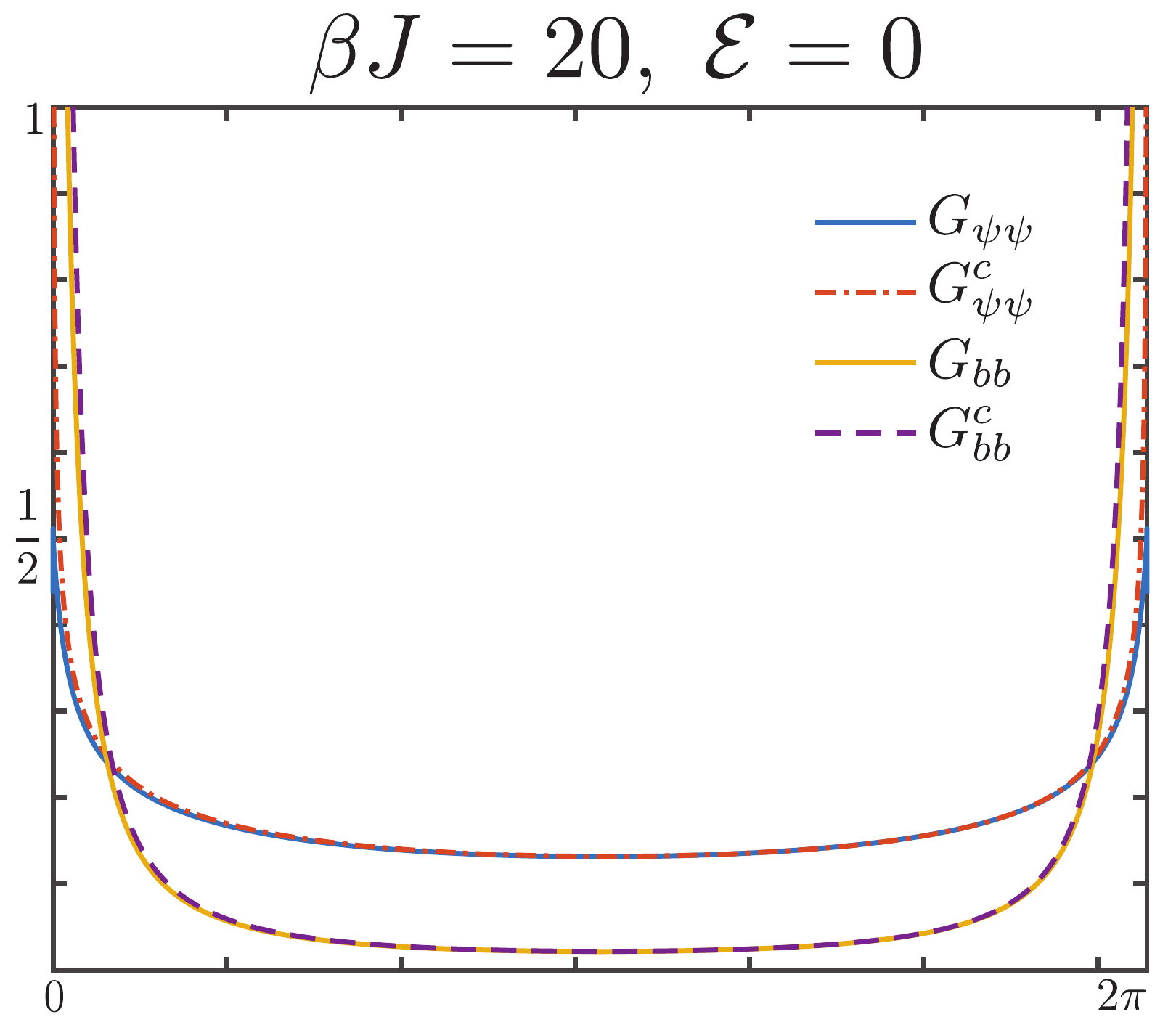}
    \includegraphics[width=0.32\textwidth]{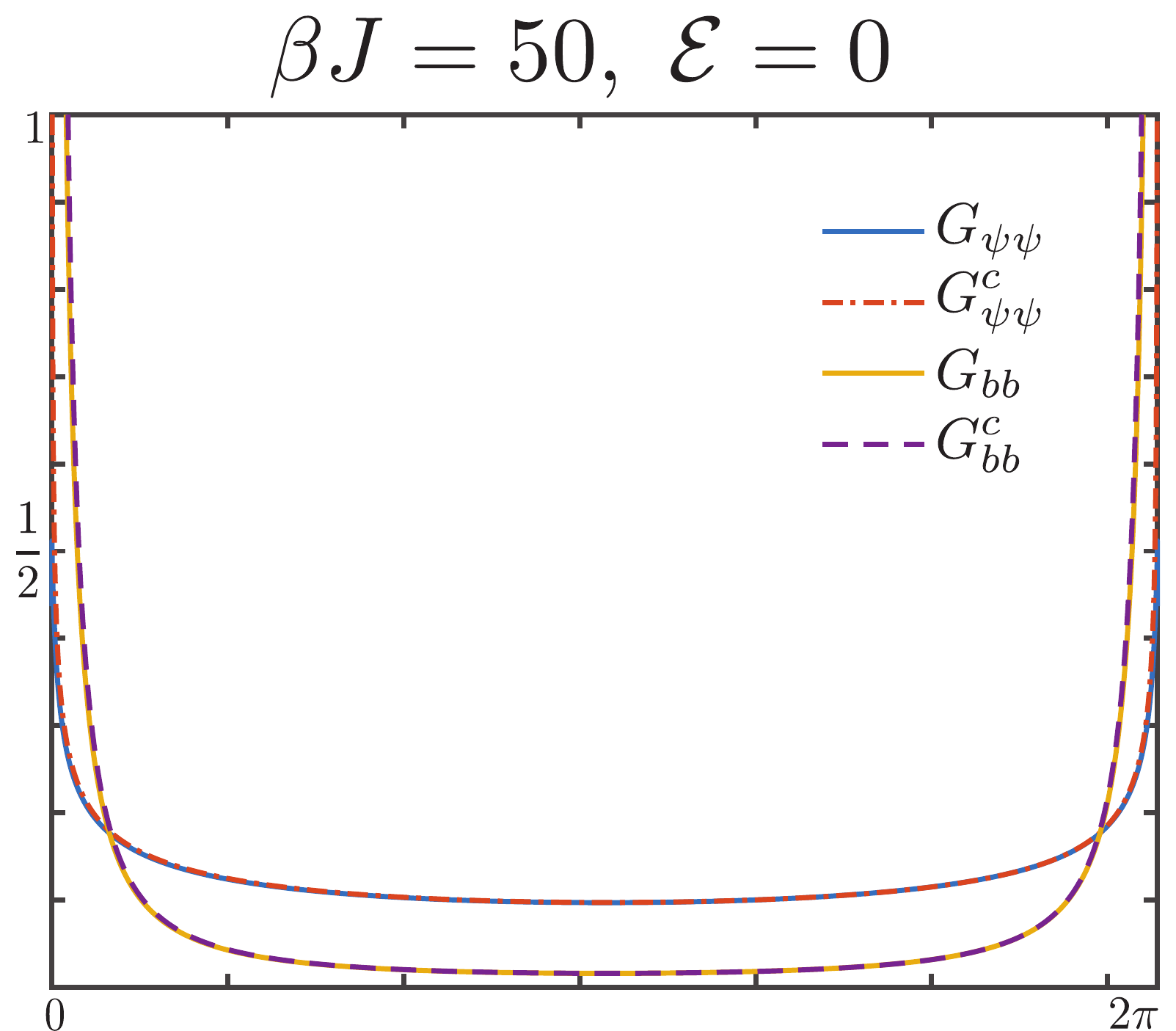}\\
    \includegraphics[width=0.32\textwidth]{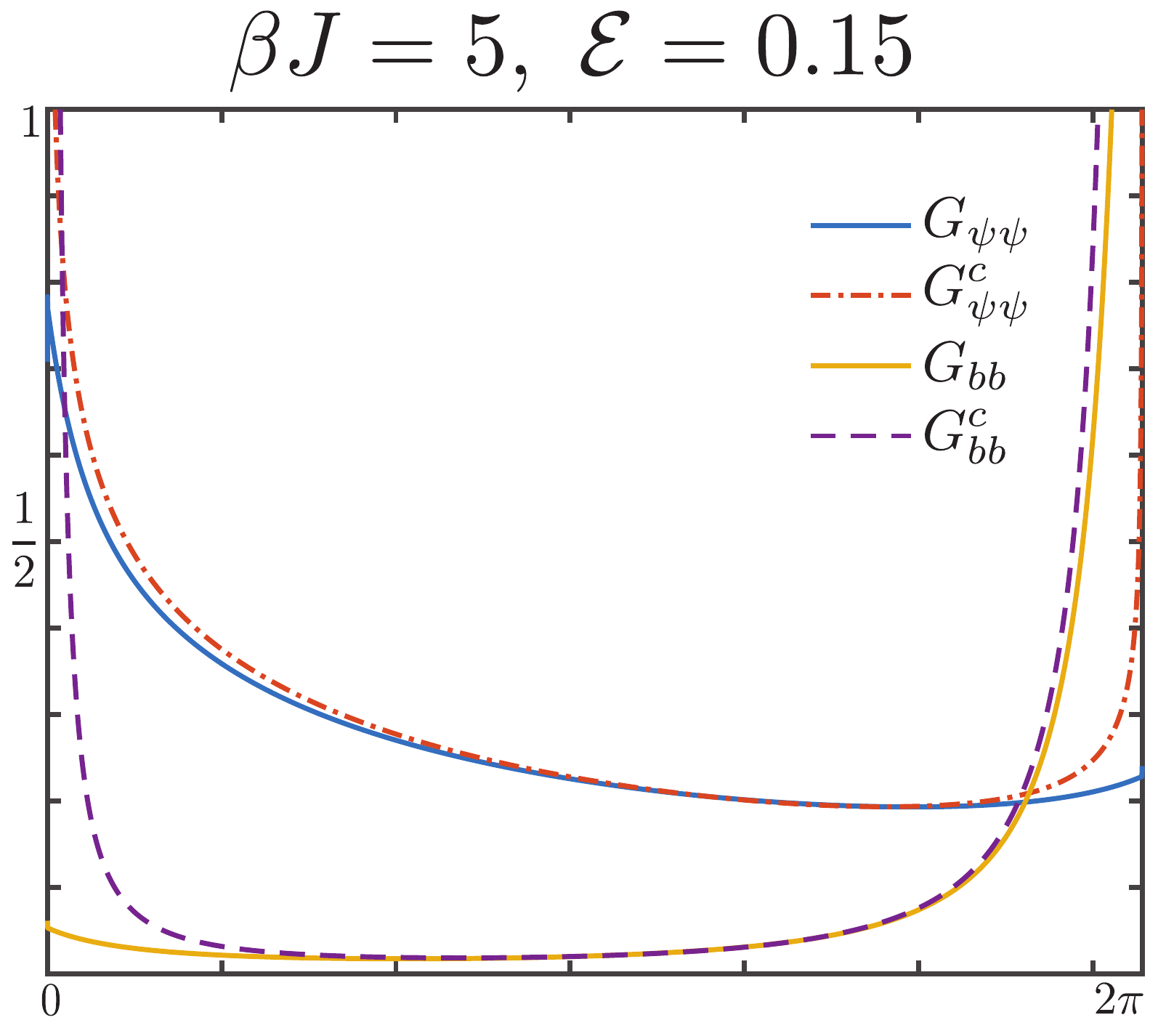}
    \includegraphics[width=0.32\textwidth]{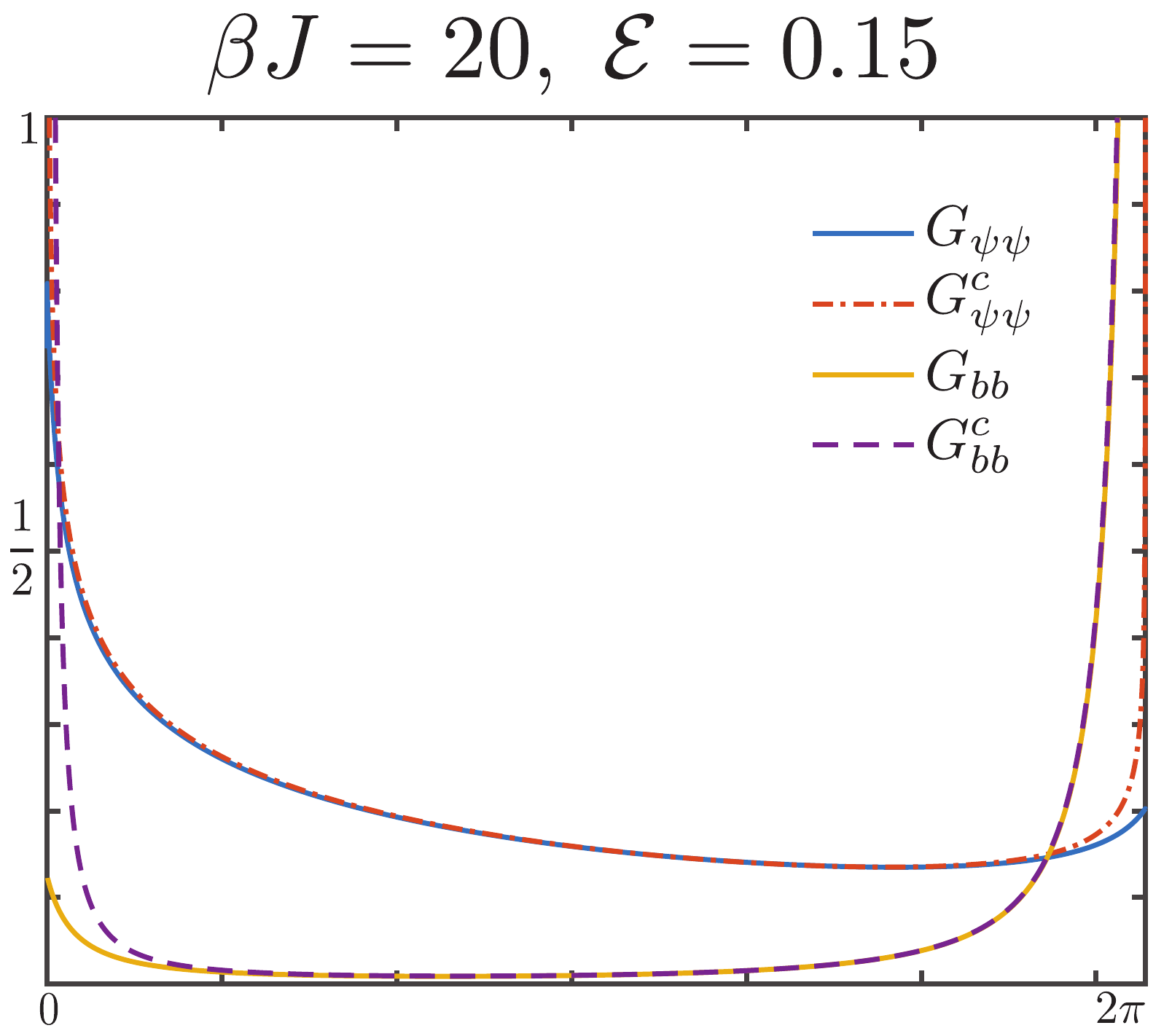}
    \includegraphics[width=0.32\textwidth]{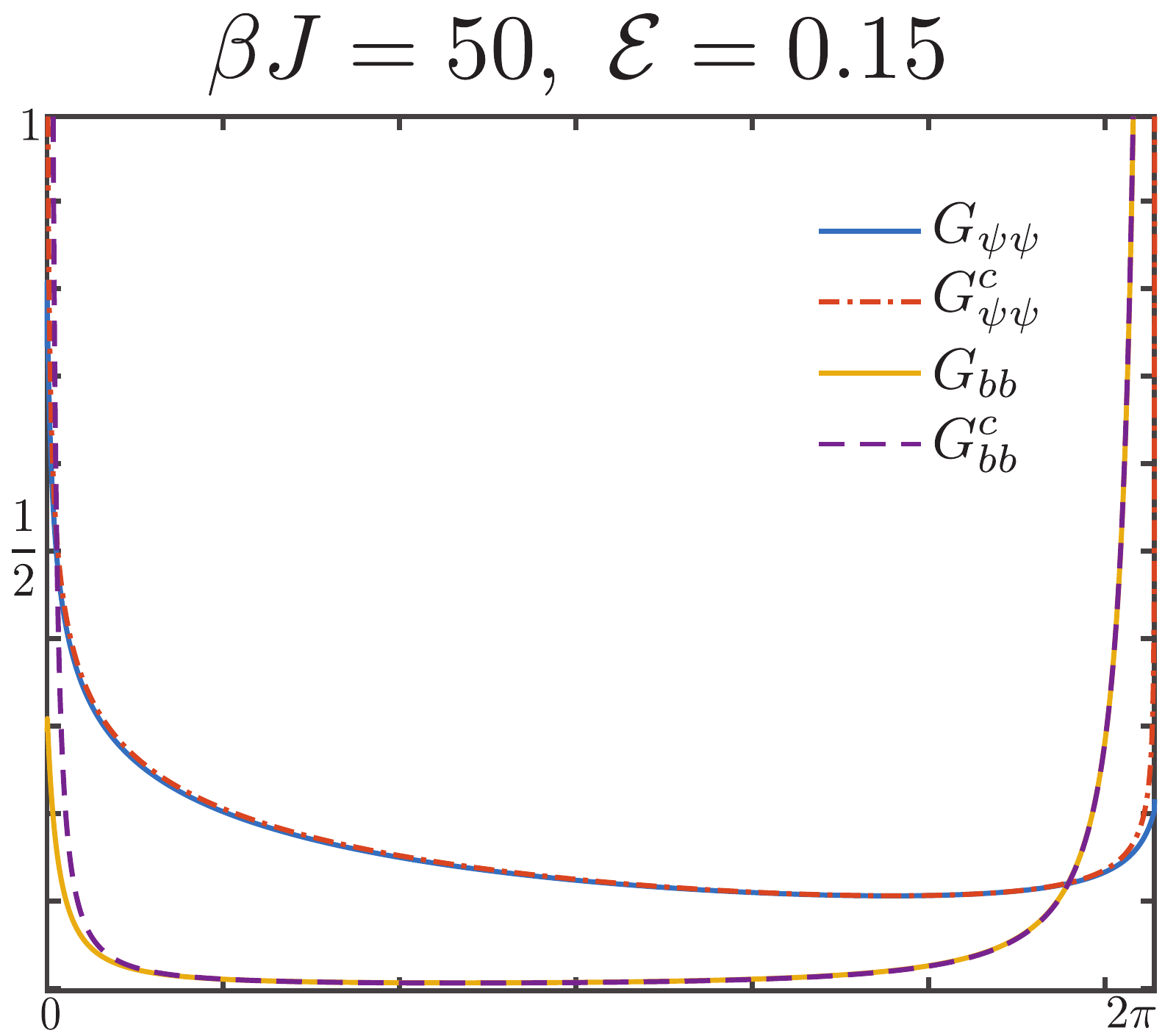}
    \caption{Numerical solutions to large $N$ Dyson Schwinger equations, where from left to right $\beta J=5, 20, 50$ and up to down $\mathcal{E}=0, 0.15.$ We plot against dimensionless quantity $\frac{2\pi\tau}{\beta}$. We also plot the best fit conformal solutions $G^c_{\psi\psi}$ and $G_{b_\psi b_\psi}^c$ and we observe good agreement up when $\tau\sim\mathcal{O}(1/J).$}
    \label{fig:DS_conformal_sol}
\end{figure}

\begin{figure}[t!]
    \centering
    \includegraphics[width=0.5\textwidth]{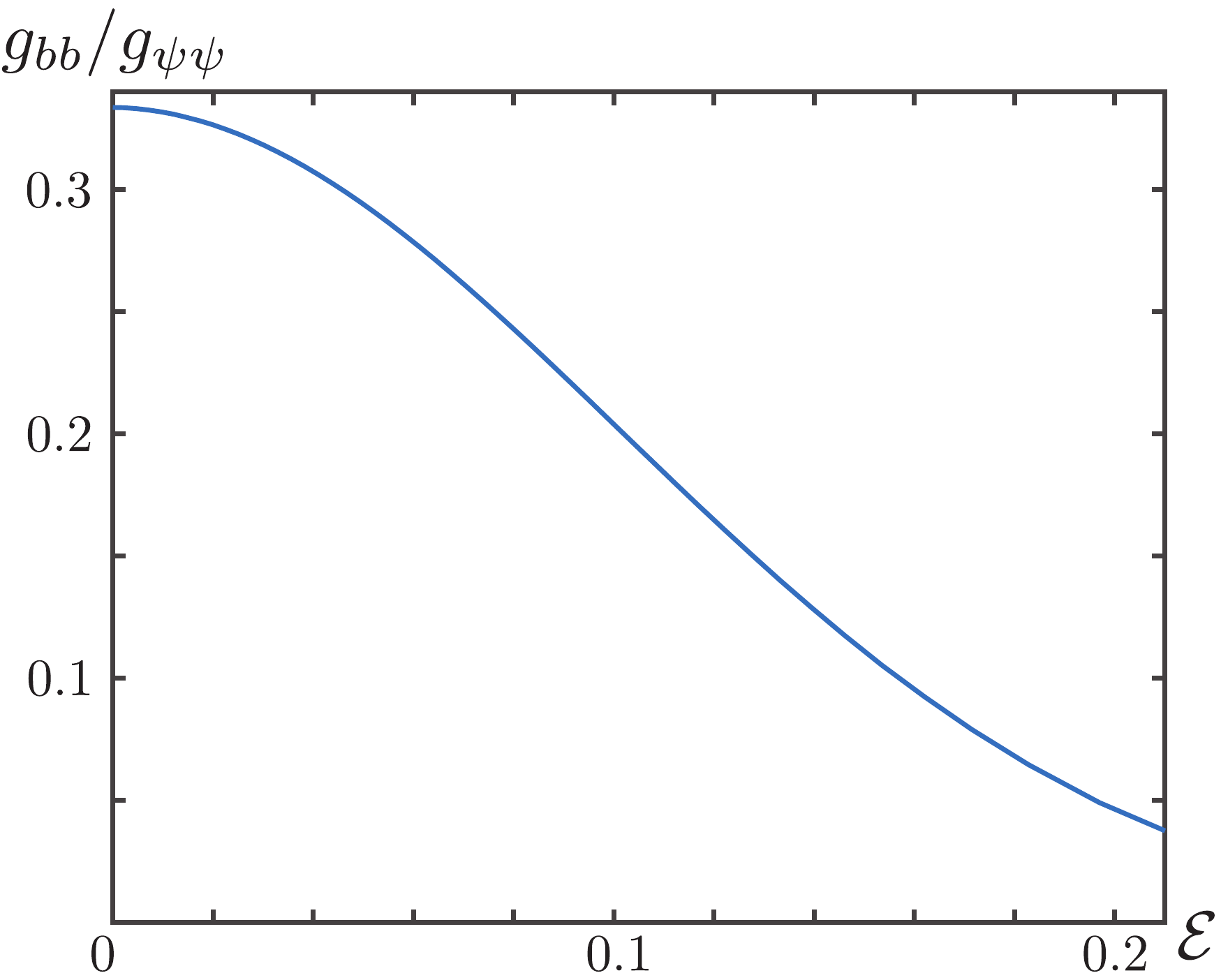}
    \caption{Plot of $g_{bb}/g_{\psi\psi}$ computed from the numerical solution for $q=3$ at different values of $\mathcal{E}.$ Note the infrared Schwinger-Dyson equations can only determine the combination $g_{\psi\psi}^{q-1}g_{bb},$ but the individual values are determined by the full solution. At $\mathcal{E}=0,$ it agrees with the supersymmetric answer $g_{bb}=2\Delta g_{\psi\psi}=\frac{1}{3}g_{\psi\psi}.$}
    \label{fig:PrefactorCurve}
\end{figure}

We can solve equations (\ref{DSeqs}) numerically by iterations. We work in Euclidean time with finite temperature, $\tau\sim \tau+\beta,$ and for numerical purpose we consider a discretized time $\tau_i=\frac{i \beta}{N_{\rm step}}.$ To describe continuous physics we require $N_{\rm step}\gg \beta J.$ We work with grand canonical ensemble and fix $\mu$. For a given $\mu$, the expectation value of the charge can be computed by
\begin{equation}
    \frac{\langle Q \rangle}{N}=\frac{1}{2N}\langle[\bar{\psi}^i,\psi^i] \rangle=\frac{1}{2}\left(G(0)-G(\beta)\right).
\end{equation}
We fit numerical solutions against the Ansatz given in equation (\ref{eq:ansatz_ft}) where we leave $g_{\psi\psi}/g_{bb}$ and $\mathcal{E}_\psi$ unfixed. Other parameters are fixed entirely by IR Schwinger Dyson equations. We take the best fitted $\mathcal{E}$ as the value $\mathcal{E}(\mu).$ Similar to the case of complex SYK, $Q(\mu)$ and $\mathcal{E}(\mu)$ both non-trivially depend on $\mu,$ and by tuning $\mu$ we may understand $Q(\mathcal{E})$. Some examples showing the result of this procedure are presented in figure \ref{fig:DS_conformal_sol}. It is evident that the conformal solution is a good approximation in the IR when $\beta J$ is large. Moreover, the UV boundary conditions also fixes the undetermined ratio $g_{bb}/g_{\psi\psi}$ as shown in figure \ref{fig:PrefactorCurve}.

\begin{figure}[t!]
    \centering
    \includegraphics[width=0.5\textwidth]{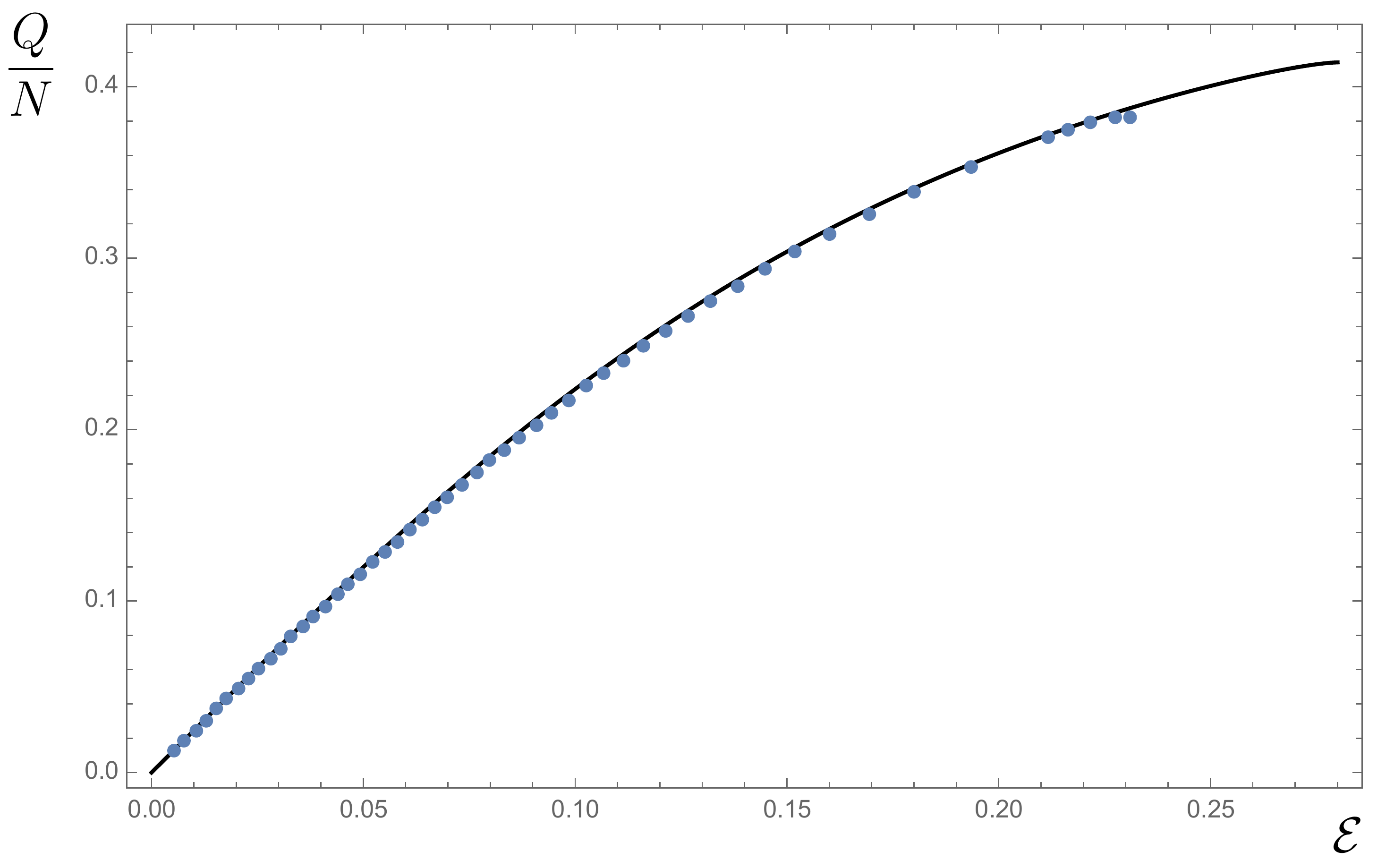}
    \caption{Plot of $Q(\mathcal{E})$ computed with the numerical solution of the Schwinger-Dyson equations for $q=3$ (blue dots). Shown in solid black line is the analytic formula \eqref{eq:LWFuetalmt}, analogous to the Luttinger-Ward relation.}
    \label{fig:QEcurve}
\end{figure}

We now turn to the determination of the charge. Similar to complex SYK, it is possible to derive an analytic relation between the charge and the spectral asymmetry. We do this in Appendix \ref{app:Luttinger}. First we define the fermionic and bosonic contribution to the total charge 
\bea
\mathfrak{q}_f(\Delta,\mathcal{E}) &=& \frac{(\frac{1}{2}-\Delta)\sinh 2\pi \mathcal{E}}{\cosh 2\pi \mathcal{E} + \cos 2 \pi \Delta}+\frac{i}{2\pi} \log \left(\frac{\cos \pi (\Delta + i \mathcal{E})}{\cos \pi (\Delta - i \mathcal{E})}\right),\label{eqn:LuttQf}\\
\mathfrak{q}_b (\Delta_b,\mathcal{E}_b) &=&\frac{(\frac{1}{2}-\Delta_b)\sinh 2\pi \mathcal{E}_b}{\cosh 2\pi \mathcal{E}_b - \cos 2 \pi \Delta_b}+\frac{i}{2\pi} \log \left(\frac{\sin \pi (\Delta_b + i \mathcal{E}_b)}{\sin \pi (\Delta_b - i \mathcal{E}_b)}\right).\label{eqn:LuttQb}
\ea
The total fermion charge is given in terms of these functions by
\beq\label{eq:LWFuetalmt}
\frac{Q}{N} = \mathfrak{q}_f(\Delta,\mathcal{E}) + (q-1) \mathfrak{q}_b(\Delta_b,\mathcal{E}_b).
\eeq
The right hand side is a function of the spectral asymmetry $\mathcal{E}$ explicitly and through the scaling dimensions $\Delta$ and $\Delta_b$. This expression is the sum of two terms. The first is the contribution to the charge from the fermion itself, while the second term is a contribution to the fermion charge from the auxiliary boson. This weird behavior is due to the fact that this zeroth-order boson is an auxiliary field that has been integrated-in to simplify the interactions. This bosonic contribution is similar to the one derived in \cite{PhysRevB.72.024534,Tikhanovskaya:2020zcw} for first-order bosons. We have verified numerically that this relation is correct, and the result is shown in figure \ref{fig:QEcurve}. In particular we can compute the charge that corresponds to the critical spectral asymmetry, and for $q=3$ it is
\beq
Q_{\rm critical} \equiv Q(\mathcal{E}_{\rm critical}) = (\sqrt{2}-1)N,
\eeq
which is smaller than the maximal allowed value $N/2$. Therefore the phase transition we find is physical.

\begin{figure}[t!]
    \centering
    \includegraphics[width=0.465\textwidth]{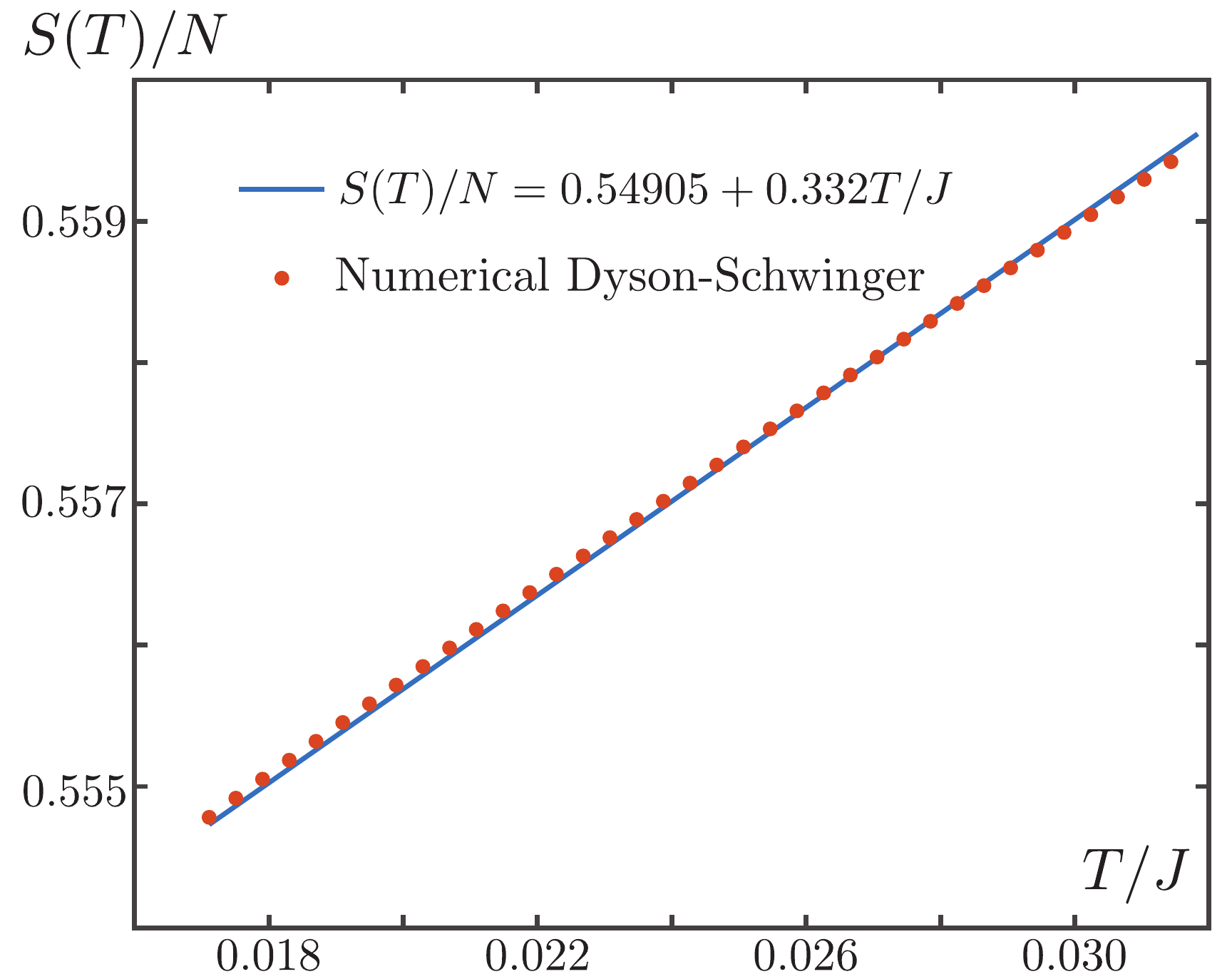}
    \caption{A demonstration that the coupling obtained via the compressibility is consistent with a direct linear fitting of the entropy $S(\beta J).$ The dots are the numerical entropy computed through numerical Dyson Schwinger solutions on a grid of size $2^{20}.$ Due to finite size effects, we can not take $T$ to be too small. We show a best fit with slope fixed to be the answer determined through compressibility. The intercept, $S_0$, is close to its analytical answer. We can also fit both the slope and the intercept, and we obtain $S(\beta J)=0.5493+0.3233/(\beta J).$ The difference is about $2.7\%$ and we conclude the answer determined both ways are consistent with each other. }   
    \label{fig: Schwarzian_coupling}
\end{figure}

At low energy and for charges small compared to $N$,  the $\mathcal{N}=2$ Schwarzian description is reliable. We can numerically determine the Super-Schwarzian coupling. Instead of directly fitting the free energy, we find it more accurate and less computationally intensive to compute the compressibility. Due to the $\mathcal{N}=2$ supersymmetry, the compressibility of the $U(1)_R$ symmetry is directly related to the super-Schwarzian coupling, as the relative coefficient between the reparametrization modes $f(\tau)$ and the $U(1)$ sigma model is fixed by supersymmetry. The bosonic part of the $\mathcal{N}=2$ Super-Schwarzian is given by \cite{Stanford:2017thb}
\begin{equation}
\label{eq:SchwarzianU(1)}
    S_b=\frac{2\pi N \alpha_S}{\beta J}\int_0^{2\pi}d\tau\left(-\text{Sch}\left(\tan\frac{f}{2},\tau\right)+ 2q^2\left(\partial_{\tau}a\right)^2\right)
\end{equation}
where $f$ is the reparametrization mode on the circle and $a(\tau)$ is the generator of the $U(1)_R$ symmetry. We have rescaled time to have period $2\pi$ for convenience. The additional factor of $q^2$ comes from normalizing the fundamental fermion to have $R$ charge 1 instead of $1/q$. On the other hand, the compressibility can be found through $U(1)_R$ sigma model action as  
\begin{equation}
    S_b\supset 2\pi \int_0^{2\pi}  \frac{ K}{2\beta} \left(\partial_{\tau}a \right)^2.
\end{equation}
We can therefore compute compressibility $K$ from the numerical Dyson-Schwinger equations and extrapolate the $\mathcal{N}=2$ Schwarzian coupling as 
\begin{equation}
    \alpha_S=\frac{1}{4 q^2} \frac{K J}{N}.
\end{equation}
To compute the compressibility, we turn on a small chemical potential, numerically solve the Schwinger-Dyson equations on a grid of size $2^{19}$ and specialize to $q=3$; the result is
\begin{equation}
    K=\left( \frac{dQ}{d\mu}\right)_{T=0}=\lim_{\mu\to 0} \frac{Q}{\mu}\approx 0.303N/J.
\end{equation}
This in term determines the Schwarzian coupling to be $\alpha_S\approx 0.00842.$ This is different from the numerical coupling of the $\mathcal{N}=0$ Schwarzian for complex SYK which is given by $\alpha_S^{\rm cSYK}=0.01418$ \cite{Maldacena:2016hyu,Gu:2019jub}. As we vary $\mathcal{E},$ we can also compute the free energy and entropy. At low temperature and zero charge, the Schwarzian theory implies a free energy which admits an expansion 
\begin{equation}
    -\beta F= -\beta E_0+ S_0+\frac{c}{2\beta},~~~ c=\frac{4\pi^2\alpha_S N}{J}\approx 0.332 \frac{N}{J}.
\end{equation}
where $E_0$ corresponds to the ground state energy and takes a non-zero value due to normal ordering when going from the Hamiltonian to the mean field action.\footnote{We thank Yingfei Gu and Pengfei Zhang for a very useful discussion on this point.} The $S_0$ is the zero temperature entropy and the $1/\beta$ term is due to the Super-Schwarzian. In figure \ref{fig: Schwarzian_coupling} we checked that the answer is consistent with a direct linear fitting against the entropy $S(\beta J).$

\begin{figure}[t!]
    \centering
    \includegraphics[width=0.465\textwidth]{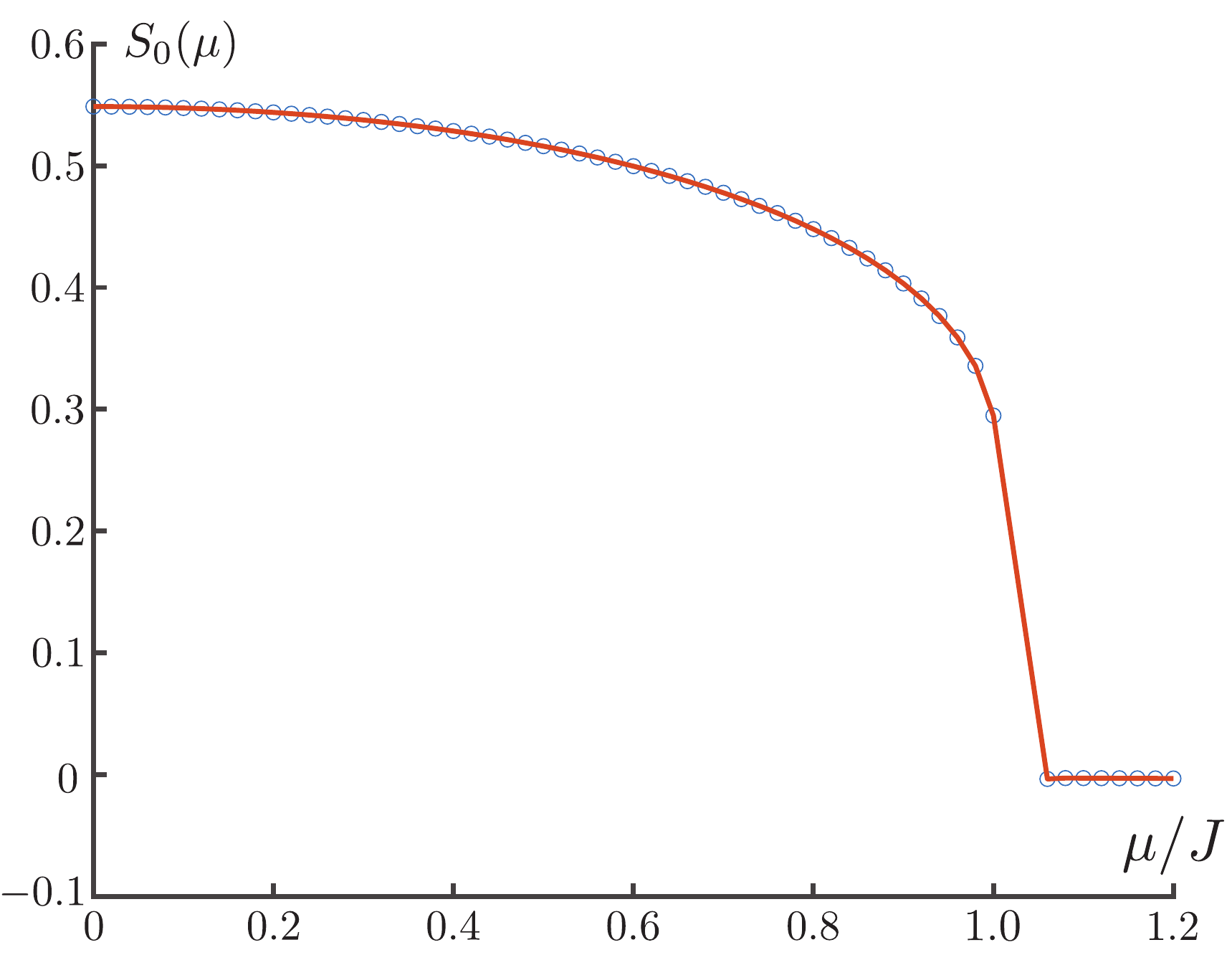}
    \caption{The entropy computed numerically as a function of the chemical potential $\frac{\mu}{J}.$ For each value of $\mu,$ we compute the free energy and entropy on a numerical grid of size $2^{25}$ at small temperatures and extrapolate the zero temperature entropy. Note at $\mu=0,$ we obtain $S_0(0)\approx 0.5484$ which is close to the value predicted by the index $\log(2\cos\frac{\pi}{6})\approx 0.5493.$ The transition happens slightly above $\mu=\mu_c=J,$ where the exponential decaying solution (\ref{eq:decaying_sol}) starts to appear. Extrapolating the precise zero temperature entropy becomes involved near the transition point. }
    \label{fig: entropy}
\end{figure}

We now explain how to use the numerical solution to compute the thermodynamic potentials. Since in the numerical procedure we fix $\mu,$ the grand potential can be computed by the on-shell action as 
\begin{equation}\label{eq:freeG}
    -\beta \Omega(\mu, T) /N= \log\left(2\cosh\left(\frac{\beta \mu}{2}\right)\right)+\sum_\omega \log\left(\frac{1+\frac{\Sigma_{\psi\psi}}{i\omega +\mu}}{1+\Sigma_{bb}}\right)-\beta \int_0^\beta d\tau ~ G_{\psi\psi}(\tau)\Sigma_{\psi\psi}(\tau),
\end{equation}
where we used the Schwinger-Dyson equations and the time translation of the solution to simplify the last term. To compute the free energy $F$, we can change to the canonical ensemble by 
\beq\label{eq:freeE}
\beta F= \beta\Omega + \beta \mu Q.
\eeq
 \begin{figure}[t!]
    \centering
    \includegraphics[width=0.465\textwidth]{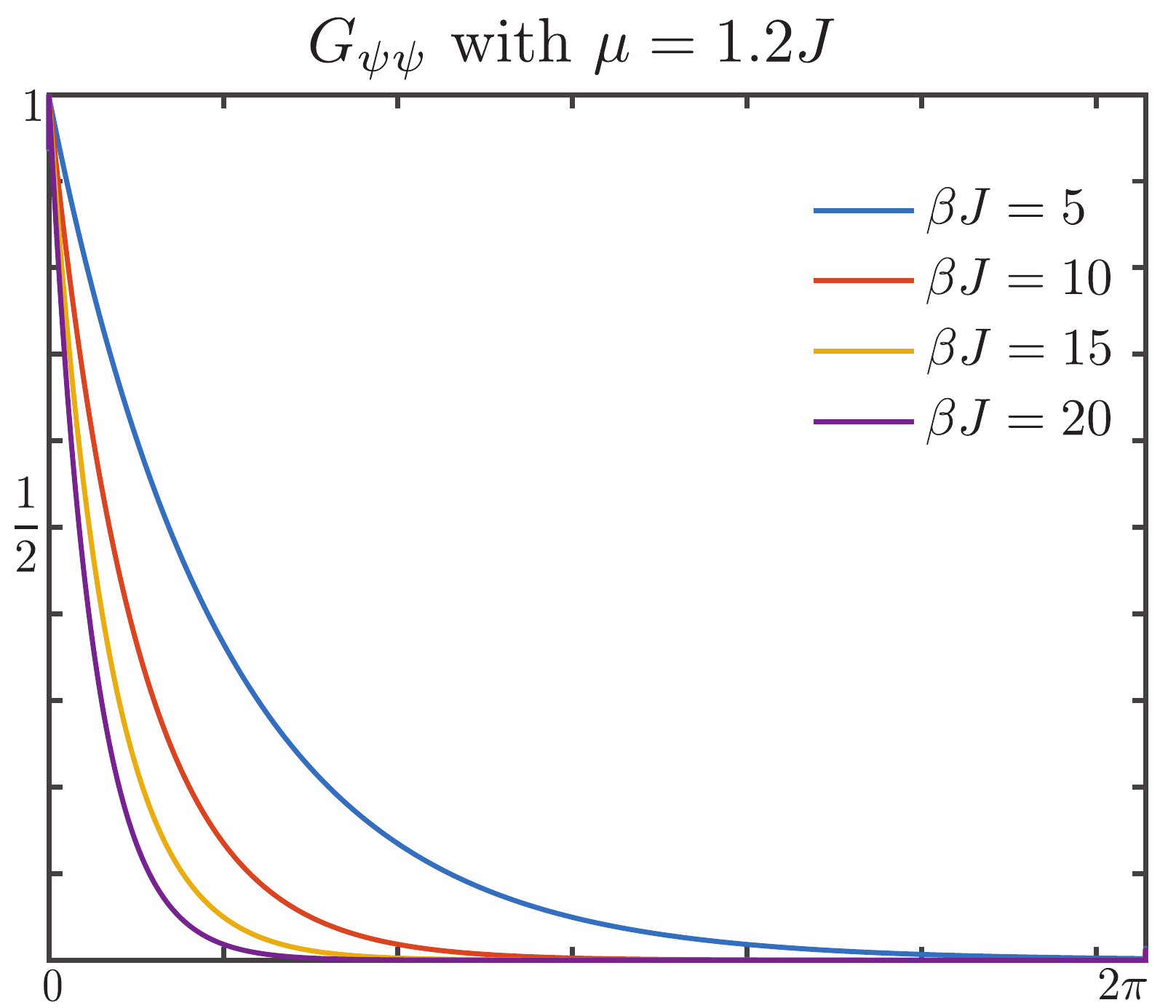}
   \includegraphics[width=0.485\textwidth]{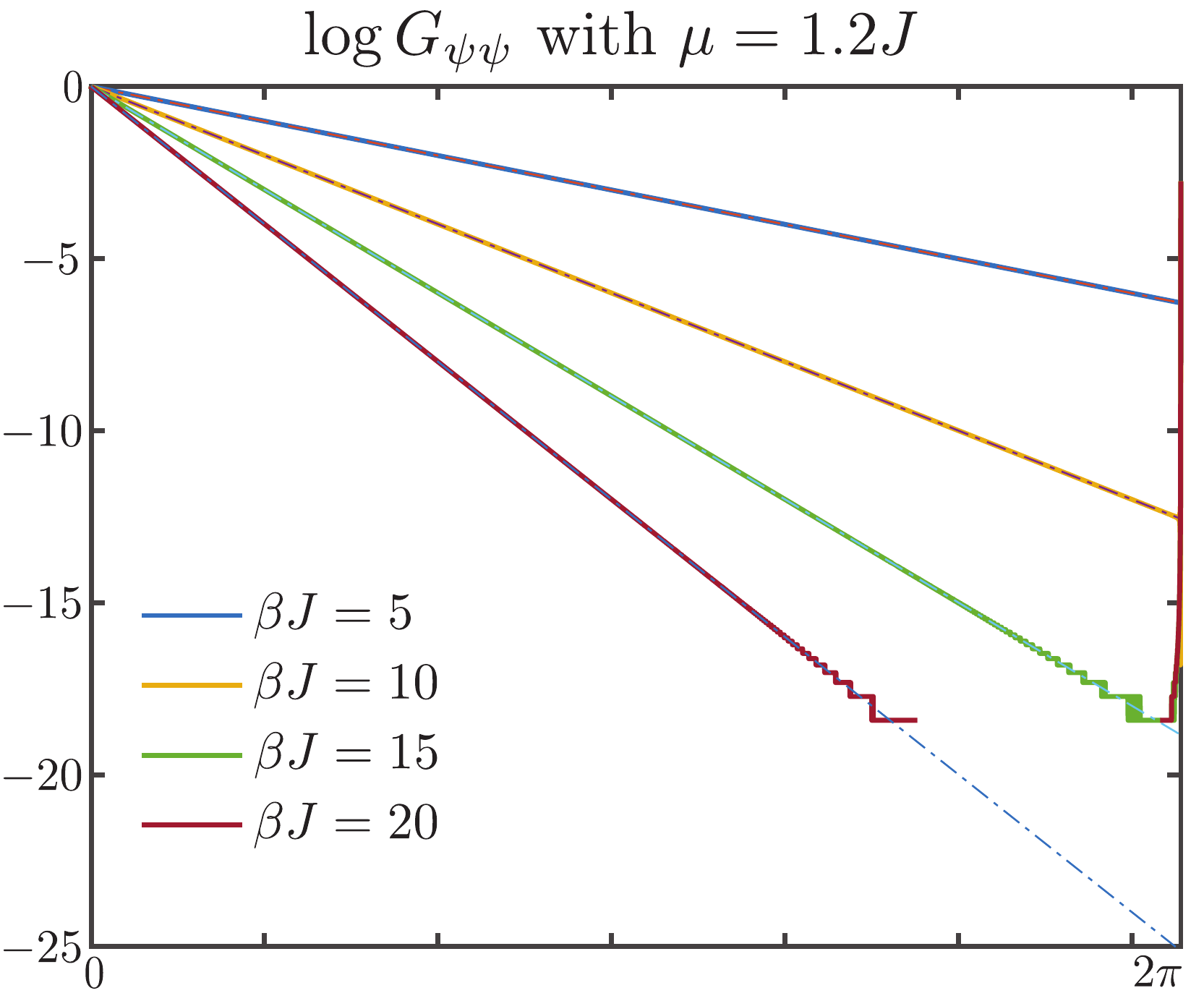}
    \caption{Left: Numerical solutions of $G_{\psi\psi}$ in the region where $\mathcal{E}>\mathcal{E}_{\text{critical}}.$ We observe exponential decay solutions. Since the solution ceases to be conformal, the infrared parameter $\mathcal{E}$ is no longer meaningful. The solutions depend on both $\beta J$ and $\mu/J.$ Right: Log plot of $G_{\psi\psi}$ at various values of $\beta J,$ where dashed lines are linear fits. We observe that the exponent is linear in $\mu$.}
    \label{fig:DS_exp_sol}
\end{figure}
For general $\mu$ we compute the free energy and entropy numerically from (\ref{eq:freeE}). We show the plot of entropy in the grand canonical ensemble against chemical potential in figure \ref{fig: entropy}. We observe a sharp transition near $\mu=J,$ where a new decaying exponential solution to the Schwinger-Dyson equations(as shown in figure. \ref{fig:DS_exp_sol}) starts to exist. The discontinuous jump in entropy is consistent with a first order transition. The transition goes from a high entropy phase to a low entropy phase, where $S_0/N$ is approximately zero. At zero temperature, such exponentially decay solutions can be found analytically to be 

\begin{equation}\label{eq:decaying_sol}
    G_{\psi\psi}\left(\tau\right)=e^{-(\mu-\mu_c)\tau}\Theta\left(\tau\right),~~ G_{bb}\left(\tau\right)=-\delta\left(\tau\right)+ e^{((q-1)(\mu-\mu_c)+J)\tau}\Theta\left(-\tau\right),
\end{equation}
where the critical chemical potential is determined to be 
\begin{equation}
    \mu_c= \frac{J}{2}(q-1). 
\end{equation}
To verify that this is a solution of the Schwinger-Dyson equations, begin by writing the ansatz $G_{\psi\psi}\left(\tau\right)=e^{-(\mu-\mu_c)\tau}\Theta\left(\tau\right),$ and use $\Sigma_{bb}(\tau)=J G_{\psi\psi}(\tau)^{q-1}$ to obtain the boson self-energy, in both position and Fourier space as
\begin{equation}
    \Sigma_{bb}\left(\tau\right)= J e^{-(q-1)(\mu-\mu_c)\tau}\Theta\left(\tau\right), ~\Rightarrow~ \Sigma_{bb}\left(\omega\right)= \frac{J}{-i \omega+(q-1)(\mu-\mu_c)}. 
\end{equation}
We can use the momentum space equation to determine $G_{bb}(\omega)$ and Fourier transform back to position space to obtain $ G_{bb}\left(\tau\right)=-\delta\left(\tau\right)+ e^{((q-1)(\mu-\mu_c)+J)\tau}\Theta\left(-\tau\right)$. The position space equation $\Sigma_{\psi\psi}(\tau)$ determines the value of $\mu_c$ since 
\begin{equation}
    \Sigma_{\psi\psi}\left(\tau\right)= J(q-1) e^{-(q-2)(\mu-\mu_c)\tau}\Theta\left(\tau\right)\left(-\delta(\tau)+e^{((q-1)(\mu-\mu_c)+J)\tau}\Theta(-\tau)\right)= -\frac{J(q-1)}{2} \delta(\tau).
\end{equation}
where we used $\Theta(\tau) \delta(\tau)=\frac{1}{2}\delta(\tau).$ We note that when $J=0$, $\mu_c$ is zero and the solution reduces to the retarded propagator of free massive fermion. The non-zero value $\mu_c$ can be thought as the minimal value required for the chemical potential to overtake the random interaction and creates a gap in the spectrum. For $q=3,$we note $\mu_c=J.$ Numerically a sharp transition occurs around this value.

\subsection{Operator spectrum}\label{sec:FuOpSpe}
In this section we will determine the scaling dimensions of bilinear operators at the nearly conformal fixed point. The operator spectrum provides important insights of the IR physics, in directions orthogonal to the reparametrization modes. The spectrum of the theory \eqref{eq:fHamiltonian} at zero charge has been worked out and extensively discussed in \cite{Fu:2016vas,Peng:2017spg, Peng:2020euz}. We generalize the analysis to non-zero chemical potential, which explicitly breaks supersymmetry. However, in the infrared, supersymmetry is only spontaneously broken by the nearly conformal solution. Since the infrared equations of motions remain supersymmetric, the superspace formalism provides a powerful tool to analyze the bilinear spectrum. The scaling dimensions of operators become continuous function of the chemical potential $\mu,$ or equivalently the spectral asymmetry parameter $\mathcal{E}.$

In particular, we will check that the spectrum does not contain an operator with a complex scaling dimension. Such a complex mode corresponds (for instance) to a bulk field below its BF bound, and thus suggests an instability. Such modes are observed in various examples in nearly conformal theory \cite{Klebanov:2018fzb,Kim:2019upg,Klebanov:2020kck}. In the presence of a complex mode, the infrared physics is not correctly described by the conformal saddle. In our case we verify that there are no complex modes in the range $0\leq \mathcal{E}< \mathcal{E}_{\rm critical},$ and thus the conformal solution exhibits a consistent spectrum. As $\mathcal{E} \rightarrow \mathcal{E}_{\rm critical}$ (the point of the phase transition), we observe the spectrum exhibits unusual features such as a continuum of states.

To determine the spectrum of bilinear operators, we find it convenient to look at the variation of large $N$ Dyson Schwinger equation, and set it to zero. To see that such a procedure determines the bilinear spectrum, we note that Dyson Schwinger equations are operator equations, and we may insert additional operators at infinity while the equations still hold. Such an insertion corresponds to a deformation 
\begin{equation}
\label{eq:SDinsertOp}
    \delta G_{\psi\psi}= \langle \bar{\psi}_i(\tau)\psi^i(0) \mathcal{O}_h(\infty) \rangle \, , \quad \delta G_{bb}= \langle \bar{b}_i(\tau)b^i(0) \mathcal{O}_h(\infty) \rangle
\end{equation}
To make such deformation non-zero, $\mathcal{O}$ must be a bilinear operator consisting of fundamental fermions and bosons. Without losing of generality, we may restrict $\mathcal{O}$ to be primary. 

Imposing $\delta G_{\psi\psi}$ satisfies the Dyson Schwinger equation provides a necessary consistency condition for the deformation to correspond to an operator insertion, and it `bootstraps' the operator spectrum. Note that this is only a necessary condition, and it is not guaranteed that all such deformation corresponds to non-trivial operators. Nontrivial operators are distinguished by their $SL(2)$ Casimir (or the supersymmetric extension) and respect the $h\rightarrow (1-h)$ symmetry which may relate different $\delta G$. 

To write the infrared Dyson Schwinger equations \eqref{eq:IRFuSD} in a more compact form, we let $\mathcal{G}(Z_1,Z_2)^T=\mathcal{G}(Z_2,Z_1)$ and let $(\star$, $\bar{\star})$ be chiral and anti-chiral convolutions, respectively. 
In this notation, the infrared equations of motion are:
\beq\label{eq:SUSYDS1}
\frac{J}{2}\mathcal{G}\star (\mathcal{G}^{q-1})^T \equiv \int dZ_2 ~\mathcal{G}(Z_1,Z_2)~ (\frac{1}{2}J\mathcal{G}(Z_3;Z_2)^{q-1}) = (\bar{\theta}_1-\bar{\theta}_3)\delta(\tau_1-\tau_3-\theta_1\bar{\theta}_1+\theta_3\bar{\theta}_3).
\eeq
\beq\label{eq:SUSYDS2}
\frac{J}{2}(\mathcal{G}^{q-1})^T\bar{\star}\mathcal{G}  \equiv \int d\bar{Z}_2 ~\mathcal{G}(Z_2,Z_1)~ (\frac{1}{2}J\mathcal{G}(Z_2;Z_3)^{q-1}) = (\theta_1-\theta_3)\delta(\tau_1-\tau_3+\theta_1\bar{\theta}_1-\theta_3\bar{\theta}_3).
\eeq
where $\mathcal{G}=\langle \bar{\Psi}_i(Z_1)\Psi^i(Z_2)\rangle$ is the anti-chiral-chiral propagator, and $D_{\theta_1}\mathcal{G}=\bar{D}_{\theta_2}\mathcal{G}=0.$ These equations are manifestively invariant under the $SU(1,1|1)$ transformation (\ref{eqn:FususytransfIR}). However, the infrared solutions with $\mathcal{E}\neq 0$ are not invariant under such transformations. Thus supersymmetry is only spontaneously broken in the infrared. In such a case, the fermionic transformations in $SU(1,1|1)$ transform the solution with zero $G_{\psi b}$ to solutions with non zero values of $G_{\psi b}.$

We vary Eq.(\ref{eq:SUSYDS1}) on both sides and take the convolution against $\mathcal{G}$ from the right. Noting $(A\star B)\bar{\star}C=-A\star (B\bar{\star}C),$ we obtain 
\beq\label{eq:kernel_def}
\begin{split}
   \delta \mathcal{G}(Z_1,Z_2)&=\frac{J(q-1)}{2}\left(\mathcal{G}\star\left(\left(\mathcal{G}^{q-2}\right)^T\delta \mathcal{G}^T\right)\right)\bar{\star} \mathcal{G}\\&= \frac{J(q-1)}{2} \int d\bar{Z}_3dZ_4 \mathcal{G}(Z_1,Z_4)\mathcal{G}(Z_3,Z_4)^{q-2} \delta \mathcal{G}(Z_3,Z_4)\mathcal{G}(Z_3,Z_2)\\&=\int d\bar{Z}_3 dZ_4 K^{\mathcal{N}=2}(Z_1,Z_2; Z_3,Z_4)\delta \mathcal{G}(Z_3,Z_4)
\end{split}
\eeq
In other words, $\delta \mathcal{G}$ must be an eigenfunction of the kernel 
\beq\label{eq:N2kernel}
K^{\mathcal{N}=2}(Z_1,Z_2; Z_3,Z_4)=\frac{J(q-1)}{2}\mathcal{G}(Z_1,Z_4)\mathcal{G}(Z_3,Z_2)\mathcal{G}(Z_3,Z_4)^{q-2},
\eeq
with eigenvalue 1. The eigenvalue equation determines the spectrum. Note the kernel can always be written in $\mathcal{N}=2$ superspace. If we assume full $SU(1,1|1)$ invariance on the super-correlator $\mathcal{G}$, the kernel can also be diagonalized directly in the superspace \cite{Bulycheva:2018qcp}. In that case, the spectrum is guaranteed to organize into $SU(1,1|1)$ supermultiplets with relative scaling dimensions fixed by symmetry. However, our conformal solution spontaneously breaks SUSY. Within a supermultiplet, the SUSY constraints on scaling dimensions no longer hold. Thus it is necessary to work in components when we diagonalize the super-kernel.

To write in components, we assume translational invariance and the supercorrelator can be expanded as: 
\begin{eqnarray}
    \mathcal{G}(Z_1,Z_2)&=&G_{\psi\psi}(\tau_1-\tau_2-\theta_1\bar{\theta}_1-\theta_2\bar{\theta}_2)+\sqrt{2}\bar{\theta}_1G_{b\psi}(\tau_1-\tau_2-\theta_2\bar{\theta}_2)\nonumber\\
    &&-\sqrt{2}\theta_2 G_{\psi b}(\tau_1-\tau_2-\theta_1\bar{\theta}_1)+2\bar{\theta}_1\theta_2 G_{bb}(\tau_1-\tau_2).
\end{eqnarray}
Now we evaluate the spectrum of the kernel Eq.(\ref{eq:N2kernel}) over the conformal saddle. In such case the vacuum carries a definite fermionic number, and 
\begin{equation}
    \mathcal{G}(Z_1,Z_2)=G_{\psi\psi}(\tau_1-\tau_2-\theta_1\bar{\theta}_1-\theta_2\bar{\theta}_2)+2\bar{\theta}_1\theta_2 G_{bb}(\tau_1-\tau_2),
\end{equation}
but the fluctuation does not:
\begin{eqnarray}
    \delta \mathcal{G}(Z_1,Z_2)&=&\delta G_{\psi\psi}(\tau_1-\tau_2-\theta_1\bar{\theta}_1-\theta_2\bar{\theta}_2)+\sqrt{2}\bar{\theta}_1\delta G_{b\psi}(\tau_1-\tau_2-\theta_2\bar{\theta}_2)\nonumber\\
    &&-\sqrt{2}\theta_2 \delta G_{\psi b}(\tau_1-\tau_2-\theta_1\bar{\theta}_1)+2\bar{\theta}_1\theta_2 \delta G_{bb}(\tau_1-\tau_2).
\end{eqnarray}
At this point we specialize to the case of $q=3$ for simplicity. Evaluating the super kernel integral in components, and matching the corresponding components in $\delta \bar{\mathcal{G}}$ we obtain
\begin{equation}
    \delta G_{\psi\psi}(\tau_{12})=-2 J\int d\tau_3 d\tau_4 \left(G_{\psi\psi}(\tau_{14})G_{\psi\psi}(\tau_{32})G_{\psi\psi}(\tau_{34})\delta G_{bb}(\tau_{34})+G_{\psi\psi}(\tau_{14})G_{\psi\psi}(\tau_{32})G_{bb}(\tau_{34})\delta G_{\psi\psi}(\tau_{34})\right)
\end{equation}
\begin{equation}
    \delta G_{\psi b}(\tau_{12})=2 J\int d\tau_3 d\tau_4 G_{bb}(\tau_{32})G_{\psi\psi}(\tau_{14})G_{\psi\psi}(\tau_{34})\delta G_{\psi b}(\tau_{34})
\end{equation}
\begin{equation}
    \delta G_{b b}(\tau_{12})=2J\int d\tau_3 d\tau_4 G_{bb}(\tau_{14})G_{bb}(\tau_{32}) G_{\psi\psi}(\tau_{34})\delta G_{\psi\psi}(\tau_{34}).
\end{equation}
We can organize the fluctuations in the bosonic and fermionic sectors as
\begin{align}\label{eq:bosonicspec}
    K_b &=2J \begin{pmatrix}-G_{\psi\psi}(\tau_{14})G_{\psi\psi}(\tau_{32})G_{bb}(\tau_{34}) & -G_{\psi\psi}(\tau_{14})G_{\psi\psi}(\tau_{32})G_{\psi\psi}(\tau_{34}) \\ G_{bb}(\tau_{14})G_{bb}(\tau_{32})G_{\psi\psi}(\tau_{34})& 0 \end{pmatrix} \, \\ K_b&*\begin{pmatrix}\delta G_{\psi\psi} \\ -\delta G_{bb}\end{pmatrix}=\begin{pmatrix}\delta G_{\psi\psi} \\ \delta G_{bb}\end{pmatrix} \, , \\
\label{eq:fermionicspec}
    K_f&=2J G_{bb}(\tau_{32})G_{\psi\psi}(\tau_{14}) G_{\psi\psi}(\tau_{34}) \, , \\
    \bar{K}_{f}&=2J G_{\psi\psi}(\tau_{32})G_{bb}(\tau_{14}) G_{\psi\psi}(\tau_{34}) \, .
\end{align}
We note that the spectrum only depends on the combination $g_{\psi\psi}^2 g_{bb},$ although individual matrix element can depend on the precise value of $g_{\psi\psi} $ and $g_{bb}.$  

To proceed we note that with non zero spectral asymmetry, the three point function of \eqref{eq:SDinsertOp} can be an arbitrary linear combination of the symmetric and antisymmetric three point functions:
\begin{equation}\label{eq:deformationAnsatz}
    \delta G_{\psi\psi}=\frac{A}{|\tau|^{2\Delta-h}}+\frac{B~\text{sgn}(\tau)}{|\tau|^{2\Delta-h}} \, , \quad \delta G_{bb}=\frac{a}{|\tau|^{2\Delta_b-h}}+\frac{b~\text{sgn}(\tau)}{|\tau|^{2\Delta_b-h}}.
\end{equation} 
Due to such a mixing, each matrix element in the bosonic and fermionic kernel should be thought as a 2 by 2 matrix. It is enough to work out the kernel with general spectral asymmetry, defined by 
\begin{equation}
\begin{split}
    &K(\{\tau,\mathcal{E}, \Delta\})=\\&\frac{\left(e^{\pi\mathcal{E}_1}\Theta(\tau_{14})-e^{-\pi\mathcal{E}_1}\Theta(\tau_{41})\right)}{|\tau_{14}|^{2\Delta_1}}\frac{\left(e^{\pi\mathcal{E}_2}\Theta(\tau_{32})-e^{-\pi\mathcal{E}_2}\Theta(\tau_{23})\right)}{|\tau_{23}|^{2\Delta_2}}\frac{\left(e^{\pi\mathcal{E}_3}\Theta(\tau_{34})-e^{-\pi\mathcal{E}_3}\Theta(\tau_{43})\right)}{|\tau_{34}|^{2\Delta_3}}
\end{split}
\end{equation}
 Explicitly, by evaluating 
 \begin{equation}
     \int d\tau_3 d\tau_4K\left(\{\tau, \mathcal{E}, \Delta\}\right)\begin{pmatrix}
     \frac{\text{sgn}(\tau_{34})}{|\tau_{34}|^{2\Delta-h}}\\\frac{1}{|\tau_{34}|^{2\Delta-h}} 
     \end{pmatrix}=K(\{\mathcal{E}, \Delta\})\begin{pmatrix}
     \frac{\text{sgn}(\tau_{34})}{|\tau_{34}|^{2\Delta-h}}\\\frac{1}{|\tau_{34}|^{2\Delta-h}} 
     \end{pmatrix}
 \end{equation}
 we can find its matrix representation in the basis $\left(\frac{\text{sgn}(\tau)}{\tau^{2\Delta-h}},\frac{1}{\tau^{2\Delta -h}}\right). $
\begin{equation}\label{eq:genericmintegral}
    K(\{\mathcal{E}, \Delta\})=\frac{1}{4\pi}\begin{pmatrix}c_f(\frac{3}{2}-\sum\Delta_i,0)(\bar{Q}-Q)& c_f(\frac{3}{2}-\sum\Delta_i,0)(P+\bar{P}) \\c_b(\frac{3}{2}-\sum\Delta_i,0)(Q+\bar{Q}) &-c_b(\frac{3}{2}-\sum\Delta_i,0)(P-\bar{P}) \end{pmatrix}
\end{equation}
where 
\begin{equation}
    c_f(\Delta,\mathcal{E})=2i \cos(\pi(\Delta +i\mathcal{E}))\Gamma(1-2\Delta) \, , c_b=2 \sin(\pi(\Delta +i\mathcal{E}))\Gamma(1-2\Delta),
\end{equation}
and we introduced the parameters
\begin{equation}
    P= c_f(\Delta_i=1,\mathcal{E}_1)c_f(\Delta_2,\mathcal{E}_2)c_f(\Delta_3,-\mathcal{E}_3),\bar{P}=c_f(\Delta_1,-\mathcal{E}_1)c_f(\Delta_2,-\mathcal{E}_2)c_f(\Delta_3,\mathcal{E}_3),
\end{equation}
\begin{equation}
    Q=c_f(\Delta_1,\mathcal{E}_1)c_f(\Delta_2,-\mathcal{E}_2)c_b(\Delta_3,\mathcal{E}_3),\bar{Q}=c_f(\Delta_1,-\mathcal{E}_1)c_f(\Delta_2,-\mathcal{E}_2)c_b(\Delta_3,\mathcal{E}_3).
\end{equation}
\begin{figure}[t!]
    \centering
    \includegraphics[width=0.495\textwidth]{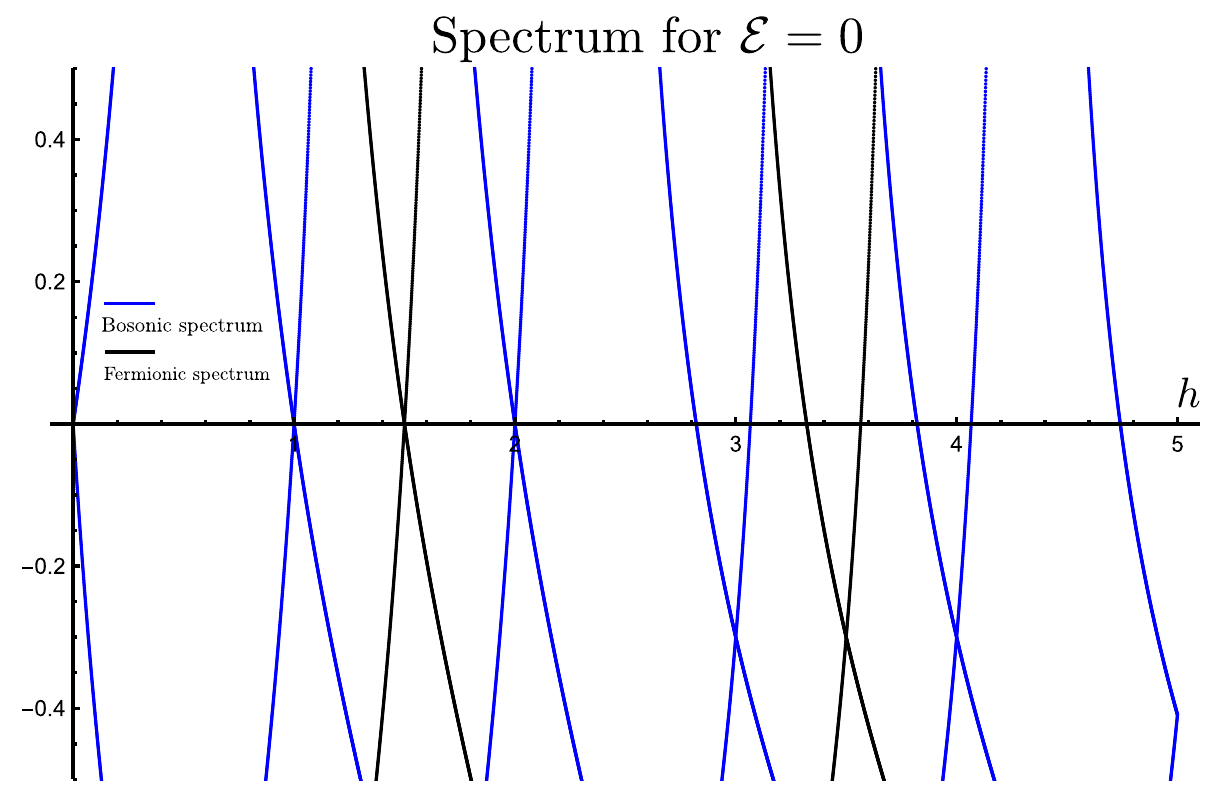}
    \includegraphics[width=0.495\textwidth]{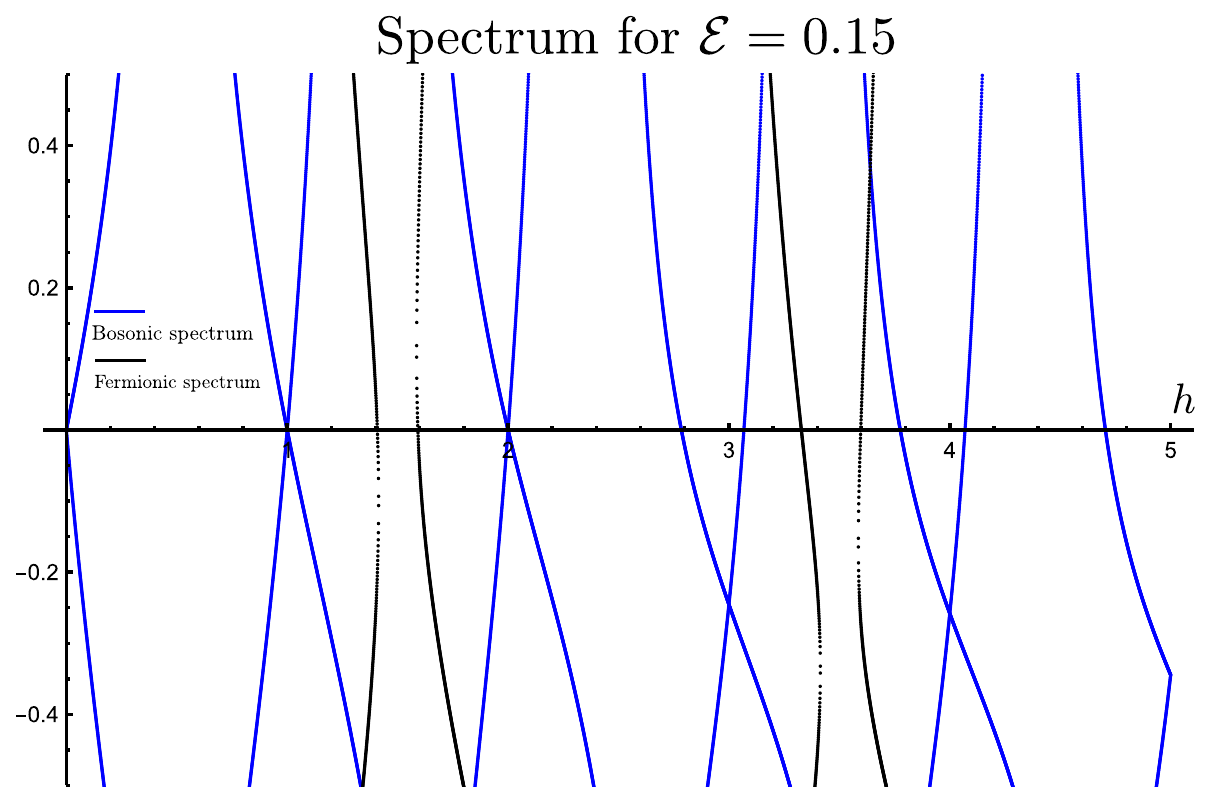}
    \caption{ The bilinear spectrum for $\mathcal{E}=0$ and $0.15$ where each intersection with horizontal axis signifies an operator with dimension corresponding to the location of the intersection. The blue curve is bosonic and the black curve is fermionic. Note $K_f$ and $\bar{K}_f$ have identical spectra, thus all fermionic curves have multiplicity 2. In addition, the presence of pairs of lines comes from the spurious doubling of the spectrum due to unphysical local symmetries. Accounting for the unphysical modes, the $\mathcal{E}=0$ spectrum possesses the $\mathcal{N}=2$ Schwarzian multiplet with 2 $h=3/2$ modes. Turning on the chemical potential leads to an IR theory with spontaneously broken supersymmetry. While the spectrum still organizes into multiplets, the $3/2$ modes are no longer protected.}
    \label{fig:KernelGeneral}
\end{figure}

We can look at the spectrum of the theory by evaluating (\ref{eq:bosonicspec})  and (\ref{eq:fermionicspec}) in terms of (\ref{eq:genericmintegral}). The bosonic and fermionic spectrum are presented in Figure~\ref{fig:KernelGeneral}, where a line crossing the axis indicates the possible presence of a operator with that dimension. The fermionic operators always carry a double degeneracy as the fermionic kernels $K_f$ and $\bar{K}_f$ have identical spectra. As explained in \cite{Fu:2016vas}, a general feature of supersymmetric SYK models is the presence of spurious modes due to new local scaling or reparametrization symmetries which act independently on each argument of the bilocals but are incompatible with the UV boundary conditions. This includes for instance a local version of the scaling discussed below \eqref{eq:FuDeltaEqn}.

At $\mathcal{E}=0,$ we observe the supermultiplet $\left (\mathbf{1}, 2 \times \mathbf{\frac{3}{2}}, \mathbf{2} \right)$ associated with the $\mathcal{N}=2$ super Schwarzian. This contains a $h=1$ mode corresponding to the R-symmetry, a pair of $h=3/2$ modes indicating the presence of a complex supercharge, and a $h=2$ mode corresponding to the Hamiltonian, or the reparametrization mode. We also see another copy of this spectrum corresponding to the spurious non-diagonal super-reparametrizations which act differently on the two arguments of the two-point function. As explained in \cite{Fu:2016vas}, the spurious reparametrization mode is not expected to produce a soft mode in the IR, since this transformation affects the UV behavior of the correlators. 

At $0< \mathcal{E} < \mathcal{E}_{\rm critical}$, we still observe two sets of reparametrizations modes together with the two $h=1$ partners and two fermionic partners. Even though the field contents follow the structure of $\mathcal{N}=2$ multiplet, the scaling dimensions do not obey the superconformal Ward identities. We find the smallest fermionic modes have $h\neq 3/2$ for $\mathcal{E}\neq 0$ due to the breaking of supersymmetry. For example at $\mathcal{E}=0.15,$ we find the four $h=\frac{3}{2}$ modes split and have scaling dimensions $h_{\pm}\approx\frac{1}{2}\pm 0.09178.$ Moreover, the first supermultiplet after the Schwarizian multiplet has dimensions $\approx(\mathbf{2.78478},2\times \mathbf{3.32804},\mathbf{3.77582})$ no longer obeying the superconformal constraints. While we do not pursue it further in this work, it would be interesting to understand the dynamics of supersymmetry breaking in theories with an emergent super-Schwarzian mode, perhaps along the lines of \cite{Sannomiya:2016mnj}.
\begin{figure}[t!]
    \centering
    \includegraphics[width=0.495\textwidth]{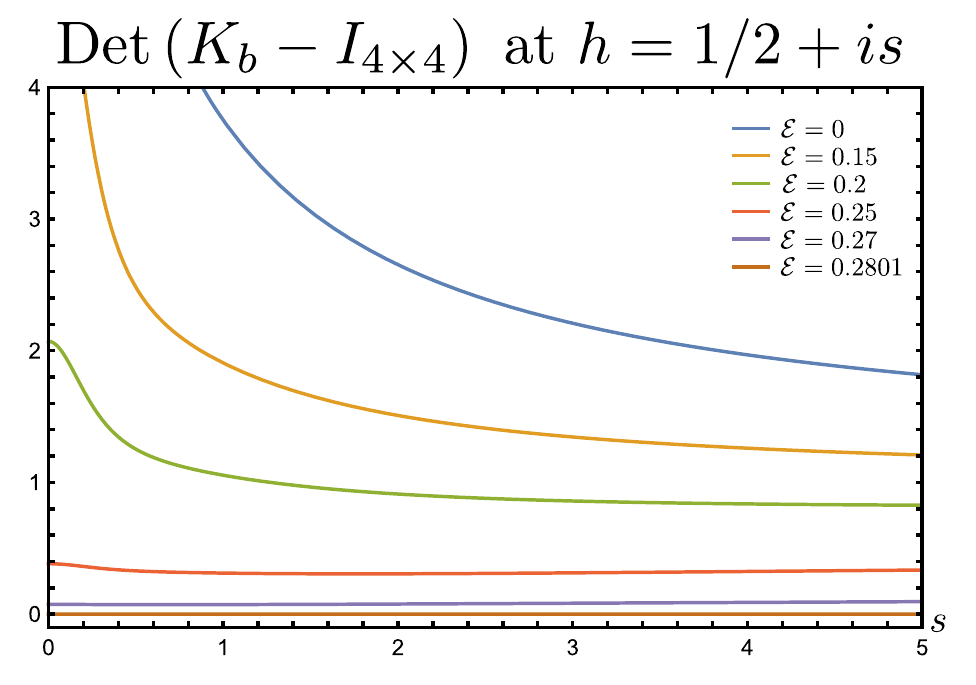}
    \caption{We plot $\text{Det}(K_b-I_{4\times 4})$ with $h=\frac{1}{2}+i s,$ where a complex mode is indicated by location of zeros. We check that there is no complex modes until $\mathcal{E}=\mathcal{E}_c, $ at which point there is also a continuous spectrum of operators in the form of $\frac{1}{2}+i s.$}
    \label{fig:complexmodes}
\end{figure}

In figure \ref{fig:complexmodes} we check for the possible complex modes, whose scaling dimension takes the form $h=\frac{1}{2}+i s$ determined by $SL(2,\mathbb{R}). $ In all admissible range of $\mathcal{E}$ we find no complex modes. In addition, we find no bosonic modes between $1<h<\frac{3}{2},$ which if present could dominate over the Schwarzian in the infrared \cite{Maldacena:2016upp, Milekhin:2021sqd}. Therefore the bilinear spectrum is free of any known problems. 

\begin{figure}[t!]
    \centering
    \includegraphics[width=0.495\textwidth]{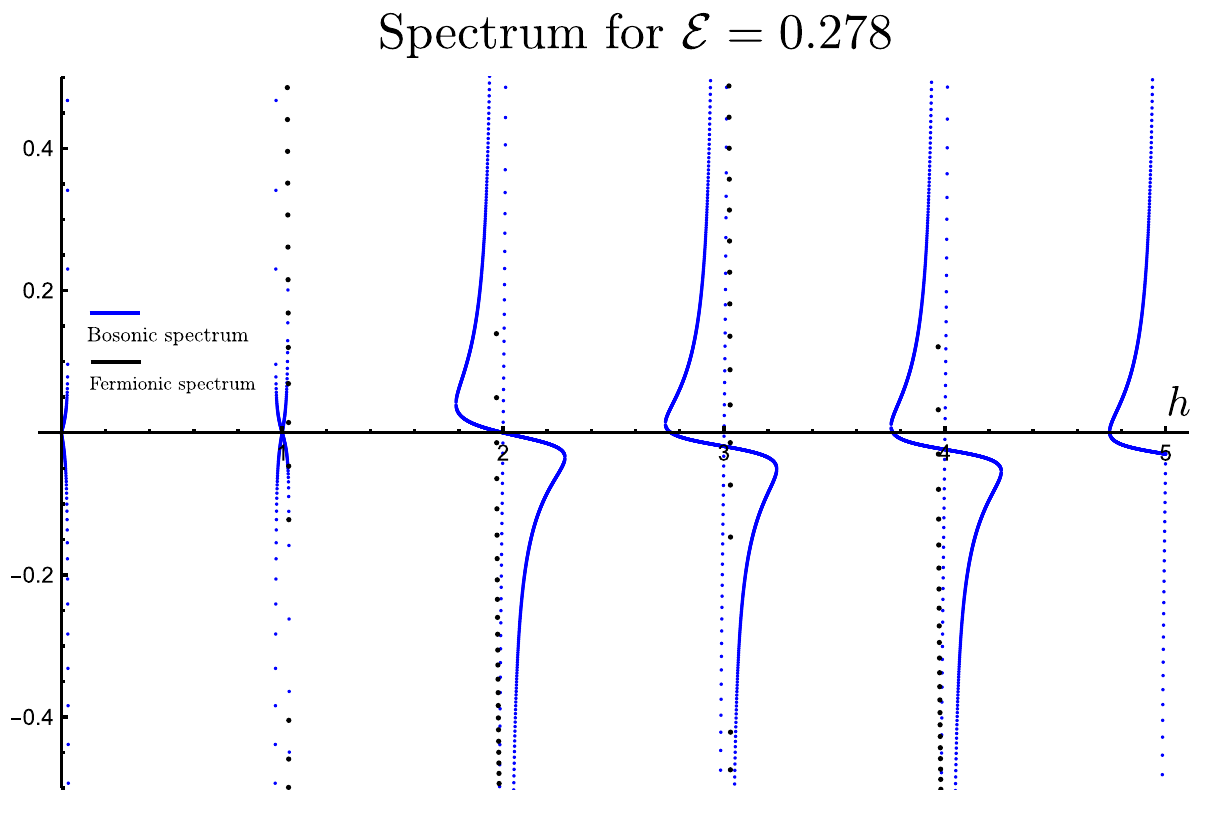}
    \includegraphics[width=0.495\textwidth]{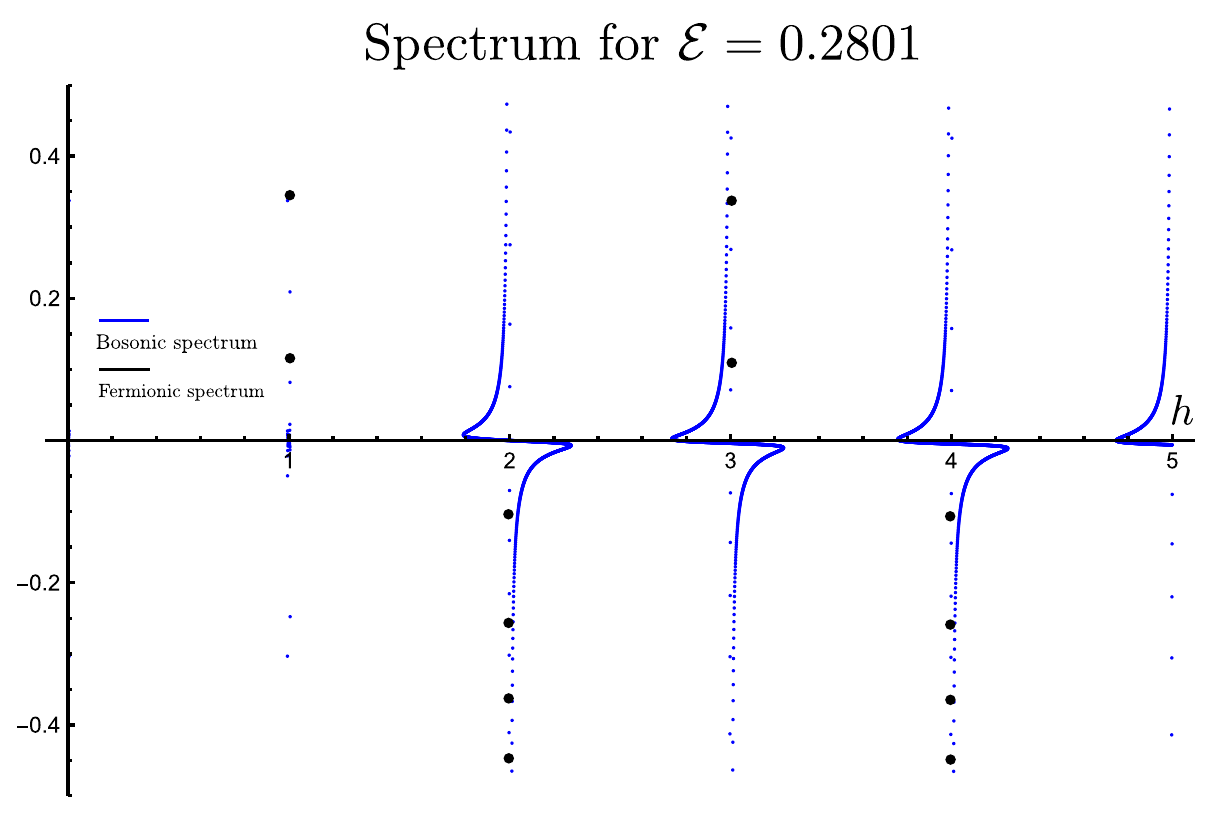}
    \caption{Some bosonic correlators near transition, with $\mathcal{E}=0.2780, 0.2801$. Fermions do not flatten out, but the bosonic eigenvalues start to flatten. At $\mathcal{E}=\mathcal{E}_{\rm critical}$, there is a continuous bosonic spectrum.}
    \label{fig:KernelNearTransition}
\end{figure}

Now we turn to the behavior close to the critical asymmetry $\mathcal{E}_{\rm critical}$. At the critical charge, although the conformal two point coefficients (\ref{eqfuconstra}) become zero, the kernel remains finite, and in fact equals the identity matrix. This gives rises to unusual feature that instead of having a discrete set of modes, a continuum of modes emerges for any real value of $h$. In figure \ref{fig:KernelNearTransition} we show the situation close but below of the transition, showing how the eigenvalue curve flattens out. Noticeably, as $\mathcal{E}\to \mathcal{E}_{\rm crticial},$ a continuum of complex modes also emerges as shown in figure \ref{fig:complexmodes}. 

\subsection{Comments on the holographic interpretation}\label{Sec:FuHoloIn}
In this section we make a proposal for a holographic interpretation of the features we found in $\mathcal{N}=2$ SYK at finite charge; we also derive a formula for the charge dependence of the zero-temperature entropy. While we do not have an exact bulk dual of the model, our interpretation has some qualitative and some quantitative features which can be compared.

We begin with the most conservative point of view. Since there is a slightly broken emergent $SU(1,1|1)$ symmetry in $\mathcal{N}=2$ SYK, and given that the dynamics is described by the $\mathcal{N}=2$ Schwarzian theory, the dual two dimensional black hole is described by $\mathcal{N}=2$ JT gravity (see for example \cite{Forste:2017apw}). This theory of gravity includes a $U(1)$ gauge field dual to the SYK electric charge. The two dimensional black hole background also includes $N$ complex fermions $\psi_{\rm bulk}$ which are dual to the $N$ SYK fermions $\psi^i$, and their bosonic partners. Following the notation in section 5 of \cite{Gu:2019jub} we take the fermions to have mass $M$ in the bulk, charge one, and they move on an electric field $\mathcal{E}$ equal to the spectral asymmetry introduced above. When this fermion is quantized with Neumann boundary conditions, the boundary two point function has the conformal form with scaling dimension $\Delta = \frac{1}{2} - \sqrt{M^2 - \mathcal{E}^2}$. We cannot use Dirichlet boundary conditions since those would have $\Delta>1/2$ which is not observed in SYK. 

Assume first that the masses of the fermions do not depend on the electric field. Then the two dimensional black hole is stable as long as the electric field is not too large $\mathcal{E}<\mathcal{E}_{\rm critical} = M$. Beyond the critical electric field the fermion develops a complex scaling dimension. When this happens there is a non-zero amplitude for Schwinger pair production that can screen the electric field and induce an instability of the two dimensional black hole. This is qualitatively similar to what we find in the analysis of the $\mathcal{N}=2$ SYK fermion as a function of the spectral asymmetry, with the only difference that $M$ has a dependence on $\mathcal{E}$\footnote{Black holes in four dimension can also have electric field dependent dimensions when placed in AdS, which plays an important role in the recent work \cite{Castro:2021wzn}.}. The numerical analysis done for $\mathcal{N}=2$ SYK indicates that the new phase after the instability is not a black hole, since it is a low entropy phase not possessing a nearly conformal symmetry.

So far the fermion coupled to JT gravity is intended to be thought of as a sector of the dual black hole to SYK. There is a different type of duality proposed in \cite{Gu:2019jub}, which we will refer to as the `spooky fermion', between some exact results in SYK and a set of fermions in the two dimensional hyperbolic disk. This proposal implies for example that the zero-temperature entropy of the complex SYK model is given by 
\beq
G_F(\mathcal{E}) = N [\log Z^F_D - \log Z^F_N].
\eeq
The right hand side involves the one-loop partition function of $N$ complex fermions with Dirichlet conditions in the boundary of the hyperbolic disk, denoted by $Z^F_D$ and $N$ ghosts fermions with Neumann boundary conditions, denoted by $Z^F_N$ and contributes with an extra minus sign. $G(\Delta,\mathcal{E})$ is the grand canonical partition function. It was shown in \cite{Gu:2019jub} that 
\beq
\frac{\partial G_F}{\partial \Delta} = - N\frac{\pi (1-2\Delta) \sin 2\pi \Delta}{\cosh 2 \pi \mathcal{E} + \cos 2 \pi \Delta}.
\eeq
Then one can compute the grand canonical partition function by integrating $\Delta$ from $1/2$ to $\Delta = 1/q$ which is the answer for complex SYK, independent of the spectral asymmetry. This identification fails for other topologies \cite{Maldacena:2018lmt}.

For $\mathcal{N}=2$ SYK we propose a similar duality at the level of the disk topology. The first difference is the presence of a bulk boson with Dirichlet boundary conditions and a bulk ghost boson with Neumann boundary conditions. Its contribution is $G_B(\Delta_b,\mathcal{E}_b) = N[\log Z^B_D - \log Z^B_N]$. The total grand potential is then
\beq
G(\Delta,\mathcal{E}) =G_F(\Delta,\mathcal{E})+G_B(\Delta_b,\mathcal{E}_b). 
\eeq
This boson is required by supersymmetry. For example at $\mathcal{E}=0$ it should satisfy $\Delta_b = \Delta +1/2$. Moreover, the charge of the boson has to be $(q-1)$ times the charge of the fermion by susy implying $\mathcal{E}_b = (q-1)\mathcal{E}$. For $\mathcal{E} \neq 0 $ susy is broken by based on the dynamics of SYK we found here we impose that $\Delta_b = 1-(q-1)\Delta$. It is possible to show that
\beq
\frac{\partial G_B}{\partial \Delta_b} = -N \frac{\pi (1-2\Delta_b) \sin 2\pi \Delta_b}{\cosh 2 \pi \mathcal{E}_b - \cos 2 \pi \Delta_b}.
\eeq
As opposed to complex SYK, for the $\mathcal{N}=2$ theory the fermion dimension is not fixed. The spooky propagator gives a way of understanding the formula \eqref{eq:FuDeltaEqn}. The grand potential depends on $\Delta$ and $\mathcal{E}$. It is reasonable to expect that the value of $\Delta$ should be such that the grand potential is extremized with respect to changes in the scaling dimension. Using that $\partial_\Delta \Delta_b=-(q-1)$ gives 
\beq\label{holoeqscaldim}
\partial_{\Delta} G (\Delta,\mathcal{E}) =0 ,~\Rightarrow \frac{ (1-2\Delta) \sin 2\pi \Delta}{\cosh 2 \pi \mathcal{E} + \cos 2 \pi \Delta}=(q-1) \frac{(1-2\Delta_b) \sin 2\pi \Delta_b}{\cosh 2 \pi \mathcal{E}_b - \cos 2 \pi \Delta_b}
\eeq
This is exactly the same equation as \eqref{eq:FuDeltaEqn}. 

We can use this picture to also derive the Luttinger-Ward relation. In order to do this start from the expression for the grand potential and apply the thermodynamic relation $Q= \frac{1}{2\pi} \partial_\mathcal{E} G$ \cite{Davison:2016ngz}. This gets contributions both from the spooky fermion and boson propagators. It is easy to check that this relation implies the Luttinger-Ward formula \eqref{eq:LWFuetalmt} which we verified by a numerical solution of the Schwinger-Dyson equations. 
\begin{figure}[h!]
    \centering
    \includegraphics[width=0.5\textwidth]{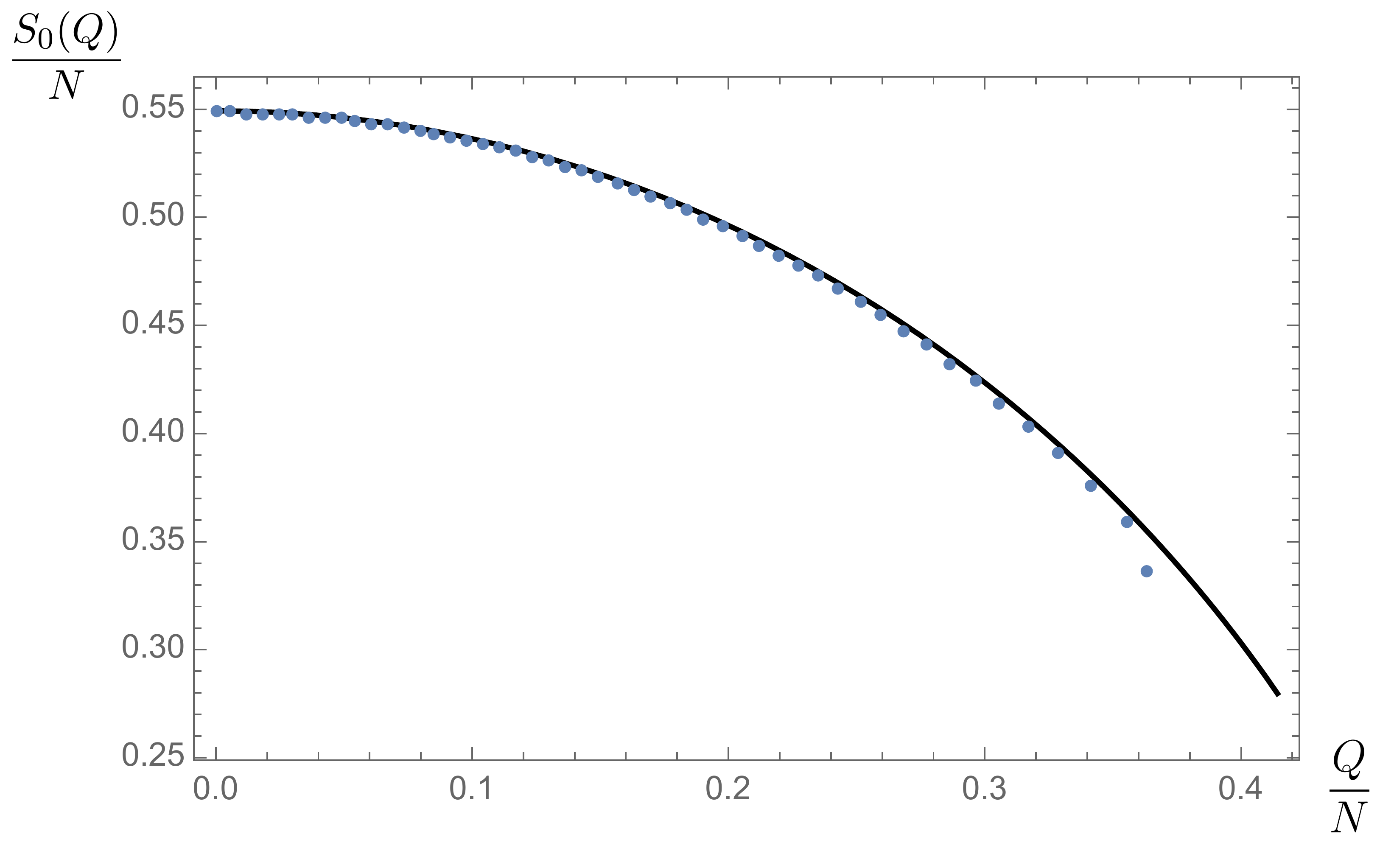}
    \caption{Plot showing the zero temperature entropy for $q=3$ $\mathcal{N}=2$ SYK, computed using the spooky fermion and boson propagators in equation \eqref{eq:fullGE} (solid black line) as a function of charge from $Q=0$ up to the critical value $Q=0.4142 N$. We also show the result from a numerical solution of the Schwinger-Dyson equations (blue dots) showing perfect agreement, at least far enough from the critical charge where the numerical procedure becomes more involved. }
    \label{fig:s0fu}
\end{figure}

Finally, we can verify the grand potential itself matches with the one obtained from the Schwinger-Dyson equation. The explicit expression is given by 
\beq\label{eq:fullGE}
G(\mathcal{E})/N = \int_{\Delta}^{1/2}dx~ \frac{\pi(1-2x)\sin 2\pi x}{\cosh 2 \pi \mathcal{E} + \cos 2 \pi x}+\int_{1-(q-1)\Delta}^{1/2}dx~ \frac{\pi(1-2x)\sin 2\pi x}{\cosh 2 \pi (q-1)\mathcal{E} - \cos 2 \pi x},
\eeq
where $\Delta$ is the fermion scaling dimension derived from \eqref{holoeqscaldim}. Using this expression we can also compute the entropy $G = S_0 (Q) - 2 \pi \mathcal{E} Q$. In figure \ref{fig:s0fu} we verify this expression coincides with the result from numerical solution of the mean field equations.  We would like to stress that $G$ cannot be simply obtained from integrating \eqref{eqn:dGdqfu} over $q$ since it is not clear what boundary conditions to use for that integral, so this provides a non-trivial check. We can verify analytically this works at $Q=0$, giving
\bea
G(\mathcal{E}=0)/N &=& \int_{\frac{1}{2q}}^{1/2}dx~ \pi(1-2x) \tan \pi x +\int_{\frac{1}{2q}+\frac{1}{2}}^{1/2}dx~ \pi(1-2x) \cot \pi x,\\
&=&\log \left( 2 \cos \frac{\pi}{2q} \right),
\ea
which coincides with the expectation from the index. This simple final answer come from non-trivial cancellations between the bosons and fermions contribution. It is possible to write down an analytic answer as a function of $\mathcal{E}$ using results of \cite{Davison:2016ngz} but it is not very illuminating.

\section{$\mathcal{N}=2$ SYK with multiple fermions} \label{sec:N2SYKNew}
In this section we will study models of $\mathcal{N}=2$ SYK with multiple complex fermions. In contrast to the model introduced in \cite{Fu:2016vas}, these new models have flavor symmetries that preserve supersymmetry. At low temperature, the IR physics is still dominated by the $\mathcal{N}=2$ Schwarzian theory with fundamental $U(1)_R$ charge equal to the charge of the supercharge. 

We will consider theories with two sets of complex fermions $\psi^i$ and $\chi^i$ with $i=1,\ldots,N$. Due to $\mathcal{N}=2$ supersymmetry, the Hamiltonian is completely specified by giving the supercharge. With more fermions, there is some arbitrariness in what we choose to be the supercharge which defines the model; in the simplest realization we will take it to be 
\begin{equation}
    \mathcal{Q}=i C_{ijk}~\psi^i\psi^j\chi^k,
    \label{eq:NuQ}
\end{equation}
where we take $C_{ijk}$ to be totally antisymmetric (this can be generalized since we only need antisymmetry in the first two indices). Then $\mathcal{Q}^2 = \bar{\mathcal{Q}}^2=0$ and $H=\{ \mathcal{Q},\bar{\mathcal{Q}}\}$. We take the complex coupling matrix to be a gaussian random distributed with variance $\langle C_{i_1 i_2 \ldots i_q} \bar{C}^{i_1 i_2 \ldots i_q}\rangle \sim J/N^{q-1}$. We will show these models also have an emergent $SU(1,1|1)$ symmetry at low temperatures, with additional features due to the existence of a new global symmetry.

The supersymmetric Lagrangian following from \eqref{eq:NuQ} has the global symmetries $U(1)_\psi \times U(1)_\chi$ which act independently on the two types of fermions $\psi$ and $\chi$. Their generators are $Q_\psi = \sum_i \bar{\psi}_i \psi^i -N/2$ and $Q_\chi = \sum_i \bar{\chi}_i \chi^i -N/2$. Both of these generators commute with the Hamiltonian but not with the supercharge. There is a linear combination that defines a supersymmetric flavor symmetry:
\beq
Q_F = Q_\psi - 2 Q_\chi ,~~~~~~[ Q_F, \mathcal{Q}] =0.
\label{eq:NewQf}
\eeq
 The existence of this flavor symmetry implies that deciding exactly what is the superconformal R-symmetry $U(1)_R$ inside the infrared superconformal group is a non-trivial problem. The simplest possibility is to pick $Q_\chi$ itself, but in general it could be any linear combination
\beq\label{eq:something2}
Q_R = Q_\chi + \alpha ~Q_F,~~~~~~[Q_R , \mathcal{Q} ] = \mathcal{Q}
\eeq
where $\alpha$ is an arbitrary real parameter. We will determine $\alpha$ from the solution of the model at low temperatures.

It will be useful for some calculations below to generalize the model to involve an arbitrary number of $\psi$ fermions in the supercharge
\begin{equation}
    \mathcal{Q}=i^{\frac{q-1}{2}} C_{i_1\ldots i_q}~\psi^{i_1}\ldots \psi^{i_{q-1}}\chi^{i_q}.
    \label{eq:NewqQ}
\end{equation}
This model also has two conserved currents $Q_\psi$ and $Q_\chi$, and a generalization of the flavor charge \eqref{eq:NewQf} is
\beq
Q_F =Q_\psi - (q-1)Q_\chi \, .
\eeq
Again, we can propose a trial low temperature R-symmetry $Q_R = Q_\chi + \alpha Q_F$, for a parameter $\alpha$ we will determine later. 

Finally, all these models again have a global discrete  $\mathbb{Z}_q$ symmetry. This is generated by $\psi \to e^{\frac{2\pi i r}{q}} \psi$ and $\chi \to e^{\frac{2\pi i r}{q}} \chi$ and it clearly commutes with the supercharge.

\subsubsection*{Calculation of the Index}
We will now compute the refined Witten index for this theory. The calculation can be done in the free fermion limit, but it will be useful to compare to the result we will find using the mean field description below. We should first define a notion of fermion number that produces cancellations between states when they do not preserve supersymmetry. We take the following definition:
\beq
(-1)^F \equiv e^{i \pi Q_\chi} 
\eeq
Since this only counts the number of $\chi$-oscillators, it is a-priori not the `true' fermion number. Nevertheless, since the supercharge involves an even number of $\psi$ fields, all states in a supermultiplet have the same $\psi$-fermion number modulo 2. Furthermore, the supercharge has fermion number one with this definition and the contribution from states in the same supermultiplet will vanish. More formally, we can say that the Hilbert space breaks up into a $\mathbb{Z}$ graded complex $\dots \overset{\mathcal{Q}}{\rightarrow} \mathcal{H}^{q_\chi-1} \overset{\mathcal{Q}}{\rightarrow}\mathcal{H}^{q_\chi} \overset{\mathcal{Q}}{\rightarrow} \mathcal{H}^{q_\chi+1} \overset{\mathcal{Q}}{\rightarrow} \dots$. This reduces mod 2 to $\mathbb{Z}_2$, the fermion number grading. There is a separate $\mathcal{Q}$-cohomology for each $q_\chi \in \mathbb{Z}$ charge.

The presence of the flavor $U(1)_F$ allows us to define the Witten index refined by the chemical potential $y$ conjugate to this charge. The index is then given by
\bea
\mathcal{I}(y) &\equiv& {\rm Tr} \Big[(-1)^F e^{-\beta H} e^{ i y Q_F} \Big],\\
&=&  \Big( 2 \cos \Big(\frac{y}{2}\Big) ~2\sin \Big(\frac{q-1}{2}y\Big) \Big)^N,
\ea
where the second line was computed in the free theory. Just like the models of \cite{Fu:2016vas}, the index vanishes with no potential; ${\rm Tr}(-1)^F =0$.

 Already at the level of the index, we may find interesting features of the model defined by \eqref{eq:NewqQ} by passing from the grand canonical (fixed $y$) to the canonical (fixed $Q_F$) ensemble. This amounts to picking a charge sector of the theory with $Q_F\to q_F$ labeling the sector (for example $q_F=0$ if we gauge the flavor symmetry). Now the index in this sector does not vanish and is given by
\bea
\mathcal{I}(q_F) &=&{\rm Tr}_{q_F}\Big[ (-1)^F e^{-\beta H} \Big],\\
&=&\int_0^{2\pi} \frac{dy}{2\pi}~ e^{- i y q_F}~ e^{N I(y)} ,~~~I(y)\equiv \log{\left(2 \cos \Big(\frac{y}{2}\Big) ~2\sin \Big(\frac{q-1}{2}y\Big)\right)}
\ea
Now we will take the large $N$ limit while keeping $q_F$ of order one. Therefore we can ignore the dependence on the flavor charge and focus on the $I(y)$ term to find the saddle point of the integral. In this limit the index can be approximated by 
\beq
\mathcal{I} \sim e^{N I(y_c)}~~~~{\rm where}~~\partial_y I(y_c)=0.
\eeq
The saddle point equation $\partial_y I(y_c)=0$ cannot be explicitly solved, but can be rewritten in a form that will be useful for a later comparison
\beq\label{eq:index}
\tan \left(\frac{y_c}{2}\right) = (q-1) \cot \left((q-1) \frac{y_c}{2}\right),~~~~\Rightarrow~~~\frac{d}{dq} \frac{\log \mathcal{I}}{N} = \frac{y_c}{2} \cot \left((q-1) \frac{y_c}{2}\right).
\eeq
To simplify the expression for the index we took a derivative of $I(y)$ with respect to $q$, which due to the saddle point equation only acts on explicit $q$ dependence. We will see below that this quantity can be identified with the zero temperature entropy $S_0$ computed from the mean field action at the particle-hole symmetry point.

We can also turn both a chemical potential for $Q_F$ and for the $\mathbb{Z}_q$ symmetry that commutes with the supercharge. The answer is given by 
\bea
\mathcal{I}(y,r) &=& {\rm Tr} \Big[ (-1)^F e^{-\beta H} e^{ i y Q_F} e^{i 2\pi \frac{r}{q} (Q_\psi+Q_\chi)} \Big],\\
&=& \Big( 2 \cos \Big(\frac{y}{2}+\frac{\pi r}{q}\Big) ~2\sin \Big(\frac{q-1}{2}y-\frac{\pi r}{q}\Big) \Big)^N
\ea
The reason that this refinement does not add much information is that it can be rewritten as 
\beq
\mathcal{I}(y,r) = (-1)^{r N}~\mathcal{I}\Big(y-\frac{2\pi r}{q},0\Big),
\eeq
and therefore its given up to a sign by the previous index. For example the fixed charge refined index is $\mathcal{I}(q_F,r) = e^{i 2 \pi r (\frac{N}{2}-\frac{q_F}{q} )}\mathcal{I}(q_F,0)$. We can use this index then to determine the $\mathbb{Z}_q$ charge of the BPS states in each fix flavor charge sector. 

\subsection{Mean field and the conformal solution}
\label{ssection:MeanFidlNew}

The derivation of the mean field action is largely identical to the one in section \ref{sec:N2SYKFu} so we will be more brief. We will again work in superspace, first introducing introduce two chiral superfields defined analogously to \eqref{eq:chiralsuperfield}:
\begin{equation}
    \Psi^i=\psi^i(\tau+\theta\bar{\theta})+\sqrt{2}\,\theta \, b_{\psi}^i \, , \quad  X^i=\chi^i(\tau+\theta\bar{\theta})+\sqrt{2}\, \theta \, b_{\chi}^i \, ,
\end{equation}
where we have again introduced auxiliary bosons $b_\psi^i$, $b_\chi^i$, and their conjugates with an explicit form obtained from the analog of \eqref{eq:susymults}. The action after integrating in the bosons can be written in superspace in terms of these chiral fields and the interaction term is 
\begin{equation}
   \mathcal{L} \supset i^{\frac{q-1}{2}} \! \! \int d\theta\left(C_{i_1i_2\dots i_q}\Psi^{i_1}\dots\Psi^{i_{q-1}} X^{i_q}\right)+ h.c.
\end{equation}
More explicitly, the $\theta $ component of $C_{i_1 i_2 \dots i_q}\Psi^{i_1}\dots\Psi^{i_{q-1}} X^{i_a}$ gives the combination $\psi^{q-1} b_\chi$ and $(q-1)b_\psi \psi^{q-2} \chi$. We introduce now the fermions two point function 
\beq
\mathcal{G}_{\Psi\Psi} (Z_1,Z_2) = \frac{1}{N}  \langle \bar{\Psi}_i(Z_1) \Psi^i(Z_2)\rangle,~~~\mathcal{G}_{XX} (Z_1,Z_2) = \frac{1}{N}  \langle \bar{X}_i(Z_1) X^i(Z_2)\rangle.
\eeq
The expansion of these superfields includes $G_{\psi\psi}(\tau_1,\tau_2)$, $G_{b_\psi b_\psi}(\tau_1,\tau_2)$, etc. To derive the mean field action we introduce a Lagrange multiplier $\Sigma_\psi(Z_1,Z_2)$ and $\Sigma_\chi (Z_1,Z_2)$, which in components are $\Sigma_\psi(Z_1,Z_2) = \frac{1}{2}\Sigma_{b_\psi b_\psi}(\tau_1-\theta_1\bar{\theta}_1,\tau_2 +\theta_2\bar{\theta}_2)+\ldots$ and similarly for $\chi$. Next we integrate out both the couplings and the fundamental superfields. The interaction term in the mean field action involves 
\beq
S \supset N \frac{J}{2} \int d\bar{Z}_1 dZ_2 ~ \mathcal{G}_{\Psi\Psi}(Z_1,Z_2)^{q-1} \mathcal{G}_{XX} (Z_1,Z_2).
\eeq
Since the procedure to find the action and its equations of motion (the Schwinger-Dyson equations) is similar to what we reviewed in section \ref{sec:FuConfSol}, we will move straight to the equations and solution. We consider time translation invariant solutions with vanishing mixed correlators between bosons and fermions.
\bea
\label{eq:UVSuperSDPsi}
&&\hspace{-1cm}\frac{1}{2}D_{\theta_3} \mathcal{G}_{\Psi\Psi}(Z_1,Z_3) \! + \! \int \! \! dZ_2 ~\mathcal{G}_{\Psi\Psi}(Z_1,Z_2) \left[ \frac{J}{2}(q-1)\mathcal{G}_{\Psi\Psi}(Z_3,Z_2)^{q-2}\mathcal{G}_{XX}(Z_3,Z_2)\right] = \delta(\bar{Z}_1-\bar{Z}_3),\\
\label{eq:UVSuperSDChi}
&&\hspace{-1cm}\frac{1}{2}D_{\theta_3} \mathcal{G}_{XX}(Z_1,Z_3) \! + \! \int \! \! dZ_2 ~\mathcal{G}_{XX}(Z_1,Z_2) \left[\frac{J}{2}\mathcal{G}_{\Psi\Psi}(Z_3,Z_2)^{q-1}\right] = \delta(\bar{Z}_1-\bar{Z}_3),
\ea
where the quantities in brackets are equal to $\Sigma_{\psi}$ and $\Sigma_{\chi}$, the superspace self-energies. 

\subsubsection*{Solutions of Schwinger-Dyson equations}
In this section, we solve the Schwinger-Dyson equations in the IR limit using the conformal ansatz, derived from the mean field action above. In position space, the component equations that determine the self energies (when off-diagonal bilinears are set to zero) are
\bea
\Sigma_{\psi\psi}(\tau)&=&J(q-1)\left(G_{\psi\psi}(\tau)^{q-2}G_{b_\chi b_\chi}(\tau)+(q-2)G_{b_\psi b_\psi}(\tau) G_{\chi\chi}(\tau)G_{\psi\psi}(\tau)^{q-3}\right),\label{eq:SDsaddle_generalq2}\\ 
\Sigma_{\chi\chi}(\tau)&=&J(q-1) G_{b_\psi b_\psi}(\tau)G_{\psi\psi}(\tau)^{q-2}, \label{eq:SDsaddle_generalq3}\\ \Sigma_{b_\psi b_\psi}(\tau)&=&J(q-1)G_{\psi\psi}(\tau)^{q-2} G_{\chi\chi}(\tau),\label{eq:SDsaddle_generalq4}\\ \Sigma_{b_\chi b_\chi}(\tau)&=&JG_{\psi\psi}(\tau)^{q-1}\label{eq:SDsaddle_generalq5}.
\ea
We can begin by picking a similar ansatz as in the previous section for all correlators 
\bea
G_{AA}(\tau)&=& \frac{g_{AA}}{|\tau|^{2\Delta_A}} \left( e^{\pi \mathcal{E}_A} \Theta(\tau) - e^{-\pi \mathcal{E}_A} \Theta(-\tau) \right),\\
G_{b_A b_A}(\tau) &=& \frac{g_{b_A b_A}}{|\tau|^{2\Delta_{b_A}}} \left( e^{\pi \mathcal{E}_{b_A}} \Theta(\tau) + e^{-\pi \mathcal{E}_{b_A}} \Theta(-\tau) \right),
\ea
where $A=\psi$ or $\chi$. We now have several different scaling dimensions $\Delta_A$, $\Delta_{b_A}$ and spectral asymmetries $\mathcal{E}_A$, $\mathcal{E}_{b_A}$ for both sets of bosons and fermions.

The self-energies can be found by solving $\Sigma_{AA}(\omega) G_{AA}(-\omega) = 1$ and $\Sigma_{b_Ab_A}(\omega) G_{b_Ab_A}(-\omega) = -1$, and the answer is the same as before for each correlator:
\bea
\Sigma_{AA}(\tau) &=& \frac{1}{g_{AA}} \frac{(1-2\Delta_{A})\sin 2 \pi \Delta_{A}}{4\pi \prod_{\pm} \cos \pi(\Delta_{A} \pm i \mathcal{E}_{A})} \frac{1}{|\tau|^{2(1-\Delta_{A})}}  
\left( -e^{\pi \mathcal{E}_{A}} \Theta(-\tau) + e^{-\pi \mathcal{E}_{A}} \Theta(\tau) \right),\nonumber\\
\Sigma_{b_Ab_A} (\tau)&=&\frac{1}{g_{b_A b_A}} \frac{(1-2\Delta_{b_A})\sin 2 \pi \Delta_{b_A}}{4\pi \prod_{\pm} \sin \pi(\Delta_{b_A} \pm i \mathcal{E}_{b_A})}\frac{1}{|\tau|^{2(1-\Delta_{b_A})}}   \left( e^{\pi \mathcal{E}_{b_A}} \Theta(-\tau) + e^{-\pi \mathcal{E}_{b_A}} \Theta(\tau) \right).\nonumber
\ea
We now follow the same steps as in the previous section to solve the rest of the equations. We begin by matching the left- and right-hand sides of equations \eqref{eq:SDsaddle_generalq2}, \eqref{eq:SDsaddle_generalq3}, \eqref{eq:SDsaddle_generalq4}, and \eqref{eq:SDsaddle_generalq5}. This again can be done in steps. We first match the spectral asymmetry of all equations. This gives the following two independent constraints 
\beq\label{eq:GeneralN2SA}
\mathcal{E}_{b_\psi} =-(q-2) \mathcal{E}_\psi - \mathcal{E}_\chi ,~~~~\mathcal{E}_{b_\chi}= -(q-1) \mathcal{E}_{\psi}.
\eeq
The interpretation of these constraints is clear since integrating out auxiliary bosons gives by $b_\psi \sim \bar{\psi}^{q-2} \bar{\chi}$ and $b_\chi \sim \bar{\psi}^{q-1}$. Therefore their spectral asymmetry should respect these relations. Moreover, there are only two $U(1)$ charges we are free to chose in our theory and therefore there should be only two independent spectral asymmetries which we take to be $\mathcal{E}_\psi$ and $\mathcal{E}_\chi$. Next, we can match the time dependence on both sides of the four equations. This gives the following two constraints on the scaling dimensions
\beq\label{eq:GeneralN2Dim}
\Delta_{b_\chi} + (q-1)\Delta_\psi = 1 ,~~~~\Delta_{b_\psi} + (q-2) \Delta_{\psi} + \Delta_\chi = 1.
\eeq
We then determine the bosonic scaling dimensions from these equations and consider $\Delta_\psi$ and $\Delta_\chi$ as independent variables. These constraints on the dimensions are also reasonable since they imply the interaction terms are marginal in the low energy effective action. We will see below that as opposed to Yukawa-like interactions without supersymmetry (examples of Yukawa interactions with first order bosons are \cite{Tikhanovskaya:2020zcw,Christos:2022lma}), the scaling dimensions will be determined from the spectral asymmetries uniquely. 

Before moving on to matching the prefactors $g_{AA}$ and $g_{b_A b_A}$ of the equations for the self-energies, we will analyze the range of allowable scaling dimensions such that the approximations made in the IR solution are self consistent. Dropping the kinetic terms requires $\Delta_{\rm fermion}>0$ and $\Delta_{\rm boson}>1/2$, which implies the following inequalities 
\beq
0<\Delta_{\psi}<\frac{1}{2(q-1)} \, ,~~~0<\Delta_\chi<\frac{1}{2}-(q-2)\Delta_\psi<\frac{1}{2}
\eeq
The bound on $\Delta_\chi$ is therefore not fixed and goes from $1/2$ to $1/(2(q-1))$ as we increase $\Delta_\psi$ within the allowed range. 

Now we can match the prefactors. This will produce self-consistency equations that uniquely determine the scaling dimensions. For each of the equations \eqref{eq:SDsaddle_generalq2}, \eqref{eq:SDsaddle_generalq3}, \eqref{eq:SDsaddle_generalq4} and \eqref{eq:SDsaddle_generalq5}, we get: 
\bea
\frac{(1-2\Delta_{\psi})\sin 2 \pi \Delta_{\psi}}{4\pi \prod_{\pm} \cos \pi(\Delta_{\psi} \pm i \mathcal{E}_{\psi})}  &=& 2(q-1) g_{\psi\psi}^{q-1}~ g_{b_\chi b_\chi} + 2(q-1)(q-2) ~g_{b_\psi b_\psi} ~g_{\chi\chi}~g_{\psi\psi}^{q-2}\label{eq:SDprefactor1}\\
\frac{(1-2\Delta_{b_\chi})\sin 2 \pi \Delta_{b_\chi}}{4\pi \prod_{\pm} \sin \pi(\Delta_{b_\chi} \pm i \mathcal{E}_{b_\chi})} &=& 2 g_{b_\chi b_\chi}~g_{\psi\psi}^{q-1} \label{eq:SDprefactor2}\\
\frac{(1-2\Delta_{\chi})\sin 2 \pi \Delta_{\chi}}{4\pi \prod_{\pm} \cos \pi(\Delta_{\chi} \pm i \mathcal{E}_{\chi})} &=& 2(q-1)~ g_{\chi\chi}~ g_{b_\psi b_\psi} ~g_{\psi \psi}^{q-2}\label{eq:SDprefactor3}\\
\frac{(1-2\Delta_{b_\psi})\sin 2 \pi \Delta_{b_\psi}}{4\pi \prod_{\pm} \sin \pi(\Delta_{b_\psi} \pm i \mathcal{E}_{b_\psi})} &=& 2(q-1) ~g_{b_\psi b_\psi}~ g_{\psi \psi}^{q-2} ~g_{\chi\chi}\label{eq:SDprefactor4}
\ea
These four equations are only consistent if the following two constraints are solved 
\bea
\frac{(1-2\Delta_{\chi})\sin 2 \pi \Delta_{\chi}}{4\pi \prod_{\pm} \cos \pi(\Delta_{\chi} \pm i \mathcal{E}_{\chi})} &=&\frac{(1-2\Delta_{b_\psi})\sin 2 \pi \Delta_{b_\psi}}{4\pi \prod_{\pm} \sin \pi(\Delta_{b_\psi} \pm i \mathcal{E}_{b_\psi})},\\
\frac{(1-2\Delta_{\psi})\sin 2 \pi \Delta_{\psi}}{4\pi \prod_{\pm} \cos \pi(\Delta_{\psi} \pm i \mathcal{E}_{\psi})} &=&\frac{(1-2\Delta_{b_\chi})\sin 2 \pi \Delta_{b_\chi}}{4\pi \prod_{\pm} \sin \pi(\Delta_{b_\chi} \pm i \mathcal{E}_{b_\chi})}
+
\frac{(q-2)(1-2\Delta_{\chi})\sin 2 \pi \Delta_{\chi}}{4\pi \prod_{\pm} \cos \pi(\Delta_{\chi} \pm i \mathcal{E}_{\chi})}
\ea
After replacing the values of scaling dimensions and spectral asymmetries for the bosons using \eqref{eq:GeneralN2Dim} and \eqref{eq:GeneralN2SA}, these two equations are sufficient to determine $\Delta_\psi(\mathcal{E}_\psi,\mathcal{E}_\chi)$ and $\Delta_\chi(\mathcal{E}_\psi,\mathcal{E}_\chi)$. We will analyze the behavior of the scaling dimensions next. First, let us point out again that using the IR solution we can only determine the following combinations of prefactors $g_{\psi\psi}^{q-2} g_{\chi\chi} g_{b_\psi b_\psi}$ and $g_{\psi\psi}^{q-1} g_{b_\chi b_\chi}$. This is due to a set of two emergent scaling symmetries, which we refer to as $\lambda_1$ and $\lambda_2$, in the IR given by 
\bea
\label{eq:NewLambdaSym1}
\lambda_1:~~&&G_{\psi\psi} \to \lambda_1 G_{\psi\psi} ,~~ G_{b_\chi b_\chi} \to \lambda_1^{1-q} G_{b_\chi b_\chi},~~G_{\chi\chi} \to \lambda_1^{2-q} G_{\chi\chi},~~G_{b_\psi b_\psi} \to G_{b_\psi b_\psi},\\
\lambda_2:~~&&G_{\psi\psi} \to G_{\psi\psi} ,~~ G_{b_\chi b_\chi} \to  G_{b_\chi b_\chi},~~G_{\chi\chi} \to \lambda_2 G_{\chi\chi},~~G_{b_\psi b_\psi} \to \lambda_2^{-1} G_{b_\psi b_\psi}.
\label{eq:NewLambdaSym2}
\ea
 Is it easy to see that these transformations can be obtained from acting with $U(1)_\psi$ and $U(1)_\chi$ independently on the two insertions appearing in the two-point function. This is necessary since otherwise a time independent transformation acting diagonally would leave the correlators invariant. These symmetries are broken in the UV and therefore the prefactors should be determined if we had access to the full solution. These transformations act on the UV correlators and we do not expect a low energy mode coming from them for the same reasons as in \cite{Fu:2016vas}.  

\subsubsection*{Supersymmetric solution}
Since we find a non-vanishing index we know the ground states preserve supersymmetry. This implies for the conformal solution that supersymmetric solutions satisfy $G_{b_Ab_A} \sim \partial_\tau G_{AA}$ for $A=\psi,\chi$, analogous to \eqref{eq:SusyDyson}. Then the bosonic scaling dimensions are given in terms of the fermionic ones
\beq
\Delta_{b_\psi} = \Delta_\psi + \frac{1}{2},~~~\Delta_{b_\chi} = \Delta_\chi + \frac{1}{2}.
\eeq
As opposed to the model studied in the previous section, this does not determine the scaling dimensions anymore from a purely dimensional analysis argument. Instead we are left with a single constraint on the fermion dimensions
\beq\label{eqn:NuSusy}
{\rm Susy}:~~~~~\Delta_\chi + (q-1) \Delta_\psi = \frac{1}{2}.
\eeq
To determine a unique solution we need to look at the spectral asymmetry. The supersymmetry relating the bosonic and fermionic correlators also imposes a matching between the spectral asymmetries $\mathcal{E}_{b_\psi} = \mathcal{E}_\psi$ and $\mathcal{E}_{b_\chi} = \mathcal{E}_\chi$. Combinging this with the relations \eqref{eq:GeneralN2SA}, one finds
\beq
{\rm Susy}:~~~~~\mathcal{E}_\chi + (q-1)\mathcal{E}_\psi = 0.
\eeq
This is not the most general solution and there are discrete configurations with complex spectral asymmetry corresponding to turning on the $\mathbb{Z}_q$ charge, analogous to \eqref{eqn:FuSusyDiscrete}. We will not discuss those solutions here.  

Using the relations above, so far supersymmetric solutions can be parametrized by $\Delta_\psi$ and $\mathcal{E}_\psi$. Given $\mathcal{E}_\psi$, we now impose the constraints coming from the prefactors in the Schwinger-Dyson equations. They give the following equation
\beq\label{eq: delta_of_eps}
\frac{\sin 2 \pi \Delta_\psi }{\cos 2 \pi \Delta_\psi + \cosh 2 \pi \mathcal{E}_\psi} = \frac{(q-1)\sin 2 \pi (q-1) \Delta_\psi}{\cosh 2 \pi (q-1) \mathcal{E}_\psi -\cos 2 \pi (q-1) \Delta_\psi} 
\eeq
This should be seen as an implicit equation determining $\Delta_\psi(\mathcal{E}_\psi)$, and all other quantities can be determined by the relations above. This is consistent with the discussion of the index, only one continuous chemical potential can be turned on while still preserving supersymmetry, the one conjugate to the UV flavor charge \eqref{eq:NewQf}. Explicit solutions are possible for small $q$. For $q=3$ it is given by
\beq 
\Delta_\psi=\frac{1}{2\pi} \arcsin\Big(\frac{1}{3}\sqrt{9-\cosh^2 2\pi\mathcal{E}_\psi}\Big).
\eeq
We will comment later on the fact that there is a critical spectral asymmetry at which the scaling dimension vanishes.

\subsubsection*{Emergent $SU(1,1|1)$ symmetry}
We will briefly point out now that the supersymmetric solution has an emergent $SU(1,1|1)$ symmetry. The full Schwinger-Dyson equations in superspace \eqref{eq:UVSuperSDPsi}, \eqref{eq:UVSuperSDChi} upon dropping the UV term are
\bea
\label{eq:NewSD1}
&& \int \! \! dZ_2 \mathcal{G}_{\Psi\Psi}(Z_1,Z_2) (J(q-1)\mathcal{G}_{\Psi\Psi}(Z_3,Z_2)^{q-2}\mathcal{G}_{XX}(Z_3,Z_2)) = 2\delta(\bar{Z}_1-\bar{Z}_3),\\
\label{eq:NewSD2}
&& \int \! \! dZ_2 \mathcal{G}_{XX}(Z_1,Z_2) (J\mathcal{G}_{\Psi\Psi}(Z_3,Z_2)^{q-1}) = 2\delta(\bar{Z}_1-\bar{Z}_3),
\ea
At late times when the UV term is ignored the equations are symmetric under the following super-reparametrizations
\bea
\mathcal{G}_{\Psi\Psi}(Z_1,Z_2) &\to& (D_{\theta_1} \theta'_1)^{2\Delta_\psi}(D_{\bar{\theta}_2} \bar{\theta}'_2)^{2\Delta_\psi}\mathcal{G}_{\Psi\Psi}(Z'_1,Z'_2),\\
\mathcal{G}_{XX}(Z_1,Z_2) &\to&  (D_{\theta_1} \theta'_1)^{2\Delta_\chi}(D_{\bar{\theta}_2} \bar{\theta}'_2)^{2\Delta_\chi} \mathcal{G}_{XX}(Z'_1,Z'_2)
\ea
as long as 
\beq\label{eqn:something}
(q-1)\Delta_\psi + \Delta_\chi = \frac{1}{2}.
\eeq
This is precisely the supersymmetric relation found above \eqref{eqn:NuSusy}. This shows that the fermions are superconformal primaries under the $SU(1,1|1)$, which is a particular case of the more general superreparametrizations above.

Since the fermions are chiral primary operators, their R-charge has to be twice their scaling dimension. This can be used to determine the parameter $\alpha$ appearing in $Q_R$ in equation \eqref{eq:something2}. For example, the $\psi$ fermion has charge $Q_\psi =1$ and $Q_\chi = 0$ and therefore $Q_R = \alpha$. The condition of $\psi$ being a chiral primary then determines the unknown coefficient:
\begin{equation}
\label{eq:alphaRmax}
    \alpha = 2\Delta_\psi \, .
\end{equation}
A similar analysis for the $\chi$ fermion gives the same $\alpha$ thanks to \eqref{eqn:something}. 

We show later the presence of an $\mathcal{N}=2$ Super-Schwarzian mode in the bilinear spectrum. For simplicity we can work at fixed $Q_F$ (we will consider relaxing this in the next section). Since the R-charge of the theory is given, up to a constant shift in the fixed $Q_F$ sector, by $Q_\chi$ this means the R-charge of the supercharge is one. The bosonic sector of the action is 
\begin{equation}
S_b=\frac{2\pi N \alpha_S}{\beta J}\int_0^{2\pi}d\tau\left(-\text{Sch}\left(\tan\frac{f}{2},\tau\right)+ 2\left(\partial_{\tau}a\right)^2\right),
\end{equation}
with $a\sim a+ 2 \pi$. We gave some details of the quantum mechanical spectrum of this theory in the introduction and emphasized it has the special feature that all BPS states have the same R-charge. This is consistent with the fact that the index in a fixed $Q_F$ sector is non-vanishing and exponentially large in $N$ (We have done some preliminary checks on these features using exact diagonalization of the model). We consider next what happens when working in an ensemble in which the flavor charge fluctuates.

\subsubsection*{Emergent local flavor symmetry}
Just as the UV $U(1)_R$ becomes a local IR symmetry in the conformal limit, but leads to the $U(1)_R$ mode of the $\mathcal{N}=2$ super Schwarzian, one might also expect that the $U(1)_F$ flavor symmetry may become a spontaneously and explicitly broken local symmetry, leading to a new physical mode (and potentially a spurious mode). In the case of $\mathcal{N}=1$ supersymmetry with a global $SO(q)$ symmetry, this situation was studied in \cite{Narayan:2017hvh} where such a Kac-Moody like enhancement and corresponding flavor mode were found. In this section, we present a generalization appropriate to our model with larger supersymmetry.

While not immediately relevant for the IR Schwinger Dyson equations and the $U(1)_F$ mode, we will briefly discuss vector multiplets and $U(1)$ gauge transformations in superspace. A vector superfield is a real superfield with the expansion
\begin{align}
    V(\tau, \theta, \bar{\theta}) = C(\tau) + \sqrt{2} \,\theta \, \lambda(\tau) - \sqrt{2} \, \bar{\theta} \, \bar{\lambda}(\tau) - \theta \bar{\theta} A(\tau) \, , \qquad V^\dagger = V \, .
\end{align}
Reality ensures the $\lambda$ are conjugates of each other and $C, A$ are real; this condition is chosen so that $e^{iV}$ is unitary. A supergauge transformation is represented by two bosonic chiral superfields $\Phi(\tau, \theta, \bar{\theta})$ and $\bar{\Phi}(\tau, \theta, \bar{\theta})$:
\begin{align}
\label{eq:FlavorGaugeTransf}
    V \rightarrow V + \bar{\Phi} - \Phi  \, , \qquad \Phi(\tau, \theta, \bar{\theta}) = \phi(\tau) + \sqrt{2}\, \theta \, \xi(\tau) + \theta \bar{\theta} \partial_\tau \phi(\tau) \, , 
\end{align}
The final term in the superfield identifies one real degree of freedom of $\phi$ with the usual gauge parameter. While we will not explore this direction further, it is interesting to note that this vector superfield would be required to introduce a supersymmetric generalization of the background chemical potential $\mu$ for the flavor symmetry.

A naive way to implement a local flavor symmetry in superspace would be to perform bi-local transformations of the form $\mathcal{G}(Z_1, Z_2) \rightarrow e^{i V(Z_1)} e^{-i V(Z_2)}\mathcal{G}(Z_1, Z_2)$ on the Schwinger Dyson equations \eqref{eq:NewSD1} and \eqref{eq:NewSD2}. However, this transformation does not take chiral superfields to chiral superfields, so the correct prescription to perform bi-local independent chiral and anti-chiral gauge transformations with the correct flavor charges using \eqref{eq:FlavorGaugeTransf}:
\begin{align}
\mathcal{G}_{\Psi\Psi}(Z_1,Z_2) &\to e^{-i \bar{\Phi}(Z_1)} e^{i \Phi(Z_2)} \mathcal{G}_{\Psi\Psi}(Z_1,Z_2) \, ,\\
\mathcal{G}_{XX}(Z_1,Z_2) &\to  e^{i (q-1) \bar{\Phi}(Z_1)} e^{-i (q-1) \Phi(Z_2)} \mathcal{G}_{XX}(Z_1,Z_2)
\end{align}
This leaves
\begin{align}
 e^{i  \left (-\bar{\Phi}(Z_1) + \bar{\Phi}(Z_3) \right )}  &\int \! \! dZ_2 \, \mathcal{G}_{\Psi\Psi}(Z_1,Z_2) [ J(q-1) \mathcal{G}_{\Psi\Psi}(Z_3,Z_2)^{q-2}\mathcal{G}_{XX}(Z_3,Z_2)] = 2\delta(\bar{Z}_1-\bar{Z}_3) \, ,\\
  e^{i (q-1) \left (\bar{\Phi}(Z_1) - \bar{\Phi}(Z_3) \right )} &\int \! \! dZ_2 \, \mathcal{G}_{XX}(Z_1,Z_2) [J \mathcal{G}_{\Psi\Psi}(Z_3,Z_2)^{q-1}] = 2\delta(\bar{Z}_1-\bar{Z}_3) \, ,
\end{align}
which is a symmetry of the equations of motion under the support of the supersymmetric $\delta$-function on the right hand side. Actually, the holomorphic dependence on $\Phi$ cancels inside the integral, and the resulting equations appear to have a full complex mode $\bar{\Phi}$ of transformations. This was already anticipated from \eqref{eq:NewLambdaSym1} which was the global scaling symmetry of the conformal answer. 

The above analysis indicates there are 2 bosonic and 2 fermionic flavor reparametrizations, but it is only the real part of the field $\phi(\tau)$ which leads to a compact $U(1)$ compatible with the UV picture. The imaginary part leads to a scale symmetry which we believe is not associated to any physical IR mode. It is less clear, however, that one fermionic mode should be physical and the other should be unphysical. This is because even if we take $\phi(\tau)$ to be purely real, the complex supersymmetry generators produce both $\xi$ and $\bar{\xi}$ fields, schematically:
\begin{align}
    \begin{matrix}
    \, & \, & \phi \in \mathbb{R} & \, & \, \\
    \, & \overset{\mathcal{Q}}{\swarrow} & \! \! \! \! \! & \overset{\bar{\mathcal{Q}}}{\searrow} & \, \\
        \xi & \, & \, & \, & \bar{\xi} \\
            \, & \overset{\bar{\mathcal{Q}}}{\searrow} & \, & \overset{\mathcal{Q}}{\swarrow} & \, \\
    \, & \, & \partial_\tau \phi & \, & \, \\
    \end{matrix}
\end{align}
We will see in the bilinear spectrum which modes actually appear in the infrared, where the global supersymmetry transformations in the diagram would be replaced with the local super-reparametrizations. 

Understanding now that there is one physical local $U(1)$ symmetry of the Schwinger-Dyson equations, we expect the breaking of this $U(1)$ leads to a new dynamical IR mode in addition to the $\mathcal{N}=2$ super Schwarzian. We already reviewed the analogous IR action in Eq.~\eqref{eq:SchwarzianU(1)} in which we wrote down the bosonic part of the $\mathcal{N}=2$ Schwarzian; the $U(1)_R$ breaking boson becomes a particle moving on the $U(1)$ group manifold (with radius given in terms of the $q$ parameter). A similar effective action appears for the $U(1)_F$ mode, but there are some important differences. First, because the $U(1)_R$ is part of the super-reparametrizations, it appears in the same multiplet as the Schwarzian. In contrast, the global flavor symmetry commutes with supersymmetry, so we expect there to be a multiplet of modes in correspondence with \eqref{eq:FlavorGaugeTransf}. Because the group element can be written in superspace, the corresponding multiplet should describe a superparticle moving on the $U(1)_F$ group manifold. While we will not derive the effective action, the situation here is similar to that in \cite{Narayan:2017hvh}. One might further guess that the collective AdS$_2$ description would include $\mathcal{N}=2$ JT gravity coupled to $U(1)_F$ $\mathcal{N}=2$ BF theory, but we leave this question for future work. 

\subsubsection*{The Zero Temperature Entropy}
We now repeat the procedure in the previous section to compute the temperature independent contribution to the partition function
\beq
\partial_q \frac{\log Z}{N} = \frac{J\beta}{2} \int G_{\psi\psi}(\tau)^{q-1} G_{b_\chi b_\chi} (\tau) \log G_{\psi\psi}(\tau) + G_{\psi\psi}(\tau)^{q-2} G_{b_\psi b_\psi}(\tau) G_{\chi\chi}(\tau)(1+(q-1)\log G_{\psi\psi}(\tau)).\nonumber
\eeq
We can evaluate the right hand side using the conformal solution. This is divergent but only the temperature independent piece is independent of the UV behavior. The answer is given by 
\beq
\partial_q \frac{\log Z}{N} = \# \beta  +   \Delta_\psi \pi^2 J (g_{\psi\psi}^{q-1}g_{b_\chi b_\chi}+(q-1)g_{\psi\psi}^{q-2}g_{\chi\chi} g_{b_\psi b_\psi})  + \mathcal{O}(\beta^{-1}),
\eeq
where again only the temperature independence piece is insensitive to the UV behavior. Since this calculation involves only the IR behavior of the action and the solution, the expression depends only on the combination that we can determine without information about the UV. Using equations \eqref{eq:SDprefactor1} to \eqref{eq:SDprefactor4}, we obtain for the temperature independent piece 
\beq
\partial_q \frac{\log Z}{N} =    \pi \Delta_\psi \left(\frac{(1-2\Delta_\chi)\sin 2\pi \Delta_\chi}{\cos 2 \pi \Delta_\chi + \cosh 2 \pi \mathcal{E}_\chi } +\frac{(2(q-1)\Delta_\psi-1)\sin 2\pi(q-1) \Delta_\psi}{\cos 2 \pi (q-1)\Delta_\psi - \cosh 2 \pi (q-1)\mathcal{E}_\psi } \right)  + \ldots ,
\eeq
If we restrict to supersymmetric configurations this simplifies to
\beq
\partial_q \frac{\log Z}{N} = \frac{\pi \Delta_\psi \sin 2 \pi (q-1)\Delta_\psi}{\cosh 2 \pi (q-1) \mathcal{E}_\psi - \cos 2 \pi (q-1) \Delta_\psi}  + \ldots ,
\eeq
Finally we note that if we define the zero temperature entropy as the contribution for $\mathcal{E}_\psi,\mathcal{E}_\chi \to 0$ then we obtain 
\beq
\frac{dS_0}{dq}=N \pi \Delta_\psi \cot \left(\pi(q-1) \Delta_\psi\right).
\eeq
When evaluated at $\mathcal{E}_\psi = \mathcal{E}_\chi = 0$, the equation for $\Delta_\psi$ obtained from the Schwinger-Dyson equation becomes 
\beq
\tan \pi \Delta_\psi = (q-1) \cot \pi (q-1) \Delta_\psi
\eeq
As anticipated above, this is exactly the same as what we found from the index. The equation for $dS_0/dq$ and the equation for $\Delta_\psi$ matches precisely with the expression \eqref{eq:index} for the index at fixed charge, after identifying $y_c \to 2 \pi \Delta_\psi$. This match can be understood as a manifestation of I-maximization in the context of quantum mechanical systems with approximate superconformal symmetry \cite{IMAX}.

\subsection{Breakdown of conformal ansatz} 

In this section we analyze what happens when we turn on the charge by increasing the spectral asymmetry. As we found in the original $\mathcal{N}=2$ SYK model, for large enough spectral asymmetry the conformal ansatz breaks down. We first analyze the supersymmetric solutions parametrized by a single variable, and then move on to the general case.

\paragraph{Supersymmetric Solution} The physics of the supersymmetric solution is very different from the case studied in section \ref{sec:N2SYKFu}. To begin with, there is a one-parameter family of solutions with varying scaling dimensions that are supersymmetric. Second, the coefficients in front of the Greens functions are also not completely determined. Using $G_{b_Ab_A}\sim \partial_\tau G_{AA}$ we can deduce that $g_{b_A b_A} = 2 \Delta_A g_{AA}$ for $A=\psi,\chi$ but now the Schwinger-Dyson equations in the IR only fix the combination $g_{\psi\psi}^{q-1} ~g_{\chi\chi}$, while $g_{\psi\psi}$ and $g_{\chi\chi}$ cannot be independently determined without incorporating the UV behavior. 
\begin{figure}[t!]
    \centering
    \includegraphics[scale=0.25]{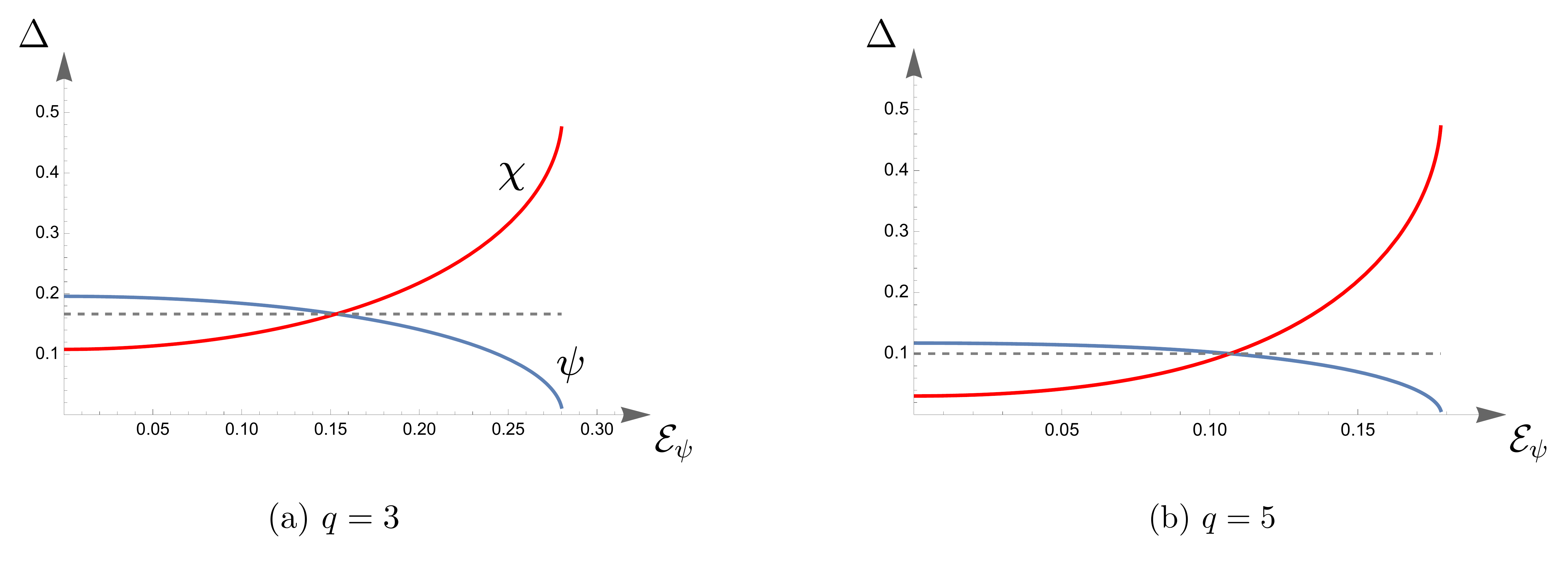}
    \caption{Fermion scaling dimensions of the supersymmetric solutions, as a function of the spectral asymmetry of $\psi$ fermion $\mathcal{E}_\psi$, for (a) $q=3$ and (b) $q=5$. We see at a finite critical value of $\mathcal{E}_\psi$ the solution becomes singular with $\Delta_\psi \to 0$ while $\Delta_\chi \to 1/2$. This point corresponds to maximal $\chi$ charge. The two curves meet at the dashed line $\Delta = 1/2q$ for a special value of $\mathcal{E}_\psi$.}
    \label{fig:SusyMulFer}
\end{figure}

Third, we find a potential breakdown of the conformal ansatz even within the supersymmetric solution. Solving the equation above numerically we find the result shown in figure \ref{fig:SusyMulFer}. There is again a critical spectral asymmetry $|\mathcal{E}_\psi| = \mathcal{E}_{\rm critical}$ such that for $|\mathcal{E}_\psi| > \mathcal{E}_{\rm critical}$ the fermion scaling dimensions are either outside the unitarity bound or complex (the other potential is determined by supersymmetry from $ \mathcal{E}_{\rm critical}$). The critical value is implicitly given by
\beq
\frac{\sinh \pi (q-1) \mathcal{E}_{\rm critical} }{ \cosh \pi \mathcal{E}_{\rm critical}} = (q-1),
\eeq
which is exactly the same as the value found in the previous section for the model defined in \cite{Fu:2016vas}. As we see in the figure above, this value decreases with increasing $q$. 

To determine whether this phase transition is physical or not we need to compute the charges and verify they are not maximal. We can use the Luttinger-Ward relation derived in the previous section to compute both $Q_\psi/N$ and $Q_\chi/N$. Each fermion charge includes a contribution from their respective bosonic partner, as explained in the previous section. The answer is given by 
\bea
Q_\psi/N &=& \mathfrak{q}_f(\Delta_\psi,\mathcal{E}_\psi) + (q-2)\mathfrak{q}_b(\Delta_{b_\psi},\mathcal{E}_{b_\psi}) +(q-1)\mathfrak{q}_b(\Delta_{b_\chi},\mathcal{E}_{b_\chi}),\label{eqnLuttNew1}\\
Q_\chi/N &=& \mathfrak{q}_f ( \Delta_\chi, \mathcal{E}_\chi) + \mathfrak{q}_b(\Delta_{b_\psi},\mathcal{E}_{b_\psi}),\label{eqnLuttNew2}
\ea
in terms of the functions defined in \eqref{eqn:LuttQf} and \eqref{eqn:LuttQb}. Using these expressions for the charges we can ask now whether the critical spectral asymmetry corresponds to an instability or not. The first observation is that $\mathfrak{q}_f(\Delta=1/2,\mathcal{E})=-1/2$ and $\mathfrak{q}_b(\Delta=1/2,\mathcal{E})=0$ regardless of $\mathcal{E}$. At the critical $\mathcal{E}=\mathcal{E}_{\rm critical}$, the scaling dimension of $\chi$ and $b_\psi$ take precisely these values $\Delta_\chi (\mathcal{E}_{\rm critical})=1/2$ and $\Delta_{b_\psi}(\mathcal{E}_{\rm critical})$. For these reasons we obtain $|Q_\chi| (\mathcal{E}_{\rm critical}) = N/2$. Since both fermion charges are bounded $0\leq |Q_\psi|, |Q_\chi|\leq N/2$ this means that there is no instability in the range. 
The previous analysis shows that in the canonical ensemble there is no phase transition since we can turn on $Q_\chi$ and $Q_\psi$ in a supersymmetric way satisfying the constraint imposed by the conformal solution. Nevertheless there can be other gapped solutions that have support in different curves in the $(Q_\psi,Q_\chi)$ plane. Then, in the grand canonical ensemble there might be phase transitions between these two sets of solutions. We expect this to be so, since at $\mathcal{E}_{\text{critical}}$, while the $Q_\chi$ is saturated, the flavor charge $Q_F$ is not. For example, for $q=3,$ at critical spectral asymmetry, $Q_F/N=\sqrt{2}<\frac{3}{2}.$ We leave a more detailed study of these possible supersymmetric phase transitions for future work.  

Finally, we observe there is always a supersymmetry preserving spectral asymmetry such that a new degeneracy appears for the two species: $\Delta_\psi = \Delta_\chi = 1/2q$. The special value of $\mathcal{E}_\psi \equiv \mathcal{E}_{\psi}^*$ is determined through the following equation 
\beq
\label{Eq:enhancepoint}
\frac{\cos \frac{\pi}{q} + \cosh 2 \pi (q-1)\mathcal{E}_\psi^*}{\cos \frac{\pi}{q} + \cosh 2 \pi \mathcal{E}_\psi^*} =(q-1)
\eeq
For example in the specific case $q=3$ gives $\mathcal{E}_\psi^* =\frac{1}{2\pi}\cosh^{-1}(3/2)=0.153..$. We can see this is consistent with the result shown in figure \ref{fig:SusyMulFer}.

\paragraph{General Behavior} In general the solution depends both on $\mathcal{E}_\psi$ and $\mathcal{E}_\chi$ independently. The behavior is similar to the one found for the supersymmetric case, when the spectral asymmetries become too large there is a breakdown of the conformal ansatz. Using the equations above, one can determine the precise region in the $(\mathcal{E}_\psi,\mathcal{E}_\chi)$ plane where the emergent conformal symmetry breaks down, although we will not attempt to do it here.

\paragraph{Non-conformal solution}
We also propose the presence of gapped exponentially decaying solutions analogous to the ones discussed in section \ref{sec:N2SYKFu} governing the low entropy phase. We can see this explicitly at zero-temperature finding the solution analogous to \eqref{eq:decaying_sol}. We find 
\begin{equation}\label{eq:non_conformal}
    G_{\psi\psi}(\tau)=e^{-\mu \tau}\Theta\left(\tau\right) \, , \, \, \,  G_{\chi\chi}(\tau)=-e^{\left(2\mu +J\right)\tau}\Theta\left(-\tau\right), 
\end{equation}
\begin{equation}
    G_{b_\psi b_\psi}\left(\tau\right)=-\delta\left(\tau\right) \, , \, \, \,  G_{b_\chi b_\chi}\left(\tau\right)=-\delta\left(\tau\right)+J e^{(2\mu+J)\tau}\Theta\left(-\tau\right). 
\end{equation}
We verified that they satisfy the full Dyson Schwinger equations \eqref{eq:SDsaddle_generalq2}-\eqref{eq:SDsaddle_generalq5}. 
We note that in contrast to \ref{eq:decaying_sol}, where the exponentially decaying solutions only exist for $\mu>\mu_c>0,$these solutions exist for any $\mu>0.$  We are able to verify this numerically.

\subsection{Operator spectrum}
\begin{figure}[t!]
    \centering
    \includegraphics[width=0.495\textwidth]{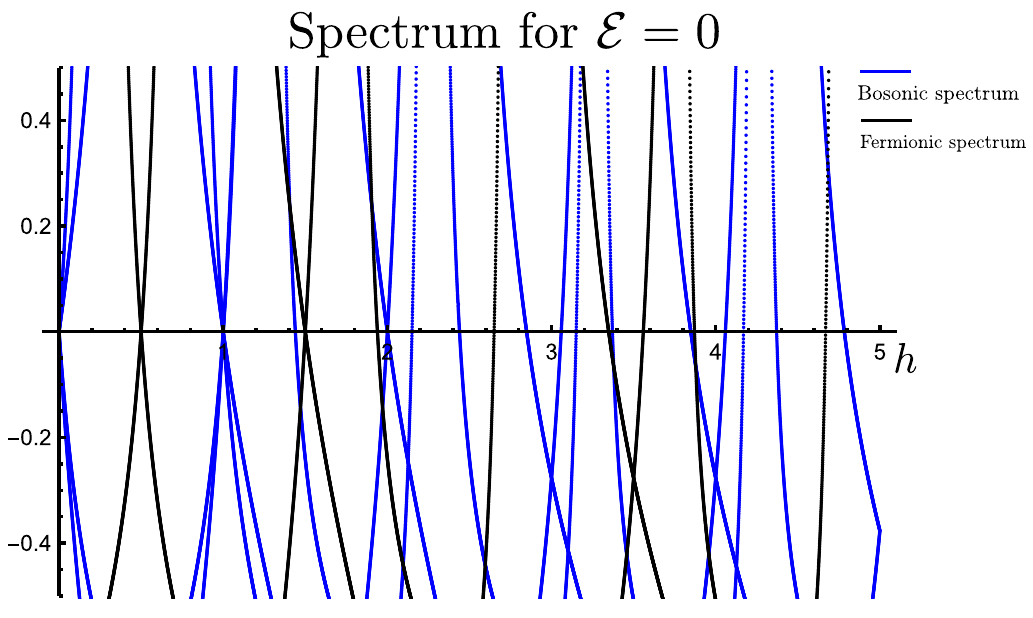}
    \includegraphics[width=0.495\textwidth]{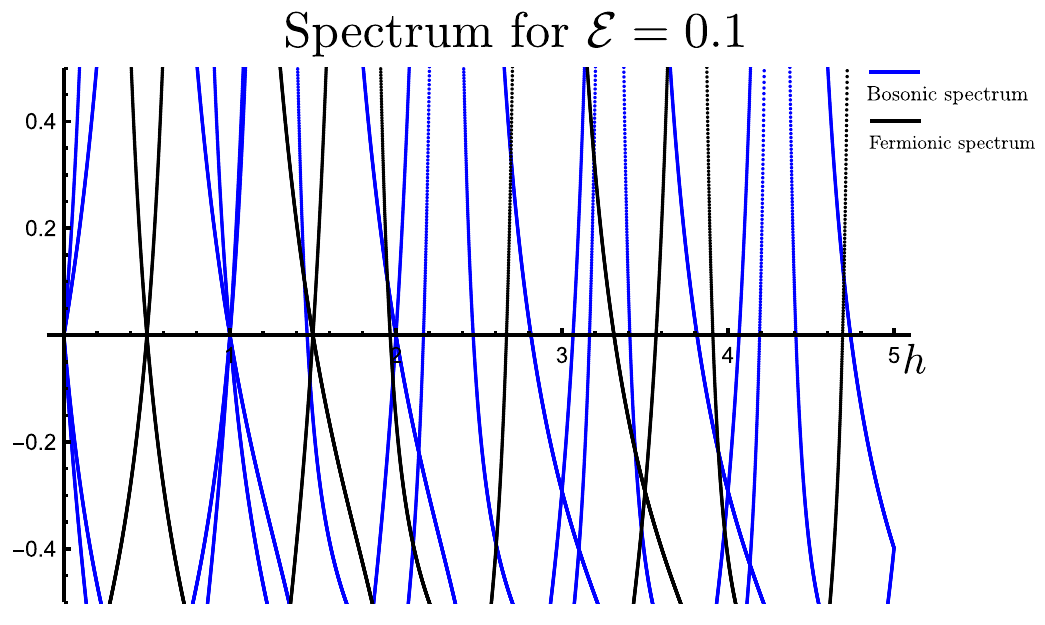}
    \caption{ The bilinear spectrum for $U(1)$ neutral sectors for $\mathcal{E}=0$ and $0.1$ where each intersection with the horizontal axis signifies an operator with dimension corresponding to the location of the intersection. The blue curve is bosonic and the black curve is fermionic. All fermionic operators are in fact doubly degenerate due to $K_f$ and $\bar{K}_f$ having identical spectrum. In addition, the presence of pairs of lines comes from the spurious doubling of the spectrum due to unphysical local symmetries. Accounting for the unphysical modes, the $\mathcal{E}=0$ spectrum possesses the $\mathcal{N}=2$ Schwarzian multiplet with two $h=3/2$ modes. Turning on the chemical potential leads to an IR theory with spontaneously broken supersymmetry. While the spectrum still organizes into multiplets, the $3/2$ modes are no longer protected.}
    \label{fig:KernelGeneral_nu}
\end{figure}

\begin{figure}[t!]
    \centering
    \includegraphics[width=0.495\textwidth]{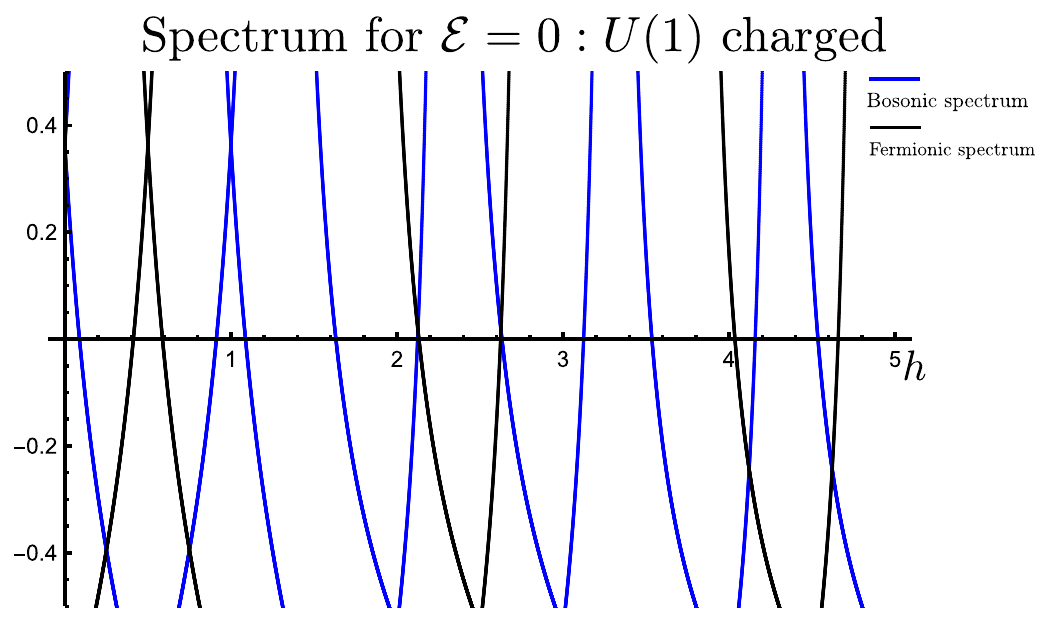}
    \includegraphics[width=0.495\textwidth]{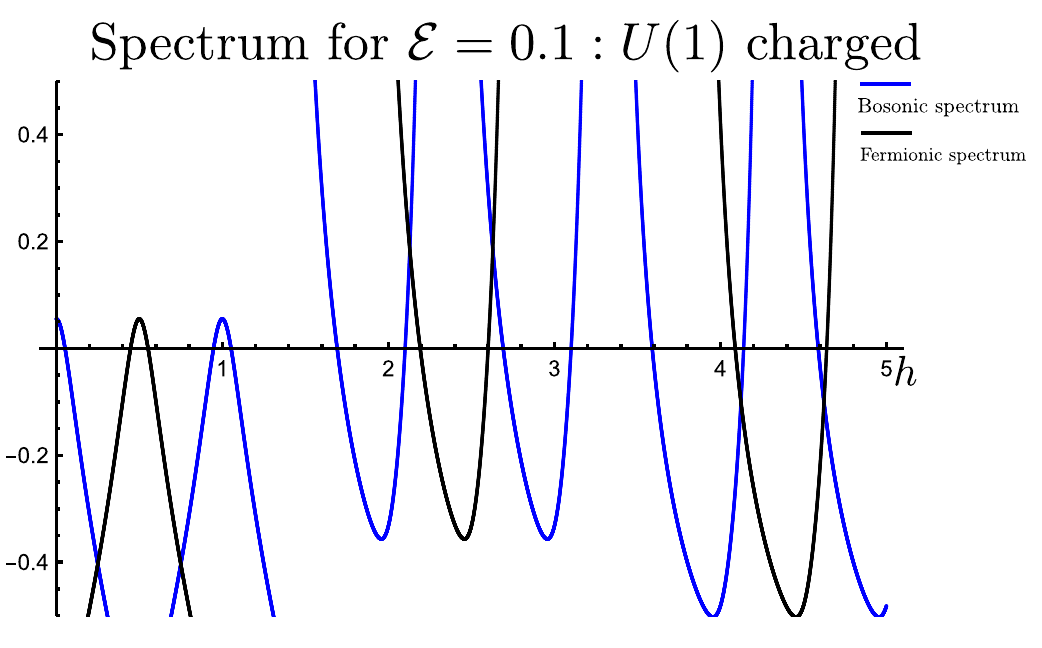}
    \caption{ Spectrum of the $U(1)$ charged sector at distinct values of $\mathcal{E}.$}
    \label{fig:KernelGeneral_nu_enhance}
\end{figure}

\begin{figure}[t!]
    \centering
    \includegraphics[width=0.7\textwidth]{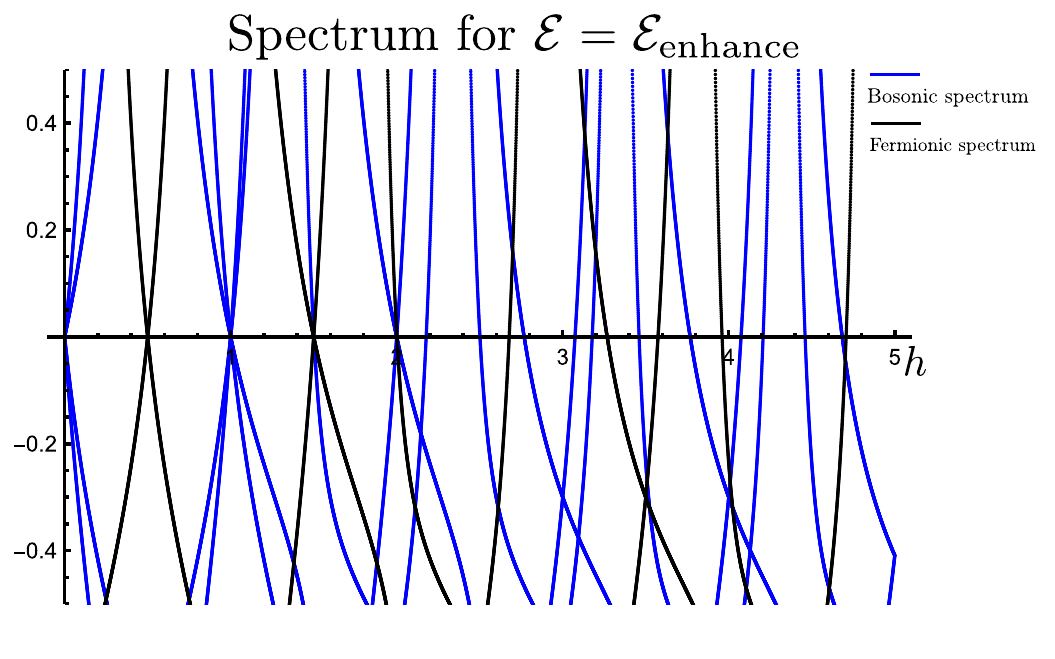}
    
    \vspace{5mm}
    
    \includegraphics[width=0.7 \textwidth]{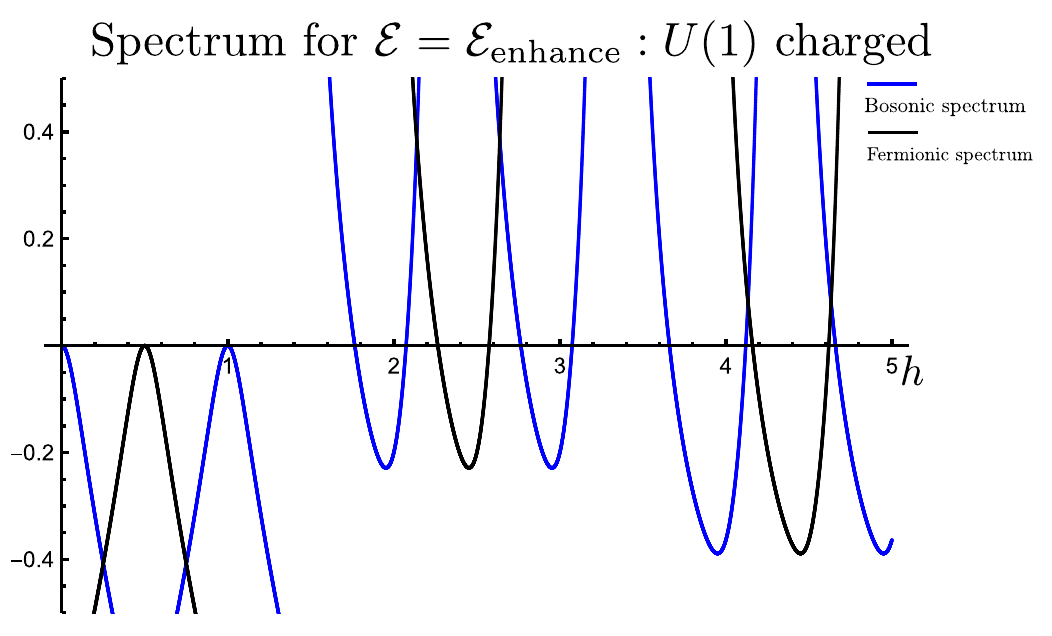}
    \caption{ Top: the bilinear spectrum for $U(1)$ neutral sectors at the enhancement point, $\mathcal{E}_{\text{enhance}}=\frac{1}{2\pi} \cosh^{-1}{\frac{3}{2}}.$ Bottom: the bilinear spectrum for $U(1)$ charged. We see that at $\mathcal{E}=\mathcal{E}_{\text{enhance}},$ there are additional double root at $h=1.$ Here, for each charged operator $\mathcal{O},$ $\bar{\mathcal{O}}$ has the same dimension. There are therefore 4 additional $h=1$ modes at the enhancement point.}
    \label{fig:KernelGeneral_nu_enhance2}
\end{figure}

In this section we will find the spectrum of bilinear operators of the model (\ref{eq:NuQ}). The calculation is similar to section \ref{sec:FuOpSpe} so we will be brief. For simplicity we specialize to the case of $q=3.$ The new feature of the model (\ref{eq:NuQ}) is that we need to define a mixed super correlator between two flavors: 
\begin{equation}
\begin{split}
    &\mathcal{G}_{\Psi X}(Z_1, Z_2)=\frac{1}{N}\langle \bar{\Psi}_i(Z_1) X^i(Z_2)\rangle=G_{\psi\chi}(t_1-t_2-\theta_1\bar{\theta}_1-\theta_2\bar{\theta}_2)+\sqrt{2}\bar{\theta}_1G_{b_\psi \chi}(t_1-t_2-\theta_2\bar{\theta}_2)\\&-\sqrt{2}\theta_2 G_{\psi b_\chi}(t_1-t_2-\theta_1\bar{\theta}_1)+2\bar{\theta}_1\theta_2 G_{b_\psi b_\chi}(t_1-t_2).
\end{split}
\end{equation}
Although the mixed super correlator is set to zero in the conformal limit, the variations can be non-trivial. For convenience we will use the correlator with subscript $\mathcal{G}_{AB}$ to specify the supercorrelator, where $A,B=\Psi$ or $X.$ In the conformal limit, the full equations of motions in superspace (without assuming $G_{\psi\chi}$ is zero) are given by 
\begin{equation}
    \mathcal{G}_{AB}\star\left(\Sigma_{BC}\right)^T=\delta_{AC}\delta(\bar{Z}_1-\bar{Z}_3)\, , \, \, \, \left(\Sigma_{AB}\right)^T\bar{\star}\mathcal{G}_{BC}=\delta_{AC}\delta(Z_1-Z_3),
\end{equation}
\begin{equation}
    \Sigma_{\Psi\Psi}=J \left(\mathcal{G}_{\Psi\Psi}\mathcal{G}_{XX}-\mathcal{G}_{\Psi X}\mathcal{G}_{X\Psi}\right) \, , \, \, \, \Sigma_{XX}=\frac{J}{2} \mathcal{G}_{\Psi\Psi}^2 \, ,\, \, \, \Sigma_{\Psi X}=-J \mathcal{G}_{X\Psi}\mathcal{G}_{\Psi\Psi},
\end{equation}
where the repeated subscripts are summed over. As in the previous model, we vary the action, perform the convolution, and evaluate on the conformal solution $\mathcal{G}_{AB}^c$; the result we obtain is 
\begin{equation}
    \delta \mathcal{G}_{\Psi\Psi}= J \left(\mathcal{G}^c_{\Psi\Psi} \star \left(\delta \mathcal{G}_{\Psi\Psi} \mathcal{G}_{XX}^c+ \mathcal{G}_{\Psi\Psi}^c \delta \mathcal{G}_{XX}\right)^T\right)\bar{\star}\mathcal{G}^c_{\Psi\Psi},
\end{equation}
\begin{equation}
    \delta \mathcal{G}_{XX}=J\left(\mathcal{G}_{XX}^c\star \left(\mathcal{G}_{\Psi\Psi}^c\delta\mathcal{G}_{\Psi\Psi} \right)^T\right)\bar{\star}\mathcal{G}_{XX}^c,
\end{equation}
\begin{equation}
    \delta \mathcal{G}_{\Psi X}=-J\left(\mathcal{G}^c_{\Psi\Psi}\star \left(\mathcal{G}^c_{\Psi\Psi}\delta\mathcal{G}_{\Psi X} \right)^T\right)\bar{\star}\mathcal{G}_{XX}^c,
\end{equation}
Along the diagonal correlators, we may define a 2 by 2 super-kernel that mixes them(where we drop the ${}^c$ superscript)
\begin{equation}
    K^{\mathcal{N}=2}_{\text{diag}}(Z_1,Z_2, Z_3, Z_4)=J\begin{pmatrix}
    \mathcal{G}_{\Psi\Psi}(Z_1,Z_4)\mathcal{G}_{\Psi\Psi}(Z_3,Z_2) \mathcal{G}_{XX}(Z_3,Z_4)& \mathcal{G}_{\Psi\Psi}(Z_1,Z_4)\mathcal{G}_{\Psi\Psi}(Z_3,Z_2) \mathcal{G}_{\Psi\Psi}(Z_3,Z_4)\\ \mathcal{G}_{XX}(Z_1,Z_4)\mathcal{G}_{XX}(Z_3,Z_2) \mathcal{G}_{\Psi\Psi}(Z_3,Z_4)& 0 
    \end{pmatrix}
\end{equation}
and the off-diagonal super kernel along the directions of the mixed super correlator
\begin{equation}
    K_{\text{off-diag}}^{\mathcal{N}=2}(Z_1,Z_2,Z_3, Z_4)=-J \mathcal{G}_{\Psi\Psi}(Z_1,Z_4) \mathcal{G}_{XX}(Z_3,Z_2)\mathcal{G}_{\Psi\Psi}(Z_3,Z_4).
\end{equation}
For supersymmetric solutions, super correlators only depend on the $SU(1,1|1)$ invariant combination $\tau_1-\tau_2-\theta_1\bar{\theta}_1- \theta_2\bar{\theta}_2- 2\bar{\theta}_1\theta_2.$ We present the bosonic and fermionic bilinear spectrum in figures \ref{fig:KernelGeneral_nu}, \ref{fig:KernelGeneral_nu_enhance} and  \ref{fig:KernelGeneral_nu_enhance2}. 

At $\mathcal{E}=0,$ the leading bosonic operator has dimension $h=1$. In fact, we observe four $h=1$ modes in the bosonic spectrum. Two of them respectively correspond to the $U(1)_R$ symmetry and $U(1)_F$ symmetry. The other two are both spurious modes, corresponding to emergent scaling symmetry in the infrared. This agrees with our analysis of the local symmetries of the model in the previous subsection \ref{ssection:MeanFidlNew}. Based on that reasoning, in the infrared we can take both $U(1)_R$ and $U(1)_F$ parameters to be complex, hence accounting for the 4 degrees of freedom. We note that the two $U(1)_F$ modes group together with two  $h=\frac{1}{2}$ modes and becomes the BPS multiplet $\left(\mathbf{\frac{1}{2}}, \mathbf{1}\right),$ whereas the two $U(1)_R$ modes group together with the $h=\frac{3}{2}$ and $h=2$ modes to form the $\mathcal{N}=2$ Super-Schwarzian multiplet $\left(\mathbf{1},2\times \mathbf{\frac{3}{2}},\mathbf{2}\right)$ and its spurious partner. Hence there are in total four $h=1$ modes and two $h=2$ modes, and four $h=\frac{3}{2}$ modes. 

A new feature of the multi-fermion model is the appearance of bosonic operator in the range $1<h<\frac{3}{2}$. At $\mathcal{E}=0$ we observe one such operator at $h\approx1.43942.$ In presence of such an operator, the $\mathcal{N}=2$ Schwarzian becomes sub-dominant in the infrared. The leading correction becomes a bilocal action that depends on the scaling dimension of such an operator: 
\begin{equation}
    S\sim \int d\tau_1 d\tau_2\left(\frac{f'(\tau_1) f'(\tau_2)}{\left(f(\tau_1)-f(\tau_2)\right)^2}\right)^h
\end{equation}
All modes with dimension in between 1 and $\frac{3}{2}$ leads to a more dominant contribution in the low temperature expansion of free energy : 
\begin{equation}
    -\beta F=-\beta E_0+S_0+\sum_{1<h<\frac{3}{2}} \frac{c_h}{\beta^{2h-2}}+\frac{c}{2\beta}+\dots
\end{equation}
Note in such a case, the infrared physics is still governed by the nearly conformal fixed point, but the conformal breaking effects due to the Schwarzian is subleading. It does not invalidate the solution \eqref{eq: delta_of_eps} but rather makes the infrared physics non-universal. One strategy to remove these extra modes (which we have not pursued further in this work) is to explicitly gauge the $U(1)_F$ in the UV. 

As we increase $0\leq\mathcal{E}_\psi<\mathcal{E}_{\text{critical}},$ since the solutions remain supersymmetric, the $\mathcal{N}=2$ Super-Schwarzian multiplet and its spurious partner remain unmodified. However the mode with dimension between $1<h<\frac{3}{2}$ is shifted. In particular, when $\mathcal{E}_\psi$ is greater than a special value $\mathcal{E}_\psi\approx 0.151391$, its dimension is shifted out of the undesired range, and the Schwarzian becomes dominant in the infrared in this range. This is slightly before the value where $\Delta_\psi= \Delta_\chi=\frac{1}{6}$, which is the value found in \eqref{Eq:enhancepoint}; $\mathcal{E}^*_\psi=\frac{1}{2\pi}\cosh^{-1}(\frac{3}{2})$.

The multi-fermion model has additional channels that are generically charged under both $U(1)$'s, along the direction of $\delta \mathcal{G}_{\Psi X}.$ Note such operators would be linear combinations of bilinears in directions of $\delta G_{\psi\chi}$ and $\delta G_{b_{\psi}b_{\chi}}.$ These operators thus have the form 
\begin{equation}\label{eq:charged_operators}
    \mathcal{J}_n=\bar{\psi}^i\partial_\tau^n\chi^i+ c \bar{b}_{\psi}^i\partial^n_\tau b_{\chi}^i,
\end{equation}
where $c$ is determined by the kernel, and each schematic form $\mathcal{J}_n$ corresponds to two distinct linear combinations. We note that although there is a bilinear operator in the range of $1<h<\frac{3}{2},$ we can not add it to the Lagrangian as it generically carries R-symmetry charge. Thus it does not change the dominance of the $\mathcal{N}=2$ Super-Schwarzian, in contrast to the mode discussed in the previous paragraph.

However, precisely at the special value of $\mathcal{E}_{\!\textrm{ enhance}}=\mathcal{E}^*_\psi=\frac{1}{2\pi}\cosh^{-1}(\frac{3}{2})$ such that $\Delta_{\psi}=\Delta_{\chi}=\frac{1}{6}$, the operators \eqref{eq:charged_operators} in fact have R-charge zero. To see that, note in the infrared, the correct R-charge \eqref{eq:something2} has $\alpha=2\Delta_\psi$ from the analysis leading to \eqref{eq:alphaRmax}, and therefore for these operators, 
\begin{equation}
    Q_R=Q_\chi+ 2 \Delta_\psi Q_F= 1 + \frac{1}{3} \times (-3)=0. 
\end{equation}

At this value of the spectral asymmetry, the leading operators $\mathcal{J}_0$ and their conjugates in the series $\mathcal{J}_n$ have dimension 1, and emerge as additional local symmetries. Therefore we observe a total of four additional $h=1$ modes to emerge, and they are only charged under $U(1)_F$. They are paired with their fermionic partners with $h=\frac{1}{2}$. The spectrum at the enhancement point is shown in figure \ref{fig:KernelGeneral_nu_enhance2}. Assuming two of these modes are physical and two are spurious, it is suggestive that the global symmetry is enhanced to $SU(2)_F$ by the addition of two new (physical) generators. We note the enhancement point is free of any known problem in the bilinear spectrum. We leave a detailed study of this symmetry enhancement and its holographic interpretation for future study.

\section{Discussion and Open Questions}\label{sec:Conclusions}
In this paper we have studied the model of \cite{Fu:2016vas} at non-zero charge and proposed new models of $\mathcal{N}=2$ SYK with interesting features. We conclude with some open questions and future directions. 

A future direction is to develop, for the models of \cite{Fu:2016vas}, a detailed picture for the bulk gravity dual description. In particular we would like to find a good model for the bulk interpretation of the low-entropy phase, with the hope of discovering new features of the instabilities of near extremal black holes in higher dimensions. Moreover, we have given a simple description of this transition in $\mathcal{N}=2$ SYK, since the fundamental SYK fermion becomes unstable. As far as we know, it is an open problem to extend this kind of picture to complex SYK, where the origin of this transition is more mysterious from the perspective of the conformal phase (some comments in this direction are made in \cite{Gu:2019jub} looking at the bilinear spectrum as a function of the charge). 

For the SYK model of \cite{Fu:2016vas} we obtained the $\mathcal{N}=2$ Schwarzian coupling at background zero charge. It would be interesting to compute the Schwarzian coupling as a function of the charge, since the coupling would be independent of the $U(1)_R$ compressibility. Instead, the low energy model would be closer to the one of complex SYK since the fermionic modes of the Super-Schwarzian would become massive. Additionally, an exhaustive analysis of the exact diagonalization of these models would help in understanding the nearly conformal sector of the spectrum, as well as the transition when this picture breaks down. Relatedly, we have not explored exhaustively the phase diagram of the $\mathcal{N}=2$ SYK model with multiple fermions as a function of an arbitrary pair of $U(1)$ charges, and even a numerical solution of the mean field equations are complicated to obtain. 

The model with multiple fermions seems to have a richer phase diagram, including the freedom to tune the charge while preserving supersymmetry. The existence of a special value of the background charge for which $\Delta_\psi = \Delta_\chi$ implies a phase of enhanced symmetry between the two fermion species; we conjecture the theory has an $SU(2)_F$ symmetry. It would be interesting to study the effective theory at this point and understand the implications for the bulk dual, which may be related to a point of enhanced symmetry for the horizon of a rotating near-BPS black hole. We have also studied the R-charge in the infrared, which differs non-trivially from the UV definition. We have found evidence that the infrared (superconformal) R-symmetry extremizes the index, presumably a `nearly conformal' version of the $\mathcal{I}$-maximization principle \cite{Benini:2015eyy,Gauntlett:2019roi,Hosseini:2019ddy,Kim:2019umc}. We will elaborate on this maximization principle in a future work \cite{IMAX}.

Finally we constructed models that realize $\mathcal{N}=2$ Schwarzian theories with fundamental R-charge one. This presents features in the spectrum that are the closest to the $\mathcal{N}=4$ Schwarzian theory describing BPS black holes in flat space \cite{Heydeman:2020hhw}, as explained in the introduction. While the $\mathcal{N}=4$, Schwarzian theory was fully solved in that work, a microscopic UV Lagrangian which realizes this Schwarzian in the infrared is currently unknown. We hope the model introduced in section \ref{sec:N2SYKNew} can help in finding such theories with $\mathcal{N}=4$ supersymmetry when considering multiple fermions transforming in representations of $SU(2)$ without the need of including second order bosons \cite{Anninos:2016szt}. We leave this for future work.

\subsection*{Acknowledgements} 
We thank I. Bah, Y. Gu, I. Klebanov, J. Maldacena, V. Narovlansky, M. Rangamani, S. Sachdev, G. Tarnopolsky, E. Witten for valuable discussions. We especially thank I. Klebanov for initial collaboration. MTH is supported by Princeton University and the Institute for Advanced Study under Grant No. DE-SC0009988 and the Corning Glass Foundation. GJT is supported by the Institute for Advanced Study and the National Science Foundation under Grant No. PHY-1911298, and by the Dipal and Rupal Patel funds. WZ was supported in part by the US NSF under Grants No. PHY-1620059 and PHY-1914860.

\appendix

\section{Luttinger-Ward relation} \label{app:Luttinger}
Here we derive the relation between the charge density of our models and the spectral asymmetry of the IR correlators. We follow the derivation presented in section 2 of \cite{Gu:2019jub}, based on similar ideas considered in Appendix C of \cite{Kitaev:2005hzj}.

We begin by deriving the Luttinger-Ward relation to the model of section \ref{sec:N2SYKFu}, introduced in \cite{Fu:2016vas}. The first step is to define a notion of ``flow'' of the green functions $G$ and $D$, introducing a bilocal conserved current $j(\tau_1,\tau_2)$. Begin by generalizing the UV contributions of the mean field action from $\delta'(\tau_1-\tau_2)-\mu \delta(\tau_1-\tau_2) \to \mu_f(\tau_1,\tau_2)$ for fermions and $\delta(\tau_1-\tau_2) \to \mu_b(\tau_1-\tau_2)$ for bosons, and defining the new self energies $\tilde{\Sigma}_{\psi\psi}(\tau_1,\tau_2) = \Sigma_{\psi\psi}(\tau_1,\tau_2) + \mu_f(\tau_1,\tau_2)$ and $\tilde{\Sigma}_{bb}(\tau_1,\tau_2) = \Sigma_{bb} (\tau_1,\tau_2) + \mu_b(\tau_1,\tau_2)$. Under this transformation we end up with a mean field action
\beq
I[G,\Sigma] = I_{IR}[G,\tilde{\Sigma}] +\int d\tau_1 d\tau_2 \left[ \mu_f(\tau_1,\tau_2) G_{\psi\psi}(\tau_2,\tau_1) + \mu_b(\tau_1,\tau_2) G_{bb}(\tau_2,\tau_1)\right],
\eeq
where $I_{IR}$ is an action such that its equation of motion are exactly the Schwinger-Dyson ones in the IR approximation. Now define the following bilocal current 
\bea
j(\tau_1,\tau_2) &=& \mu(\tau_1,\tau_2)G_{\psi\psi}(\tau_2,\tau_1) + (q-1)\mu_b(\tau_1,\tau_2)G_{bb}(\tau_2,\tau_1) - (1\leftrightarrow 2) ,\\
&=& \tilde{\Sigma}_{\psi\psi}(\tau_1,\tau_2) G_{\psi\psi}(\tau_2,\tau_1) + (q-1) \tilde{\Sigma}_{bb}(\tau_1,\tau_2) G_{bb}(\tau_2,\tau_1) - (1\leftrightarrow 2).
\ea
Now we can exploit the symmetries of the IR equations to say something about this bilocal current. The IR equations coming from $I_{IR}$ are invariant under the local $U(1)$ transformations
\beq
G_{\psi\psi}(\tau_1,\tau_2) \to e^{i \lambda(\tau_1)} e^{-i \lambda(\tau_2) } G_{\psi\psi}(\tau_1,\tau_2),~~~~G_{bb}(\tau_1,\tau_2) \to e^{ i (q-1) \lambda(\tau_1) } e^{- i (q-1) \lambda(\tau_2)} G_{bb}(\tau_1,\tau_2).
\eeq
Imitating the derivation in \cite{Gu:2019jub} we can use this symmetry to show that, evaluated on a classical solution of the mean field action, the bilocal current satisfies the local conservation through the vanishing of $\int_{-\infty}^{+\infty} j(\tau_1,\tau_0) d\tau_1 = 0$. Then we can define the charge as 
\beq
\tilde{Q} = \int_{-\infty}^{\tau_0} d\tau_1 \int_{\tau_0}^\infty d\tau_2 ~j(\tau_1,\tau_2).
\eeq
We can verify using the UV behavior of the Green functions that this coincides with the charge of the $\psi$ fermion when we choose $\mu_f \to \delta'(\tau_1-\tau_2)-\mu \delta(\tau_1-\tau_2)$ and $\mu_b\to \delta(\tau_1-\tau_2) $, then we get
\beq
\tilde{Q} = - \int_{-\infty}^\infty d\tau \tau \left( \delta'(\tau)G_{\psi\psi}(-\tau) + (q-1) \delta(\tau) G_{bb}(-\tau)\right),
\eeq
after some simplification. Following \cite{Gu:2019jub} we can check that this notion of charges matches with the expectation
\bea
\tilde{Q} &=& \frac{G_{\psi\psi}(0^+)+G_{\psi\psi}(0^-)}{2} +{\rm constant},\\
&=& \frac{Q}{N} +{\rm constant}.
\ea
To obtain this expression we defined the value of $G_{\psi\psi}(0)$ as the average and we set $\tau D(\tau)|_{\tau \to 0}$ to a constant to be determined below by consistency. The last step of the calculation is to perform the integral in the IR and match the UV answer above. This was already done in section 2.2.3 of \cite{Gu:2019jub} so we can simply quote the answer for the fermion in equation \eqref{eqn:LuttQf}. The answer for the boson is simply given by a shift $\mathcal{E} \to \mathcal{E}+i/2$ since this removes extra minus signs that appear in fermion correlators. The final answer for the boson contribution is given in equation \eqref{eqn:LuttQb}. This was of deriving the bosonic contribution has an ambiguity from the analytic continuation of the logarithm. We pick a sheet such that for $\mathcal{E}_b=0$ the contribution to the charge vanishes as well. The total answer for the fermion charge is 
\beq\label{eq:LWFuetalmtAppendix}
\frac{Q}{N} = \mathfrak{q}_f(\Delta,\mathcal{E}) + (q-1) \mathfrak{q}_b(\Delta_b,\mathcal{E}_b).
\eeq
This fixes also the constant in the expression for $\tilde{Q}$ above, demanding that when $\mathcal{E}=0$ the total fermion charge $Q$ must vanish. 

Now we can generalize this to the models of section \eqref{sec:N2SYKNew}. In this case we have two $U(1)$ symmetries and therefore we expect two Luttinger-Ward relations relating $\mathcal{E}_\chi$ and $\mathcal{E}_\psi$ to $Q_\chi$ and $Q_\psi$. The two symmetries act on the correlators as 
\bea
&&G_{\psi\psi}(\tau_1,\tau_2) \to e^{i \lambda(\tau_1)} e^{-i \lambda(\tau_2) } G_{\psi\psi}(\tau_1,\tau_2),~~~G_{\chi\chi}(\tau_1,\tau_2) \to  G_{\chi\chi}(\tau_1,\tau_2),\nonumber\\
&&G_{b_\psi b_\psi}(\tau_1,\tau_2) \to e^{i(q-2) \lambda(\tau_1)} e^{-i(q-2) \lambda(\tau_2) } G_{b_\psi b_\psi}(\tau_1,\tau_2),~G_{b_\chi b_\chi}(\tau_1,\tau_2) \to e^{ i (q-1) \lambda(\tau_1) } e^{- i (q-1) \lambda(\tau_2)} G_{b_\chi b_\chi}(\tau_1,\tau_2)\nonumber,
\ea
and the other symmetry is
\bea
&&G_{\psi\psi}(\tau_1,\tau_2) \to G_{\psi\psi}(\tau_1,\tau_2),~~~G_{\chi\chi}(\tau_1,\tau_2) \to e^{i \lambda(\tau_1)} e^{-i \lambda(\tau_2) } G_{\chi\chi}(\tau_1,\tau_2),\nonumber\\
&&G_{b_\psi b_\psi}(\tau_1,\tau_2) \to e^{ i  \lambda(\tau_1) } e^{- i \lambda(\tau_2)} G_{b_\psi b_\psi}(\tau_1,\tau_2),~~~~G_{b_\chi b_\chi}(\tau_1,\tau_2) \to  G_{b_\chi b_\chi}(\tau_1,\tau_2)\nonumber.
\ea
Now it is straightforward to generalize the previous derivation. In order to do this we introduce two bilocal currents $j(\tau_1,\tau_2)$ one for each symmetry and evaluate it in both the UV and IR. The final answer for the two charges is given in equations \eqref{eqnLuttNew1} and \eqref{eqnLuttNew2}. We have fixed ambiguities regarding the UV behavior of the boson two-point function by demanding that at zero spectral asymmetry the charge should vanish, similar to the case in the previous section.

\section{Kernels with multiple fermions in components}
We can decompose the 2 by 2 super-kernel components wise. The fermionic ones give 
\begin{equation}
    \begin{pmatrix}\delta G_{\psi b}\\ \delta G_{\chi B}\end{pmatrix}=
    \begin{pmatrix}2J G_{\psi\psi}(t_{14})G_{b_{\psi}b_{\psi}}(t_{32})G_{\chi\chi}(t_{34}) & 2J G_{\psi\psi}(t_{14})G_{b_{\psi}b_{\psi}}(t_{32})G_{\psi\psi}(t_{34})\\ 2J  G_{\chi\chi}(t_{14})G_{b_{\chi}b_{\chi}}(t_{32})G_{\psi\psi}(t_{34}) & 0\end{pmatrix}\begin{pmatrix}\delta G_{\psi b}\\ \delta G_{\chi B}\end{pmatrix}
\end{equation}
The bosonic components consist of two distinct sectors, the neutral sector and the charged sector. The charged sector can be further decomposed into the bosonic and fermionic parts
\begin{equation}
    \begin{pmatrix}\delta G_{\psi\chi}\\ \delta G_{b_\psi b_\chi}\end{pmatrix}=2J\begin{pmatrix}
    G_{\psi\psi}(t_{14})G_{\chi\chi}(t_{32})G_{b_\psi b_\psi}(t_{34}) & G_{\psi\psi}(t_{14})G_{\chi\chi}(t_{32})G_{\psi \psi}(t_{34})\\ -G_{b_\psi b_\psi}(t_{14})G_{b_\chi b_\chi}(t_{32})G_{\psi \psi}(t_{34})& 0
    \end{pmatrix}\begin{pmatrix}\delta G_{\psi\chi}\\ \delta G_{b_\psi b_\chi}\end{pmatrix}
\end{equation}
\begin{equation}
\delta G_{\psi b_\chi}=-2J G_{\psi\psi}(t_{14})G_{b_\chi b_\chi}(t_{32})G_{\psi\psi}(t_{34})\delta G_{\psi b_\chi},
\end{equation}
\begin{equation}
\delta G_{b_\psi \chi}=-2J G_{b_\psi b_\psi}(t_{14})G_{\chi \chi}(t_{32})G_{\psi\psi}(t_{34})\delta G_{b_\psi \chi},
\end{equation}
Finally the neutral sector for the bosonic correlators is given by 
\begin{equation}
{\tiny    \begin{pmatrix}
    -2J G_{\psi\psi}(t_{14})G_{\psi\psi}(t_{32})G_{b_\chi b_\chi}(t_{34}) & 
    -2J G_{\psi\psi}(t_{14})G_{\psi\psi}(t_{32})G_{b_\psi b_\psi}(t_{34}) &
    -2J G_{\psi\psi}(t_{14})G_{\psi\psi}(t_{32})G_{\chi \chi}(t_{34})    & 
    -2J G_{\psi\psi}(t_{14})G_{\psi\psi}(t_{32})G_{\psi \psi}(t_{34})\\
    -2J G_{\chi\chi}(t_{14})G_{\chi\chi}(t_{32})G_{b_\psi b_\psi}(t_{34})& 
    0 & 
    -2J G_{\chi\chi}(t_{14})G_{\psi\psi}(t_{32})G_{\chi \chi}(t_{34}) & 
    0\\
    2J G_{b_\psi b_\psi}(t_{14})G_{b_\psi b_\psi}(t_{32})G_{\chi \chi}(t_{34}) &
    2J G_{b_\psi b_\psi}(t_{14})G_{b_\psi b_\psi}(t_{32})G_{\psi \psi}(t_{34}) &
    0 &
    0\\ 
    2J G_{b_\chi b_\chi}(t_{14})G_{b_\chi b_\chi}(t_{32})G_{\psi \psi}(t_{34})& %
    0&0&0
    \end{pmatrix}}
\end{equation}\normalsize

\bibliographystyle{utphys2}
{\small \bibliography{Biblio}{}}

\end{document}